\documentclass[11pt]{article}
\usepackage[DIV12]{typearea}
\usepackage{color}
\usepackage{amsmath,amssymb}
\usepackage{amsfonts}  
\usepackage{mathrsfs}
\usepackage{graphicx} 
\usepackage{subfigure}
\usepackage{bbm}
\usepackage{longtable}
\usepackage[small]{caption2} 
\usepackage{cite}
\usepackage{pifont}
\usepackage{enumerate} 
\usepackage{pdflscape}
\usepackage{hhline}  
\usepackage{multirow} %
\usepackage{booktabs} 
\usepackage{tabulary}
\usepackage[compat=1.1.0]{tikz-feynman}
\usepackage{nicefrac}
\usepackage{braket}
\usepackage{comment}
\usepackage{xspace}

\newcommand{\I}{\mathrm{i}}

\newcommand{\SO}[1]{\ensuremath{\mathrm{SO}(#1)}}
\newcommand{\SU}[1]{\ensuremath{\mathrm{SU}(#1)}}
\newcommand{\U}[1]{\ensuremath{\mathrm{U}(#1)}}
\newcommand{\Z}[1]{\ensuremath{\mathbbm{Z}_{#1}}} 
\newcommand{\ZxZ}[2]{\ensuremath{\mathbbm{Z}_{#1}\times\mathbbm{Z}_{#2}}} 

\newcommand{\rep}[1]{\ensuremath{\boldsymbol{#1}}}
\newcommand{\crep}[1]{\ensuremath{\boldsymbol{\overline{#1}}}}
\newcommand{\maG}{\ensuremath{\mathcal{G}} }

\newcommand{\CP}{\ensuremath{\mathcal{CP}}\xspace}
\newcommand{\x}{\ensuremath{\times}}
\newcommand{\vev}[1]{\ensuremath{\langle{#1}\rangle}}
\newcommand{\id}{\ensuremath{\mathbbm{1}}}
\newcommand{\coloneq}{\ensuremath{\mathrel{\mathop:}=}}

\usepackage{xcolor}
\definecolor{darkgreen}{HTML}{109930}
\definecolor{pink}{rgb}{0.858, 0.188, 0.478}

\newcolumntype{R}[1]{>{\raggedleft\arraybackslash}p{#1}} 

\allowdisplaybreaks

\usepackage[pdftex,hidelinks,colorlinks=true,citecolor=darkgreen]{hyperref}
\hypersetup{
    pdftitle = {Higgs-portal dark matter from non-supersymmetric strings},
    pdfauthor = {Cervantes, Perez-Figueroa, Perez-Martinez, Ramos-Sanchez}
}

\begin{document}

\date{}
\title{
  \vskip 2cm
  {\bf\huge Higgs-portal dark matter from non-supersymmetric strings}\\[0.8cm]
}
\author{
 {\bf\normalsize
  Esau Cervantes$^1$\footnote{\texttt{esau.cervantes@ncbj.gov.pl}},
  Omar P\'erez--Figueroa$^2$\footnote{\texttt{omar\_perfig@ciencias.unam.mx}}, 
  Ricardo P\'erez--Mart\'inez$^3$\footnote{\texttt{ricardo.perezmartinez@uadec.edu.mx}}}\\
 {\bf\normalsize
  and
  Sa\'ul~Ramos--S\'anchez$^2$\footnote{\texttt{ramos@fisica.unam.mx}}
 }\\[5mm]
 {\it\normalsize $^1$National Centre for Nuclear Research, Pasteura 7, 02-093 Warsaw, Poland,}\\[5mm]
 {\it\normalsize $^2$Instituto de F\'isica, Universidad Nacional Aut\'onoma de M\'exico,}\\
 {\it\normalsize POB 20-364, Cd.Mx. 01000, M\'exico}\\[5mm]
 {\it\normalsize $^3$Facultad de Ciencias F\'isico-Matem\'aticas, Universidad Aut\'onoma de Coahuila,}\\
 {\it\normalsize Edificio A, Unidad Camporredondo, 25000, Saltillo, Coahuila, M\'exico}
}

\maketitle 

\thispagestyle{empty}

\vskip 1cm
\begin{abstract}
Large classes of non-supersymmetric string models equipped with standard-model 
features have been constructed, but very little of their phenomenology
is known. Interestingly, their spectra exhibit scalar fields whose only
couplings to observed particles is through a multi-Higgs sector. On the other
hand, bottom-up models with Higgs portals offer still an acceptable framework 
for dark matter. We explore realizations of such Higgs portals in promising 
heterotic orbifold models without supersymmetry. We find that a sample model 
includes Higgs vacua that are stable at one-loop, in which the Higgs sector is
compatible with particle-physics observations and a scalar can account
for the measured dark matter abundance. In such vacua, interesting constraints 
on the masses of the dark matter candidate and the heavy Higgs sector are uncovered.
These compelling results are not limited to string models, as they can be embedded
in similarly motivated bottom-up schemes.
\end{abstract}

\clearpage
\newpage

\section{Introduction}

In the pursuit for connecting string theory with observations, the absence of
low-energy supersymmetry (SUSY) motivates the exploration of four-dimensional (4D)
non-SUSY string constructions. These can arise from the ten-dimensional (10D) 
tachyon-free heterotic string theory with no spacetime SUSY and gauge group 
\SO{16}\x\SO{16}~\cite{Gross:1984dd, Dixon:1986iz, AlvarezGaume:1986jb}.
Efforts along these lines, include free fermionic constructions~\cite{Florakis:2021bws,Ashfaque:2015vta,Faraggi:2020fwg,Faraggi:2020wld,Faraggi:2020hpy},
Calabi-Yau~\cite{Blaszczyk:2015zta} and coordinate-dependent 
compactifications~\cite{Abel:2015oxa,Abel:2017vos},
non-SUSY vacua of Gepner models~\cite{Aoyama:2020aaw,Aoyama:2021kqa},
and Abelian orbifold compactifications of the heterotic string~\cite{Dixon:1985jw,Dixon:1986jc,Ibanez:1987xa,Ibanez:1986tp}.

More generally, heterotic orbifolds have shown to lead to phenomenologically
viable models both with SUSY~\cite{Ibanez:1987sn,Kobayashi:2004ya,Buchmuller:2005jr,Forste:2006wq,Lebedev:2006kn,Kim:2007mt,Lebedev:2008un, MayorgaPena:2012ifg,Nilles:2014owa,Olguin-Trejo:2018wpw,Baur:2019iai,Parr:2020oar,Nilles:2020gvu,Baur:2022hma}, 
and without SUSY~\cite{Blaszczyk:2014qoa}. In these works, it is shown that 
non-SUSY heterotic orbifolds yield large classes of models with the gauge group 
and matter content of the standard model (SM), including tachyon freedom at 
perturbative level~\cite{Perez-Martinez:2021zjj}. (Open questions in this 
and similar kinds of models have been discussed 
in~\cite{Abel:2015oxa,Abel:2017vos,Abel:2017rch,Satoh:2015nlc,GrootNibbelink:2017luf,Itoyama:2019yst,Aoyama:2020aaw,Aoyama:2021kqa,Koga:2022qch}.) 
This naturally motivates phenomenological studies of non-SUSY heterotic 
orbifold models. One finds, among other features, that they exhibit couplings 
between multiple Higgs fields and some complex scalars that are neutral under 
the SM gauge group. Interestingly, these ingredients are the cornerstone of 
so-called {\it Higgs portals}~\cite{Patt:2006fw}.

Higgs portals are extensions of the SM in which one additional field can account for 
the origin of dark matter (DM)~\cite{Kolb:2017jvz,Cosme:2018nly,Arcadi:2019lka,Arcadi:2021mag,Azatov:2022tii} 
and alleviate the stability of the Higgs vacuum (see e.g.~\cite{Hiller:2022rla}), 
among other cosmological features (see e.g.~\cite{Lebedev:2021xey} for a review 
of Higgs portals in cosmology). It is known that these scenarios 
are viable candidates to describe the dynamics of DM and observable physics. Moreover, 
precision or alternative analyses of simple Higgs portals for DM seem to suggest 
that the strongest observational constraints might be relaxed by additional 
considerations~\cite{Arroyo:2016wal,Casas:2017jjg,Hardy:2018bph}. In this sense, stringent observational
bounds on these frameworks are untightened in the presence of two or more Higgs 
doublets~\cite{Chang:2017gla,Bell:2017rgi}. This implies that non-SUSY heterotic orbifolds
might be naturally equipped with successful DM phenomenology based on Higgs portals.

From a bottom-up perspective, multi-Higgs models are known to exhibit phenomenologically 
appealing features for particle physics too. In particular, two-Higgs-doublet models 
(2HDM) exhibit many interesting properties~\cite{Branco:2011iw} leading to successful 
compatibility with particle physics, see e.g.~\cite{Chowdhury:2015yja}. Some 2HDM with
Higgs-portal DM have been studied~\cite{Drozd:2014yla,Han:2020ekm}, where a scalar
DM candidate seems to be preferred~\cite{Cai:2013zga} and some extant particle-physics 
anomalies can be explained~\cite{Bandyopadhyay:2017tlq}. Introducing a complex scalar 
in this case leads to Higgs portals satisfying DM, flavor and collider
constraints~\cite{Dutta:2022xbd}. The complexity of the model and its phenomenology 
grows very fast in the general case of $N>2$ Higgs doublets, but it can still yield 
results that can be confronted with observations, see e.g.~\cite{Grossman:1994jb,Ivanov:2010ww,Ivanov:2010wz,Keus:2013hya,Bento:2017eti,Ivanov:2017dad}
and references therein.

In this work, we aim at providing a proof of principle of the existence of satisfactory 
Higgs-portal DM models arising from non-SUSY heterotic orbifolds. We shall study whether 
explicit 4D effective field theories arising from \Z2\x\Z4 heterotic orbifolds
endowed with basic properties of the SM, can also yield Higgs-portal scenarios,
where i) the relic density of a DM candidate complies with current observational
bounds~\cite{Planck:2018vyg}, ii) the Higgs vacuum is stable, and iii) it produces
the right values for Higgs mass and vacuum expectation value (VEV). If fulfilled,
these conditions should also lead to some predictions on the mass of the DM candidate
and the possible heavy Higgs spectrum.

The content of this paper is organized as follows. In order to fix our notation and set
the general aspects of our framework, we provide first a brief review of 
the relevant features of orbifold compactifications of the non-SUSY heterotic 
string in section~\ref{sec:non-susy} and Higgs portal DM in section~\ref{sec:portals}.
Then, we study the general features of sample string-derived models with 
Higgs portals and six (section~\ref{sec:sixHiggses}) 
and two (section~\ref{sec:twoHiggses}) Higgs doublets. With the help of
numeric algorithms, we analyze the phenomenology of the sample stringy 2HDM in 
section~\ref{sec:results}, which then leads to our conclusions in 
section~\ref{sec:conclusions}. We devote the appendices to the details of 
the matter spectra of our models, and some relevant computations.

\section{Non-supersymmetric orbifold compactifications}
\label{sec:non-susy}
Let us now spend some words on the framework that represents the source 
of the models we study in this work.

A possible origin of (multi-)Higgs portals for scalar DM is the 10D 
non-supersymmetric  heterotic string with gauge group 
$\SO{16}\x\SO{16}$~\cite{Dixon:1986iz,AlvarezGaume:1986jb} compactified 
on Abelian toroidal orbifolds. Its massless, tachyon and anomaly-free 
spectrum consists of 240 gauge bosons, 256 spinors and 256 cospinors. 
In addition, the massless gravity sector includes the dilaton, the graviton, 
and the Kalb-Ramond field.   

In order to make contact with our 4D Universe, we can compactify six spatial 
dimensions of this theory on a 6D toroidal orbifold. These orbifolds
can be defined as the quotient of a 6D torus divided by a set of its
isometries or, equivalently, as the quotient of $\mathbbm{R}^6$ and 
a space group $\mathbbm{S}$, which is generated by the elements of a 
rotational point group $P\,\subset\,\mathrm{O}(6)$, and some translations $\mu$. 
Although the absence of supersymmetry allows for a wider class 
of orbifold compactifications,\footnote{There are 7103 admissible choices
of $P$, see~\cite[section 5]{Fischer:2012qj}.} for simplicity we consider 
here models arising only from the 138 Abelian geometries used in our recent
work~\cite{Perez-Martinez:2021zjj}, based on the classification of 
ref.~\cite{Fischer:2012qj}. In the notation of this reference, orbifold 
geometries (i.e.~space groups $\mathbbm S$) are labeled by the \Z{N} or 
\Z{N}\x\Z{M} point group followed by two numbers which label a compatible 
toroidal lattice, and a set of translations $\mu$.

Modular invariance of string theory demands embedding $\mathbbm S$ into 
the 16D gauge degrees of freedom, which also guarantees anomaly freedom of the 
resulting effective theory in 4D. In orbifolds with \Z{N}\x\Z{M} point group,
the embedding can be parametrized by eight 16D vectors: two shift vectors $V_1$ and 
$V_2$ and six discrete Wilson lines $W_\alpha$, $\alpha=1,\ldots,6$. These are subject 
to conditions to ensure modular invariance and compatibility with the geometry associated 
with the space group~\cite{Ploger:2007iq,Blaszczyk:2014qoa}. Once a space group
and its gauge embedding are chosen, there are standard techniques used to 
arrive at their associated low-energy effective 4D field theory, with a specific 
gauge group $\mathcal{G}_{\text{4D}}$ and a massless 
spectrum~\cite{Ramos-Sanchez:2008nwx,Vaudrevange:2008sm,Blaszczyk:2014qoa}.

We are interested in SM-like models, which we define by the following properties: 
i) the SM gauge group $\mathcal{G}_{\text{SM}}:=\SU3_c\x\SU2_L\x\U1_Y$ is included in
$\mathcal{G}_{\text{4D}}=\mathcal{G}_{\text{SM}}\x\mathcal{G}'\x[\U1']^n$, where 
$\mathcal{G}'$ has rank $12-n$ and is a product of non-Abelian gauge factors, and
ii) the massless spectrum consists of the SM particles plus a number of several kinds 
of exotic vectorlike fermions and scalars. Modular invariance guarantees that 
at most one $\U1'$ exhibits chiral anomalies~\cite{Casas:1987us,Kobayashi:1996pb}, 
which are cancelled by the Green-Schwarz mechanism~\cite{Green:1984sg}. Once
this (pseudo-)anomalous $\U1'$ is identified, we build all other gauge Abelian 
symmetries to be orthogonal to it and hence non-anomalous. In particular, 
in our models the $\U1_Y$ hypercharge is non-anomalous and
chosen to be compatible with \SU5 grand unification, which can be achieved by
the proper normalization of the hypercharge generator, as was explained in detail long 
ago~\cite{Ibanez:1993bd} and has been extensively used in the construction of promising 
heterotic orbifold models ever since
\cite{Buchmuller:2005jr,Buchmuller:2005sh,Buchmuller:2006ik,Lebedev:2006kn,Lebedev:2008un,Nilles:2009yd,Olguin-Trejo:2018wpw,Perez-Martinez:2021zjj}.
(For additional details, see our example in appendix~\ref{sec:Ygenerator1}.)

Recently, we performed a large search of SM-like models from the $\SO{16}\x\SO{16}$ 
heterotic string by using our own non-SUSY version of the \texttt{orbifolder}. One 
of the main results in that work was the identification of SM-like models that exhibit
either no fermions or scalars in the exotic sector, as reported in~\cite[table 8]{Perez-Martinez:2021zjj}. 
We called such promising constructions {\it almost SM} models. In those models, which
always include harmless scalar SM singlets, if they lack exotic fermions, some scalar 
leptoquarks appear; further, models without exotic scalars include extra vectorlike pairs
of lepton doublets and down-type quarks. In 505 out of the 547 models identified as  
{\it almost SM} the minimal number of Higgses is six.            
Outside the set of {\it almost SM}, i.e.\ 
including additional exotic states, there do exist non-SUSY compactifications with less 
Higgs doublets. For example, there are 3,192 models with two Higgses, and only 13 models with
one Higgs. Due to the large number of exotics in the models with one Higgs
doublet, they are less phenomenologically appealing than the non-SUSY orbifold
models with two or more Higgs fields.

For our study of DM Higgs portals arising from non-SUSY heterotic orbifolds, we
take here models based on the space group \Z2\x\Z4 (2,4). We choose this
geometry only because most of the phenomenologically viable models arise from
the point group \Z2\x\Z4 and the geometry (2,4) is especially fruitful, 
see~\cite[table 1]{Perez-Martinez:2021zjj}. In section~\ref{sec:sixHiggses}
we shall discuss the properties of one {\it almost SM} based on this geometry
and endowed with six Higgs doublets. As we will see, fully analyzing such a 
model will be too complicated, although some interesting features can be
captured under certain simplifications. Hence, it is worthwhile to also choose
a sample model based on the same geometry, but that includes two Higgs doublets.
As we will see in section~\ref{sec:twoHiggses}, the picked model exhibits vectorlike 
scalar leptoquarks and down-type quarks.

\section{Higgs portals}
\label{sec:portals}
In order to fix our notation and provide some of the elements used in our discussion, 
let us describe the main features of Higgs portals in the context of DM models.

The simplest scenario of Higgs portals includes a new scalar field $S$, which couples
to a Higgs doublet only quadratically. If $S$ is real, it requires the introduction of 
an {\it ad hoc} \Z2 symmetry, under which $S$ is odd, to stabilize the vacuum while 
guaranteeing the portal. Alternatively, $S$ can be a complex scalar. Whereas this adds 
up a new degree of freedom, note that all matter fields emerging from string theories 
are complex, so this is the only scheme we realize in string-derived constructions.

The potential of the SM model (or a SM-like model) that includes the field $S$ can be
split as
\begin{equation}
  V ~=~ V_Y + V_\phi + V_S + V_{\phi S}\,,
\end{equation}
where $V_Y$ includes all Yukawa couplings, $V_\phi$ contains all Higgs field(s) self-interactions, 
$V_S$ depends only on the DM candidate $S$, and $V_{\phi S}$ provides the interactions
between $S$ and the Higgs field(s) that determine the Higgs portal.
In the simplest case, with only one Higgs doublet $\phi$, the renormalizable $S$-dependent
contributions to the SM potential take the form
\begin{subequations}
\label{eq:couplings1S}
\begin{eqnarray}
\label{eq:VS}
  V_S(S)             &=& \mu_S^2 |S|^2 + \lambda_S \left(|S|^2\right)^2\,,\\
\label{eq:Vportal1HS}
  V_{\phi S}(\phi,S) &=& \lambda_{\phi S}|\phi|^2|S|^2\,,
\end{eqnarray}
\end{subequations}
where $|S|^2:=S^*S$ and $|\phi|^2:=\phi^\dagger\phi$, as usual. In general, we take 
$\mu_S,\lambda_S>0$ in $V_S$ to ensure that $S$ does not develop a vacuum expectation 
value (VEV) and, hence, remains a DM candidate as it does not mix with the Higgs.
Note that $V_S+V_{\phi S}$ depends only on the norm of $S$ and not on its phase,
as if $S$ was charged under an additional \U1 symmetry. In the models we 
study, this is indeed the case, as we shall see in sections~\ref{sec:sixHiggses}
and~\ref{sec:twoHiggses}. However, such symmetries are eventually broken and 
the pseudoscalar associated with the phase of $S$ develops some large mass
and can contribute to DM (see e.g.~\cite{Gonderinger:2012rd}). For the sake 
of simplicity, we assume here that the mass of this field is too large (of order
$M_\mathrm{Pl}$, the scale of \U1 breakdown) and thus decoupled from observations.
Hence, our scenario is effectively equivalent to a Higgs portal with a real
scalar DM candidate given by $|S|$.

There exist some models arising from string theory endowed with only one Higgs doublet, but 
they typically include additional exotic matter that makes them less appealing~\cite[Table 8]{Perez-Martinez:2021zjj}. 
So, we are compelled to study models with a larger number of Higgs doublets. In such a case, 
the Higgs-portal potential reads in general
\begin{equation}
\label{eq:VportalNHS}
  V_{\phi S}(\phi_i,S) ~=~ \sum_i \lambda_{iS}|\phi_i|^2|S|^2+
                           \sum_{i\neq j}\lambda_{ijS}\phi_i^\dagger\phi_j |S|^2\,,\\
\end{equation}
where $i,j$ label the Higgs doublets and $|\phi_i|^2:=\phi_i^\dagger\phi_i$. In the dynamics
of the DM candidate $S$, we must still consider~\eqref{eq:VS}. We assume that 
$\lambda_{ijS}=\lambda_{jiS}$, and that the couplings in eqs.~\eqref{eq:couplings1S} 
and~\eqref{eq:VportalNHS} are real.

The associated freeze-out relic density $\Omega_\mathrm{DM}$ of $|S|$ can be directly computed for
models based on Higgs portals. One important assumption in those computations is that 
DM thermalizes, which is only possible for large reheating temperatures. One can verify 
that in these scenarios $\Omega_\mathrm{DM} h^2 \propto \vev{\sigma v_r}^{-1}$. The 
thermally-averaged cross section $\vev{\sigma v_r}$ is determined by the DM relative
velocity $v_r$ and the annihilation cross section $\sigma$ of DM scalars to SM fermions, 
via the Higgs portal. From eq.~\eqref{eq:VportalNHS} it is clear that $\sigma$ is in 
general controlled by the product of the couplings $\lambda_{iS},\lambda_{ijS}$ and the 
Higgs VEV(s). Even though analytic expression of $\Omega_\mathrm{DM}$ can be obtained at 
tree level in some examples, in general cases and at 1-loop it becomes challenging. 
Instead, one can use the program \texttt{micrOMEGAs}~\cite{Belanger:2018ccd}, which
computes the DM relic density at tree and 1-loop level and contrasts it with observations
for models defined in a specific format that can be obtained from other computational tools.

In the standard calculation of the relic density only the self-coupling $\mu_S$ (and other
couplings associated with the DM-candidate mass) in eq.~\eqref{eq:VS} is relevant. 
However, self-interactions lead to the self-thermalization of the dark sector. 
If thermodynamic equilibrium is not achieved or maintained within 
the dark sector, the standard way to calculate the DM relic density is not valid
and other couplings may turn to be relevant, see e.g.~\cite{Hryczuk:2022gay}. We
assume here the simplest scenario, so that $\lambda_S$ cannot be constrained by
DM data, but only by perturbativity bounds.

\section{Multi-Higgs portals from string theory}
\label{sec:sixHiggses}
In this section we consider a non-SUSY model with six Higgs doublets in which DM can arise 
from a scalar Higgs portal (similar bottom-up DM approaches may not require extra fields~\cite{Alakhras:2017bbe}).
As mentioned earlier, models with six Higgs fields are abundant in string compactifications with
no SUSY. This leads to a scenario with a very rich Higgs sector with many still missing 
charged and uncharged Higgs particles, and complicated mixing matrices. Clearly, this also 
increases the complexity of Higgs portals with the Higgs sector, making it too challenging
to study in detail the model.
Nonetheless, we discuss it as a means a) to observe the generic properties 
of Higgs portals in these models, and b) to motivate the discussion of stringy models
with a smaller Higgs sector, which nevertheless share the interesting properties of their
more generic partner models with six Higgses. Many features discussed here will be
also relevant in our simpler model with two Higgses in section~\ref{sec:twoHiggses}.

We choose the model arising from compactifying the \SO{16}\x\SO{16} heterotic string 
on the $\ZxZ{2}{4}\,(2,4)$ orbifold geometry (see~\cite{Fischer:2012qj} for details 
of the geometry), whose gauge embedding is defined by the shift vectors and Wilson lines
\begin{subequations}
\begin{eqnarray}
V_1 &=& (0, 0, 0, 0, 0, \tfrac{1}{2}, \tfrac{1}{2}, 2),(0, 0, 0, 0, \tfrac{1}{2}, \tfrac{1}{2}, \tfrac{1}{2}, \tfrac{1}{2})\,,\\
V_2 &=& (-\tfrac{3}{8}, -\tfrac{1}{8}, -\tfrac{1}{8}, -\tfrac{1}{8}, \tfrac{3}{8}, \tfrac{1}{8}, \tfrac{9}{8},-\tfrac{3}{8}),(-\tfrac{5}{8}, -\tfrac{1}{8}, \tfrac{1}{8}, \tfrac{1}{8}, -\tfrac{1}{8}, -\tfrac{1}{8}, \tfrac{1}{8}, \tfrac{5}{8})\,,\\
W_3 = W_4 = W_6 &=&(0, 0, 0, 0, 0, 0, 0, 0),(-\tfrac{1}{2}, 0,-\tfrac{3}{2}, 0, -\tfrac{3}{2},1,-1,\tfrac{3}{2})\,,\\
W_5 &=& (-\tfrac{7}{4}, -\tfrac{1}{4}, -\tfrac{1}{4}, \tfrac{3}{4}, \tfrac{7}{4}, -\tfrac{3}{4}, \tfrac{9}{4}, -\tfrac{7}{4}),(-\tfrac{7}{4}, \tfrac{5}{4}, -\tfrac{5}{4}, \tfrac{9}{4}, -\tfrac{3}{4}, \tfrac{7}{4}, \tfrac{7}{4}, \tfrac{3}{4})\,,
\end{eqnarray}
\end{subequations}
and $W_1=W_2=0$. The resulting 4D gauge group is
\begin{equation}
\maG_{\text{4D}} ~=~ \maG_{\text{SM}} \x \maG'\x [\U1']^8\,,\qquad\text{where}\quad 
\maG' ~=~ \SU3_\text{flavor}\x\left[\SU2\x\SU2\right]_\text{hidden}
\label{eq:g4d-6Higgs}
\end{equation}
and one $\U1'$ is anomalous. Note that there is an additional \SU3 gauge factor in $\mathcal G'$
that acts as a flavor gauge symmetry, as some of the quark and lepton fields as well as
the Higgs fields are charged under this symmetry.
The complete massless fermionic and scalar spectra are shown, respectively, 
in table~\ref{tab:complete4902f} and~\ref{tab:complete4902s} of appendix~\ref{ap:6Hspectrum}. 
Table~\ref{tab:tabmodel4902} displays a summary of the spectrum of our model. We omit here
quantum numbers under the hidden gauge group, as SM fields are not charged under it; we also 
avoid $\U1'$ charges for simplicity.
In table~\ref{tab:tabmodel4902} and throughout this paper we adopt the usual notation for 
fermions. In particular, $\mathrm{L},\mathrm{R}$ denote left and right fermion 
projections, i.e.\ $\psi_{\mathrm{L},\mathrm{R}} = P_{\mathrm{L},\mathrm{R}}\psi$, where 
$P_{\mathrm{L},\mathrm{R}}=(\id\mp\gamma^5)/2$; further, $\overline{\psi}=\psi^\dagger\gamma^0$.
\begin{table}[t!]
\centering
{
   \begin{tabular}{|rlc|c|rlc|}
   \cline{1-3}\cline{5-7} 
   \#  & Fermionic irrep.   & Label &&  \#  & Scalar irrep. & Label \\
   \cline{1-3}\cline{5-7}
     5 & $(\rep{1},\rep{2};\rep{1})_{\nicefrac{-1}{2}}\phantom{A^{A^A}}$ & $\ell_{\mathrm{L},i}$ 
&&   2 & $(\rep1, \rep2; \rep3)_{\nicefrac{1}{2}}$                 & $\phi_i$ \rule{0pt}{2.5ex}\\
     2 & $(\rep1, \rep2; \rep1)_{\nicefrac{1}{2}}$                 &  $\overline{\ell}'_{\mathrm{L},i}$
&&  1 & $(\rep1, \rep1; \rep1)_{0}$                                & $S$ \\ 
     1 & $(\rep1, \rep1; \rep3)_{1}$                               &  $\overline{e}_\mathrm{L}$
&&   51  &    $(\rep1, \rep1; \rep1)_{0}$      	                   & $s_i$  \\ 
     1 & $(\rep3, \rep2; \rep3)_{\nicefrac{1}{6}}$                 &  $q_\mathrm{L}$
&&   8  &    $(\rep1, \rep1; \rep3)_{0}$                           & $s'_i$  \\ 
   \cline{5-7}
     1 & $(\crep3, \rep1; \rep3)_{\nicefrac{-2}{3}}$                 &  $\overline{u}_\mathrm{L}$
 & \multicolumn{4}{c}{}\\ 
     7 & $(\crep3, \rep1; \rep1)_{\nicefrac{1}{3}}$                 & $\overline{d}_{\mathrm{L},i}$ 
 & \multicolumn{4}{c}{}\\
     4 & $(\rep3, \rep1; \rep1)_{\nicefrac{-1}{3}}$                 & $d'_{\mathrm{L},i}$
 & \multicolumn{4}{c}{}\\
   69 & $(\rep1, \rep1; \rep1)_{0}$                 & $\nu_{\mathrm{R},i}$ 
 & \multicolumn{4}{c}{}\\
     12 &   $(\rep1, \rep1; \crep3)_{0}$            & $\nu'_{\mathrm{R},i}$ 
 & \multicolumn{4}{c}{}\\
     2 & $(\rep1, \rep1; \rep3)_{0}$                & $\nu''_{\mathrm{R},i}$
 & \multicolumn{4}{c}{}\\ 
    \cline{1-3}  
    \end{tabular}
    \caption{Summary of the massless spectrum of a non-supersymmetric string model with 6 Higgses.
         We display the quantum numbers w.r.t.\ the gauge factors $\mathcal{G}_\text{SM}\x\SU3_\text{flavor}$,
         where the hypercharge is indicated as subindex. The complete spectrum is shown in table~\ref{tab:complete4902f} 
         and~\ref{tab:complete4902s} of appendix~\ref{ap:6Hspectrum}, for fermions and scalars respectively.
         Note that the two Higgs multiplets are also charged under $\SU3_\text{flavor}$, leading to a total of
         6 Higgs doublets. The multiplicity of some observable fermions is also determined by
         this symmetry. $\mathrm{L}$ and $\mathrm{R}$ denote left and right fermion projections 
         and a bar over a fermion indicates the Dirac conjugate. It is easy to verify that
         the model is free of chiral gauge anomalies, as expected~\cite{Casas:1987us,Kobayashi:1996pb}.
         \label{tab:tabmodel4902}}
}
\end{table} 
 
Before studying the Higgs portal, let us discuss some aspects of the Yukawa sector
of this model. We see that in the fermion sector there are two (four) vectorlike pairs 
of lepton doublets (down-quark singlets). These exotics can be decoupled from low-energy 
physics if some SM singlets $s_i$ develop large VEVs and couplings of the type
$s_i \ell_{\mathrm{L},j} \overline\ell'_{\mathrm{L},k}$ ($s_i \overline{d}_{\mathrm{L},j} d'_{\mathrm{L},k}$) 
are admitted by all symmetries of the model. This generates a large effective mass for
these states. As an example, let us analyze the masses of exotic lepton doublets.
Considering the invariant couplings arising from the details shown in appendix~\ref{ap:6Hspectrum}, 
the lowest-order mass matrix for the leptonic sector is given by\footnote{In 
table~\ref{tab:complete4902s} there are six copies of 
$s_{15}$. Here we display only one of them.}
\begin{equation}
\label{mass_L}
M_\ell ~=~ \begin{pmatrix} 
0 & 0 & 0 & c_1 & c_2 s_4^* \\ 0 & c_3 s_{15} & c_4 s_{15} & 
c_5 s_1^* & 0 
\end{pmatrix}\,,
\end{equation}
such that 
\begin{equation}
\overline{\ell}'\,M_\ell\,\ell + \text{h.c.}~\subset~\mathcal L\,.
\end{equation}
Here we defined 
$\overline{\ell}':= \left(\overline{\ell}'_{\mathrm{L} ,1},\,\overline{\ell}'_{\mathrm{L} ,2}\right)$
and
$\ell := \left(\ell_{\mathrm{L} ,1},\,\ell_{\mathrm{L} ,21},\,\ell_{\mathrm{L} ,22},\,\ell_{\mathrm{L} ,3},\,\ell_{\mathrm{L} ,4}\right)^\intercal$.
The coefficients $c_i$ in eq.~\eqref{mass_L} are some coupling constants 
that are assumed to be real and of order unity for simplicity. One can readily show 
that $M_\ell$ has (full) rank two. The two nontrivial eigenvalues of $M_\ell M_\ell^\intercal$ 
correspond to the squared masses of the massive exotic lepton fields $\overline{\ell}'_{\mathrm{L},i}$
and the two massive linear combinations of $\ell_{\mathrm{L},i}$.

In order to have specific states and masses, we must assume a sufficiently general
and simple singlet VEV configuration. We consider that all singlets have vanishing VEVs
except for 
\begin{equation}
\label{eq:VEVconfig}
\{s_1,\,s_2,\,s_3,\,s_4,\,s_6,\,s_{11},\,s_{16},\,s_{18},\,s_{23}\}\,,
\end{equation}
which develop real and nontrivial VEVs. At this point, this vacuum configuration is just 
{\it ad hoc}.\footnote{Fixing these (and other) VEVs would require to solve the longstanding question of 
the dynamics of all moduli (including these singlets) in string model building, which is  
not the goal of this work. Note that, due to the various charges of $s_i$ (see table~\ref{tab:complete4902s}), 
these fields develop a perturbative and non-perturbative potential of their own, which has first 
to be fully determined and then minimized.}
Yet note that the precise values of the VEVs is irrelevant for our discussion. To simplify the notation, we 
use just $s_i$ for the VEVs $\langle s_i\rangle$. So, in eq.~\eqref{mass_L} we take $s_{15}\to 0$. 
Hence, we obtain the squared masses
\begin{equation}
\frac{1}{2}\left(c_1^2+c_5^2 s_1^2 + c_2^2 s_4^2 \pm \sqrt{(c_1^2 + c_5^2 s_1^2 + c_2^2 s_4^2 )^2-4 c_2^2 c_5^2 s_1^2 s_4^2}\right)\,,
\end{equation}
for two massive exotic lepton doublets, which turn out to be built by linear combinations
of $\ell_{\mathrm{L},3}$ and $\ell_{\mathrm{L},4}$ (see table~\ref{tab:complete4902f}).
The massless physical $\ell_\mathrm{L}$ eigenstates of $M_\ell^\intercal M_\ell$ are 
given by $\ell_{\mathrm{L},1}$, $\ell_{\mathrm{L},21}$ and $\ell_{\mathrm{L},22}$.
We relabel the massless eigenstates as $\ell_{\mathrm{L} ,1} \to \ell_{\mathrm{L} ,1}$, 
$\ell_{\mathrm{L} ,21} \to \ell_{\mathrm{L},2}$ and $\ell_{\mathrm{L} ,22}\to \ell_{\mathrm{L},3}$.
The physical down-quark mass eigenstates, labeled form now on as 
$\overline d_{\mathrm{L} ,1}$, $\overline d_{\mathrm{L} ,2}$ 
and $\overline d_{\mathrm{L} ,3}$, are obtained analogously. For the details of the
corresponding computations, see appendix~\ref{ap:ferm_mass}.

Considering the charges of SM states given in appendix~\ref{ap:6Hspectrum},
one can easily build the potential of the model. In particular, taking into 
account the symmetric singlet arising from the contraction $\rep{3}\otimes\crep{3}$ 
under $\SU3_\mathrm{flavor}$, we find that, at leading order, the Yukawa potential 
of the physical states is given by
%
\begin{equation}
\begin{split}
V_Y& \supset \mathcal{Y}_{e,1}^i \, \overline{e}_{\text{L},I}\phi_{iI}^\dagger \ell_{\mathrm{L},1} 
+  \mathcal{Y}_{e,2}^i \, \overline{e}_{\text{L},I}\phi_{iI}^\dagger \ell_{\mathrm{L},2} 
+  \mathcal{Y}_{e,3}^i \, \overline{e}_{\text{L},I}\phi_{iI}^\dagger \ell_{\mathrm{L},3} 
+ \mathcal{Y}_{\nu,13}^i \, \overline{\nu}_{\text{L},13I}' \phi_{iI} \ell_{\mathrm{L},2}  \\
& + \mathcal{Y}_{\nu,26}^i \, \overline{\nu}_{\text{L},26I}' \phi_{iI} \ell_{\mathrm{L},3}
+ \mathcal{Y}_{u}^i \, \epsilon^{IJK} \overline{u}_{\text{L},I}\phi_{iJ}q_{\mathrm{L},K}
+ \mathcal{Y}_{d, 1}^i \, \overline{d}_{\mathrm{L},1} \phi_{iI}^\dagger q_{\mathrm{L},I} 
+ \mathcal{Y}_{d, j}^i \, \overline{d}_{\mathrm{L},j} \phi_{iI}^\dagger q_{\mathrm{L},I} 
+ \text{h.c.}\,,
\end{split}
\label{eq:obssinglet}
\end{equation}
where  $I,J,K=1,2,3$ are $\SU3_\mathrm{flavor}$ indices whereas $i=1,2$ runs 
over the multiplicity of the Higgs and $j=2,3$ over the indices of down-quark fields. 
Further, $\epsilon^{IJK}$ is the Levi-Civita symbol, and $\mathcal{Y}_{e,1}^i$ and
$\mathcal{Y}_{d,1}^i$ are dimensionless coefficients. Furthermore,
\begin{subequations}
\begin{eqnarray}
\mathcal{Y}_{d,j}^i &=& \zeta_{d,j}^i s_{18} s_6^* s_{16}^*\,,\\
\mathcal{Y}_{e,2}^i   &=& \zeta_{e,2}^i\,s_{23}s_{11}^*s_{16}^* + \,\tilde\zeta_{e,2}^i s_{18} s_6^* s_{16}^* 
\qquad\text{and}\qquad 
\mathcal{Y}_{e,3}^i ~=~ \zeta_{e,3}^i\,s_{4}s_{18} s_{6}^*\,,
\end{eqnarray} 
\end{subequations}
with $\zeta_{d,j}^{i}, \zeta_{e,j}^i$ and $\tilde\zeta_{e,2}^i$ constants. 
As usual, summation over repeated indices in eq.~\eqref{eq:obssinglet} is understood. 
Notice that we allow for $s_{23}\neq 0$, in order to have a tool that may explain
the observed mass differences between the down-quark and charged lepton sectors.
The leading-order Yukawa potential~\eqref{eq:obssinglet} does not provide a complete 
framework for flavor physics. In particular, it does not produce masses for all
quarks. However, higher-order couplings with singlets $s_i$ can yield the missing
masses and possibly fit the flavor data. Since our focus is on the scalar sector, 
hereafter only the couplings to the heaviest generation of fermions are considered
relevant.

\subsection{Higgs sector and Higgs portals}

Let us now discuss some aspects of the scalar sector of this model. 
Using the charges in table~\ref{tab:complete4902s}, we find that there are
singlets that satisfy the conditions of required by DM candidates in 
Higgs-portal scenarios, as discussed in section~\ref{sec:portals}. For simplicity,
we consider only the scalar field $S$ of table~\ref{tab:complete4902s}. Note that
$s_{10}$ has identical quantum numbers and is thus, in principle, a second DM candidate.
In the spirit of~\cite{Dutta:2022xbd}, we ignore this possibility here.
The Higgs-portal interactions of $S$ and the Higgs fields $\phi_i$,
$i=1,2$, are given at leading order, as in eq.~\eqref{eq:VportalNHS}, by
\begin{align}
V_{\phi S}(\phi_1, \phi_2, S) ~=~ \lambda_{1S} \left|\phi_1\right|^2 |S|^2 
+ \lambda_{2S} \left|\phi_2\right|^2 |S|^2 
+ \left[\lambda_{12S} (\phi_1^\dagger \phi_2)|S|^2 + \text{h.c.}\right]\,,
\label{eq:hportal} 
\end{align}
where $\phi_i^\dagger\,\phi_j = \phi_{i,I}^\dagger\,\phi_{j,I}$, 
with summation over $I=1,2,3$ implied.

Since we have six Higgs fields due to their transformation under $\SU3_\mathrm{flavor}$, 
after electroweak symmetry breakdown, besides the lightest (already) observed Higgs boson
and the three degrees of freedom that are `eaten up' by the gauge bosons, our model includes
20 additional effective degrees of freedom in the Higgs sector. Although in principle the 
number of free parameters due such large Higgs sector can be very large, the gauge flavor
symmetry reduces it.
The self-interactions of the Higgs sector are given at leading order by
\begin{equation}
\begin{split}
V_\phi(\phi_1, \phi_2) &=~ \mu^2_{11}\left|\phi_1\right|^2 + \mu^2_{22} \left|\phi_2\right|^2 
                + \lambda_{1} \left|\phi_1\right|^4 + \lambda_{2} \left|\phi_2\right|^4  
                + \lambda_{12} \phi_1^\dagger \phi_1 \phi_2^\dagger \phi_2 \\
                &+ \left[ \mu^2_{12} \phi_1^\dagger \phi_2 
                + \lambda_5 (\phi_2^\dagger\phi_1)^2 
                + \lambda_6 (\phi_2^\dagger\phi_1)\left|\phi_1\right|^2 
                + \lambda_7 (\phi_2^\dagger\phi_1)\left|\phi_2\right|^2 
                + \text{h.c.}\,\right]\,.
\label{eq:VH6Higgs}
\end{split}
\end{equation}
To express $V_\phi$ in terms of the $\SU3_\mathrm{flavor}$ components, we must
determine the two singlets arising from the product 
\begin{equation}
\left(\rep{3} \otimes \rep{3}\otimes \crep{3} \otimes \crep{3}\right)_{\rep1}~=~
\left(\crep{3}\otimes \rep{3}\right)_{\rep1} \oplus \left(\rep{6}\otimes \crep{6}\right)_{\rep1}
\end{equation}
for the quartic terms in eq.~\eqref{eq:VH6Higgs}. For all these terms but $|\phi_1|^2 |\phi_2|^2$ 
one of the singlets vanishes. In the exceptional case the invariant products are proportional 
to (see e.g.~\cite{Park:2013fda})
\begin{equation}
\frac{1}{\sqrt2}\left(\phi_{1I}^\dagger \phi_{1I} \phi_{2J}^\dagger \phi_{2J} - \phi_{1I}^\dagger \phi_{2I}\phi_{2J}^\dagger \phi_{1J}\right)
\qquad\text{and}\qquad
\frac{1}{2}     \left(\phi_{1I}^\dagger \phi_{1I} \phi_{2J}^\dagger \phi_{2J} + \phi_{1I}^\dagger \phi_{2I}\phi_{2J}^\dagger \phi_{1J}\right)\,.
\end{equation}
Hence, the terms in the potential~\eqref{eq:VH6Higgs} are explicitly given by
(up to some factors that are absorbed in the coupling parameters)
\begin{equation}
\label{eq:VHcomponents}
\begin{split}
V_\phi(\phi_1, \phi_2) &=
                \mu^2_{11} \left|\phi_{1I}\right|^2 + \mu^2_{22} \left|\phi_{2I}\right|^2 
                + \lambda_1 \left|\phi_{1I}\right|^2 \left|\phi_{1J}\right|^2 
                + \lambda_2 \left|\phi_{2I}\right|^2 \left|\phi_{2J}\right|^2 \\
                &+ \lambda_3\left|\phi_{1I}\right|^2 \left|\phi_{2J}\right|^2 
                + \lambda_4 \phi_{1I}^\dagger \phi_{2I}\phi_{2J}^\dagger \phi_{1J}\\ 
                &+ \left[ \mu^2_{12} \phi_{1I}^\dagger \phi_{2I} 
                + \lambda_5 \phi_{2I}^\dagger \phi_{1I} \phi_{2J}^\dagger \phi_{1J} 
                + \lambda_6 \phi_{2I}^\dagger \phi_{1I} \left|\phi_{1J}\right|^2
                + \lambda_7 \phi_{2I}^\dagger \phi_{1I} \left|\phi_{2J}\right|^2
                + \text{h.c.} \right].
\end{split}
\end{equation}
Here, we have defined $|\phi_{iI}|^2:=\phi_{iI}^\dagger \phi_{iI}$, with fixed $i=1,2$,
and summation over $I=1,2,3$ implied. We also defined for future convenience
$\lambda_3:=\frac{\lambda_{12}}{2}(1+\sqrt2)$ and $\lambda_4:=\frac{\lambda_{12}}{2}(1-\sqrt2)$.
Notice that the hermicity of the potential implies $\mu_{11}^2=(\mu_{11}^2)^*$, 
$\mu_{22}^2=(\mu_{22}^2)^*$, $\lambda_1 = \lambda_1^*$, $\lambda_2 = \lambda_2^*$, 
and $\lambda_{12} = \lambda_{12}^*$. There are thus five real and four 
complex parameters, hence altogether 13 parameters in the Higgs sector. Note
that demanding \CP conservation, as we will do henceforth, implies that the four 
complex parameters also become real, reducing the counting to nine real parameters in this sector. 

To discuss some aspects of the electroweak symmetry breakdown in this model, recall that all 
$\phi_{iI}$ are doublets of $\text{SU}(2)_\text{L}$. Following the standard convention 
$Q = Y + T_3$, we can expressed them as
\begin{equation}
\phi_{iI} = \begin{pmatrix} \phi_{iI}^{+} \\ \phi_{iI}^0 \end{pmatrix}\,,
\label{eq:parametrization}
\end{equation}
where only the uncharged components acquire VEVs $v_{iI}$. Close to the vacua
defined by the VEVs, the resulting Higgs perturbations can be parametrized by
\begin{equation}
\label{eq:vevparametrization}
\phi_{1I} \longrightarrow 
\begin{pmatrix} 
\phi_{1I}^{+} \\ \frac{1}{\sqrt{2}}(v_{1I} + \sigma_{1I})
\end{pmatrix} 
\qquad\text{and}\qquad 
\phi_{2I} \longrightarrow 
\begin{pmatrix} \phi_{2I}^{+}\\ \frac{1}{\sqrt2}(v_{2I}+\sigma_{2I}) 
\end{pmatrix}\,.
\end{equation}
Hence, in the Higgs vacuum we naturally obtain the tadpole contributions
\begin{equation}
 V_\phi(\sigma_{1I}, \sigma_{2I}) ~\supset~ 
     \frac{1}{2}\left(v_{1I} \zeta  + v_{2I} \gamma \right) \sigma_{1I}
   + \frac{1}{2}\left(v_{1I} \gamma + v_{2I} \kappa \right) \sigma_{2I}\,,
\end{equation}
where we defined
\begin{subequations}
\label{eq:tadpoleTerms}
\begin{eqnarray}
\zeta  &:=& \mu_{11}^2 + \lambda_1 v_{1J} v_{1J} + \frac{1}{2}\lambda_{3}v_{2J}v_{2J} + \lambda_6 v_{1J}v_{2J}\,,\\
\gamma &:=& \mu_{12}^2 + \frac12\lambda_6 v_{1J}v_{1J} + \frac12\lambda_7v_{2J}v_{2J}  + 
            \frac12\left(\lambda_{4} + 2\lambda_5\right) v_{1J}v_{2J}\,,\\
\kappa &:=& \mu_{22}^2 + \frac{1}{2}\lambda_{3}v_{1J}v_{1J} + \lambda_2 v_{2J} v_{2J} + \lambda_7 v_{1J}v_{2J}\,. 
\end{eqnarray}
\end{subequations}
Note that $\lambda_3$ and $\lambda_4$ are not independent here.
Assuming linear independence of the components of the fields $\sigma_1$ and $\sigma_2$ 
leads to the tadpole-cancellation conditions\footnote{Note that this also
ensures that the expansion around the VEV is a critical point. The minimum
condition is satisfied when we demand the mass matrices to be positive definite.} 
\begin{equation}
\label{eq:tadpoleCancel}
 v_{1I} \zeta  + v_{2I} \gamma ~\stackrel{!}{=}~ 0
 \qquad\text{and}\qquad
 v_{1I} \gamma + v_{2I} \kappa ~\stackrel{!}{=}~ 0
 \qquad\text{for}\quad I=1,2,3\,.
\end{equation}

Since further explicit computations can become insurmountable, we make a number of
strong {\it ad hoc} simplifying assumptions. We will work in the so-called Higgs basis,\footnote{The 
angle that mixes the doublets $(\phi_{1I}, \phi_{2I}^\dagger)^\intercal$ is the same 
angle that rotates the charged components. Note, however, that there are many more mixings
in this multifield case, which we do not consider.} where $v_{2I} = 0$ for $I=1,2,3$. In addition, 
we impose the hierarchy $v_{11}\gg v_{12}\gg v_{13}$, by introducing the new parameter 
$\epsilon\ll 1$, such that $\epsilon\, v_{11} = v_{12}$, $\epsilon\, v_{12} = v_{13}$ and
$\epsilon^2 =  v_{13}/v_{11} \sim 0$. A similar hierarchy has been studied 
in~\cite{Keus:2013hya,Khater:2021wcx} in models with three Higgs doublets.
Our last simplification is to assume that $v_{1I}$ are real as the complex phase of 
each VEV can be absorbed by the phase of the uncharged fields.

With these assumptions, our discussion simplifies somewhat. In particular, in the 
Higgs basis, the tadpole-cancellation conditions~\eqref{eq:tadpoleCancel} become
\begin{equation}
\label{eq:cond11}
\zeta  ~=~ \mu_{11}^2 + \lambda_1 v_{1J} v_{1J} ~\stackrel{!}{=}~ 0\qquad\text{and}\qquad 
\gamma ~=~ 2\mu_{12}^2 + \lambda_6 v_{1J}v_{1J} ~\stackrel{!}{=}~ 0 \,,
\end{equation}
which allow for simple restricting relations among some of the Higgs parameters and 
the remaining VEVs.

To proceed further, we will expand each uncharged scalar field
in its real and imaginary parts corresponding to real scalar fields, 
e.g.\ $\sigma_{iI} = \rho_{iI} + \I\eta_{iI}$.
Under our assumptions, the resulting $6\x6$ symmetric mass matrix 
for $(\rho_{11},\rho_{12},\rho_{13},\rho_{21},\rho_{22},\rho_{23})^\intercal$
takes the form
\begin{equation}
M_{\rho}^2 = \begin{pmatrix} \Omega_1 & \epsilon\,\Omega_1 & 0 & P & \epsilon\,P & 0 
\\ \epsilon\,\Omega_1 & 0 & 0 & \epsilon\,P & 0 & 0
\\ 0 & 0 & 0 & 0 & 0 & 0 \\ P & \epsilon\,P & 0 & \Omega_2 + \xi/2 & \epsilon\,\Omega_2 & 0
\\ \epsilon\,P & 0 & 0 & \epsilon\,\Omega_2 & \xi/2 & 0
\\ 0 & 0 & 0 & 0 & 0 & \xi/2\end{pmatrix}\,,
\label{eq:Mphi} 
\end{equation}
where the coefficients are given in eq.~\eqref{entries_Mphi} of our appendix~\ref{ap:6H}. 
The squared eigenmasses are
\begin{equation}
m_{h, H}^2 ~=~ \frac{1}{2} 
      \left(2\Omega_1 + 2\Omega_2 + \xi\mp\sqrt{(4P)^2 + \widetilde{\Omega}^2}\right)\,, \qquad 
m_{G_i}^2 ~=~ 0\,, \qquad
m_{D_i}^2 ~=~\xi\,,
\label{eq:eigenmphi}
\end{equation}
with $i=1,2$ and $\widetilde\Omega:=2(\Omega_2 - \Omega_1) + \xi$.
The states in eq.~\eqref{eq:eigenmphi} correspond to the four massive Higgs states 
$h,H,D_{1,2}$ and two massless Goldstones $G_{1,2}$. The lightest eigenstate $h$ 
is identified with the observable Higgs boson, whose mass is constrained to be 
$m_{h} = 125.25(17)\,\text{GeV}$~\cite{ParticleDataGroup:2022pth}. $H$ and the 
degenerate states $D_{1,2}$ are heavier Higgs fields, which have not been
detected yet. Hence, their masses are considered predictions of these kind of 
models.

We require that the nonvanishing squared eigenmasses be positive, which also 
ensures that the minimum is stable. The mass eigenstates are given by
\begin{equation}
\label{eq:eigenvphi}
\begin{aligned}
h   &=~ n_+\left( -r_+\rho_{11} -\epsilon r_+\rho_{12} + 4P\rho_{21} + 4P\epsilon\rho_{22}\right)\,,\\
H   &=~ n_-\left( -r_-\rho_{11} -\epsilon r_-\rho_{12} + 4P\rho_{21} + 4P\epsilon\rho_{22}\right)\,,\\
D_1 &=~ -\epsilon\rho_{21} + \rho_{22}\,,                              & D_2 &=~ \rho_{23} \,,\\
G_1 &=~ -\epsilon \rho_{11} + \rho_{12}\,, & G_2 &=~ \rho_{13}\,,
\end{aligned}
\end{equation} 
where $r_{\pm}:=\widetilde\Omega\pm\sqrt{(4P)^2+\widetilde\Omega^2}$ and
$n_\pm:=\left(2(4P)^2+2\widetilde\Omega \,r_{\pm}\right)^{-\nicefrac12}$.

On the other hand, the quadratic terms of the pseudoscalar fields $\eta$ are  
\begin{align}
V_\phi(\eta_{21},\eta_{22},\eta_{23})~\supset~
\left(\frac{1}{2} \xi \delta_{IJ} + \frac{1}{2} \omega_{2,IJ} - \frac12\kappa_{2,IJ}\right) \eta_{2I} \eta_{2J}\,,
\label{eq:quadratic3}
\end{align}
where $\omega_{2,IJ}$ and $\kappa_{2,IJ}$ are given in eq.~\eqref{eq:ctes}.
There are three Goldstone bosons $\eta_{11}$, $\eta_{12}$ and $\eta_{13}$. 
We denote the massive states as $\tilde{\eta}_1$, $\tilde{\eta}_2$ and $\tilde{\eta}_3$. 
Their squared eigenmasses are
\begin{equation}
m_{\tilde{\eta}_1}^2 ~=~ \xi + \Omega'_2 - \left|\Omega'_2\right| \,,\qquad 
m_{\tilde{\eta}_2}^2 ~=~ \xi + \Omega'_2 + \left|\Omega'_2\right| 
\qquad\text{and}\qquad 
m_{\tilde{\eta}_3}^2 ~=~ \xi\,.
\end{equation}
Notice that either $m_{\tilde{\eta}_1}^2$ or $m_{\tilde{\eta}_2}^2$ will be equal 
to $m_{\tilde{\eta}_3}^2$ depending on the sign of $\Omega_2'$, so that two 
states are degenerate. The mass eigenstates are
\begin{equation}
\tilde{\eta}_1 = N_-(q_- \eta_{21} + \eta_{22}) \,, \qquad 
\tilde{\eta}_2 = N_+(q_+ \eta_{21} + \eta_{22}) \,, \qquad 
\tilde{\eta}_3 = \eta_{23}\,,
\end{equation}
where 
\begin{equation}
q_{\pm} \coloneq \frac{1\pm\text{sgn}(\Omega_2')}{2\epsilon}\,,\qquad 
N_\pm \coloneq (q_\pm^2 + 1)^{-\frac{1}{2}}\qquad\text{and}\qquad \Omega_2' \coloneq 
\frac{1}{2}v_{11}^2\left(\frac12\lambda_{4} - \lambda_5 \right)\,.
\end{equation}
{Admittedly, this is a very rich chargeless Higgs spectrum. One can in principle
follow analogous steps for the charged Higgs fields and gain a full picture of the 
Higgs sector.

\vskip 2mm
On the bright side, we observe that the general scenario of a six-Higgs model 
based on non-SUSY heterotic orbifold compactifications leads to a rich Higgs sector 
with a plausible scalar DM candidate built in a Higgs portal, which seems to be a
generic feature. Since the model includes 9 free parameters within the Higgs 
sector and 14 if we also count those of the Higgs-portal sector demanding \CP 
conservation, it appears quite simple to dial these parameters to fit Higgs and DM 
observable constraints. On a less positive note, even with the strong {\it ad hoc}
assumptions we imposed, it is precisely the richness of the Higgs sector, with its 
large chargeless and charged Higgs mass spectrum, what challenges the predictivity 
of the model. On the one hand, this introduces plenty of unobserved particles; on 
the other, studying their couplings to the DM candidate $S$ requires an analysis 
that seems at best very difficult. Of course, addressing the most general case, 
with no simplifications, would further lead to a scenario which can only be studied 
numerically and which most likely will face severe computational constraints.
It seems then natural, as a starting point, to turn to study string models with a 
smaller number of Higgses, even though the price to pay shall be to have to deal 
with additional exotic particles.

\section{A stringy realization of Higgs portals with two Higgses}
\label{sec:twoHiggses}

Let us now discuss a SM-like model arising from heterotic orbifolds with
geometry $\Z2\x\Z4$ (2,4) that shares some properties with the previous
model, but presents a simpler scalar sector with only two Higgs doublets. As
in the previous section, we shall assume \CP conservation.
The model is defined by the shift vectors and Wilson lines
\begin{subequations}
\begin{eqnarray}
V_1 &=& (-\tfrac{5}{4}, -\tfrac{1}{4}, \tfrac{1}{4}, \tfrac{1}{4}, \tfrac{1}{4}, \tfrac{1}{4}, \tfrac{1}{4}, \tfrac{1}{4}),(-\tfrac{7}{4}, -\tfrac{7}{4}, -\tfrac{1}{4}, -\tfrac{1}{4}, -\tfrac{1}{4}, -\tfrac{1}{4}, \tfrac{1}{4}, \tfrac{1}{4})\,,\\
V_2 &=& (\tfrac{5}{8}, -\tfrac{3}{8}, -\tfrac{7}{8}, -\tfrac{1}{8}, -\tfrac{1}{8}, \tfrac{1}{8}, \tfrac{1}{8}, \tfrac{5}{8}),(-\tfrac{7}{8}, -\tfrac{3}{8}, -\tfrac{1}{8}, \tfrac{1}{8}, \tfrac{1}{8}, \tfrac{7}{8}, -\tfrac{7}{8}, \tfrac{5}{8})\,,\\
W_3 = W_4 = W_6 &=& (-\tfrac{7}{4},\tfrac{5}{4},\tfrac{5}{4},\tfrac{3}{4},\tfrac{3}{4},-\tfrac{3}{4},\tfrac{1}{4},\tfrac{5}{4}),(\tfrac{1}{4},\tfrac{1}{4},\tfrac{7}{4},\tfrac{1}{4},\tfrac{3}{4},\tfrac{5}{4},\tfrac{9}{4},\tfrac{9}{4})\,,\\
W_5 &=& (0,0,0,0,0,0,0,0),(0,1,-2,1,1,1,2,-2)\,,
\end{eqnarray}
\end{subequations}
and $W_1=W_2=0$. The resulting 4D gauge group is
\begin{equation}
\mathcal{G}_{\text{4D}} ~=~ \mathcal{G}_{\text{SM}} \x \mathcal{G}'\x [\U1']^8\,,
\qquad\text{where}\quad 
\maG' ~=~ \SU2_\text{flavor}\x\left[\SU3\x\SU2\right]_\text{hidden}
\label{eq:g4d}
\end{equation}
and one $\U1'$ is anomalous. The massless spectrum with respect to 
$\mathcal{G}_{\text{SM}}\x\SU2_\text{flavor}$ is shown in table~\ref{tab:spectrumSM-m5115}, 
where we follow the same notation as in the previous section. 
The detailed spectrum under the full 4D gauge group $\mathcal{G}_{\text{4D}}$ 
in eq.~\eqref{eq:g4d} is presented in tables~\ref{tab:fullf} and~\ref{tab:fulls} 
of appendix~\ref{appsect:spect2HDM} for the fermion and scalar particles, respectively.
We observe that this model exhibits two Higgs doublets, and an exotic sector including
some scalar leptoquarks and vectorlike down quarks, all of which can develop masses as
some scalars spontaneously break the symmetries of the hidden sector.

\begin{table}[t!]
\begin{center}
{
   \begin{tabular}{|rlc|c|rlc|}
   \cline{1-3}\cline{5-7} 
   \#  & Fermionic irrep.   & Label &&  \#  & Scalar irrep. & Label \\
   \cline{1-3}\cline{5-7}
     3 & $(\rep1, \rep2; \rep1)_{\nicefrac{-1}{2}}$                  & $\ell_{\mathrm{L},i}$ 
&&   2 & $(\rep1, \rep2; \rep1)_{\nicefrac{1}{2}}$                   & $\phi_i$ \rule{0pt}{2.5ex}\\
     1 & $(\rep1, \rep1; \rep2)_{1}$                                 & $\overline{e}_\mathrm{L}$ 
&&   1 & $(\rep1, \rep1; \rep1)_{0}$                                 & $S$     \\ 
     1 & $(\rep1, \rep1; \rep1)_{1}$                                 & $\overline{e}_{\mathrm{L},3}$ 
&&   107  &   $(\rep1, \rep1; \rep1)_{0}$                            & $s_i$  \\ 
     1 & $(\rep3, \rep2; \rep2)_{\nicefrac{1}{6}}$                   & $q_\mathrm{L}$ 
&& 8 &  $(\rep1, \rep1; \rep2)_{0}$                                  & $s'_i$  \\ 
     1 & $(\rep3, \rep2; \rep1)_{\nicefrac{1}{6}}$                   & $q_{\mathrm{L},3}$ 
&& 2 &  $(\crep3, \rep1; \rep2)_{\nicefrac{1}{3}}$                   & $x{_i}$ \\
   \cline{5-7}
     1 & $(\crep3, \rep1; \rep2)_{\nicefrac{-2}{3}}$                 & $\overline{u}_\mathrm{L}$ 
 & \multicolumn{4}{c}{}\\
     1 & $(\crep3, \rep1; \rep1)_{\nicefrac{-2}{3}}$                 & $\overline{u}_{\mathrm{L},3}$ 
 & \multicolumn{4}{c}{}\\
     5 & $(\crep3, \rep1; \rep1)_{\nicefrac{1}{3}}$                  & $\overline{d}_{\mathrm{L},i}$ 
 & \multicolumn{4}{c}{}\\
     2 & $(\rep3, \rep1; \rep1)_{-\nicefrac{1}{3}}$                  & $d'_{\mathrm{L},i}$ 
 & \multicolumn{4}{c}{}\\
   131 & $(\rep1, \rep1; \rep1)_{0}$                                 & $\nu_{\mathrm{R},i}$ 
 & \multicolumn{4}{c}{}\\
   14 & $(\rep1, \rep1; \overline{\rep2})_{0}$                       & $\nu'_{\mathrm{R},i}$ 
 & \multicolumn{4}{c}{}\\ 
    \cline{1-3}  
    \end{tabular}
    \caption{Massless spectrum for the SM-like model with two Higsses with quantum numbers under $\maG_\text{SM}\x\SU2_{\text{flavor}}$, 
          where the hypercharge is displayed as subscripts. $\mathrm{L}$ and $\mathrm{R}$ denote 
          left and right fermion projections and a bar over a fermion indicates the Dirac conjugate. The fermion families in
          $\ell_{\mathrm{L},i}$ and $\overline{d}_{\mathrm{L},i}$ are distinguished by a $\U1'$ charge. The complete spectrum under 
          $\maG_{\text{4D}} = \maG_{\text{SM}} \x \maG'\x \U1'^8$, where $\maG_{\text{SM}} = \SU3_c \x \SU2_L \x \U1_Y$ and $\maG' = 
          \SU2_{\text{flavor}}\x\SU3\x\SU2$, is presented in table~\ref{tab:fullf} and~\ref{tab:fulls} of appendix~\ref{appsect:spect2HDM}, 
          for the fermions and scalars, respectively. As in the previous model, chiral fermions do not yield anomalies,
          as expected from its string nature~\cite{Casas:1987us,Kobayashi:1996pb}.
          \label{tab:spectrumSM-m5115}}
}
\end{center}
\end{table} 

The renormalizable $\mathcal{G}_{\text{4D}}$-invariant Yukawa couplings are given 
at leading order by 
\begin{align}\label{eq:VYuk2}
\begin{split}
V_Y(\phi_i,\ell_{\mathrm{L},i},...) & = \mathcal{Y}_{e, 33}^i \,
\overline{e}_{\text{L},3} \phi^\dagger_i \ell_{\mathrm{L},3}  
+ \mathcal{Y}^j_{\nu,ik}\,\epsilon^{ab} 
\overline{\nu}_{\text{L},14i} \phi_{ja} \ell_{\mathrm{L},kb} \\ 
& + \mathcal{Y}^i_{d, 33}\,\overline{d}_{\mathrm{L},3}\phi^\dagger_i q_{\mathrm{L},3} 
+ \mathcal{Y}^i_{u,IJ} \epsilon^{ab}\,
\overline{u}_{\text{L},I}\phi_{ia} q_{\mathrm{L},Jb} + \text{h.c.} \quad
\text{with} \quad \mathcal{Y}^i_{u, IJ} = \mathcal{Y}^i_{u} \epsilon_{IJ}\,, 
\end{split}
\end{align}
where $i,j=1,2$ and $k=1,2,3$ are multiplicity indices, $a,b=1,2$ are $\SU2_L$ indices 
and $I,J=2,3$ are $\SU2_\text{flavor}$ indices. Further Yukawa couplings appear at 
higher orders suppressed by scalar singlet VEVs. Note that $\overline{d}_{\mathrm{L},3}$
is one of the mass eigenstates of the model, as discussed in appendix~\ref{app:Md2HDM}.
On the other hand, the interaction potential for the scalar leptoquarks $x_i$ 
with SM fermions is given by
\begin{equation}
\label{eq:leptoquark-int}
\begin{split}
V_x(x_i,...) & =  
\Gamma_{q\ell}^i \,\epsilon^{IJ}\epsilon^{ab} \overline{q}^\mathrm{C}_{\text{R},Ia} x_{i,J} \ell_{\text{L},3b} 
+ \Gamma^i_{ue_3}\,\delta^{IJ} \overline{u}^\mathrm{C}_{\text{L},I} x_{i,J} e_{\text{R},3} \\ 
& +  \Gamma^i_{u_3e}\, \delta^{IJ} \overline{u}^\mathrm{C}_{\text{L},3} x_{i,I} e_{\text{R},J} 
+ \Gamma^\nu_{ij}\,\overline{d}^\mathrm{C}_{\text{L},i} (x_{1,c} + x_{2,c}) \delta^{cd} \nu'_{\text{R},12}{_{jd}} \\  
& +  \Gamma^i_{qq_3}\,\delta^{IJ}\epsilon^{ab} \overline{q}^\mathrm{C}_{\text{R},Ia}x^\dagger_{i,J} q_{\text{L},3b}   
+ \Gamma^i_{ud_3}\,\epsilon^{IJ} \overline{u}^\mathrm{C}_{\text{L},I} x^\dagger_{i,J} d_{\text{R},3} 
+ \text{h.c.}\,,
\end{split}
\end{equation}
where the superscript $\mathrm C$ stands for charge conjugation, such that 
$\overline{q}^\mathrm{C}_\text{R}$ is a left handed fermion. Further, $i,j=1,2,3,4$, 
$a,b=1,2$ are $\SU2_L$ indices, and $I,J=1,2$ are $\SU2_\mathrm{flavor}$ indices. We 
give explicitly the gauge invariant $\SU2_\mathrm{flavor}$ components, whose detailed
charges are given in table~\ref{tab:fullf} of appendix~\ref{appsect:spect2HDM}. 
We omit here a detailed analysis of the flavor sector of the model and assume that
the couplings of leptoquark interactions~\eqref{eq:leptoquark-int} are suppressed with
respect to all other interactions because we are rather interested in the properties 
of the Higgs sector. However, note that this feature must be studied with great 
care as it might lead to undesirable observations~\cite{Dorsner:2012nq,Becirevic:2016oho,Patel:2022wya}.

Let us now study the details of the Higgs sector. Fortunately, as we shall see, 
the results here turn out to be special cases of our previous discussion in the model 
with six Higgses. The only difference is that, in this case, the two Higgs doublets
\begin{equation}
\phi_1 = \begin{pmatrix} \phi_1^+ \\ \phi_1^0 \end{pmatrix}\,\quad\text{and}\quad\,\phi_2 
= \begin{pmatrix} \phi^+_2 \\ \phi^0_2 \end{pmatrix},
\label{eq:Higgsesn}
\end{equation}
are singlets under $\maG'$. To start with, we must now determine its potential.

Taking into account all charges given in
appendix~\ref{appsect:spect2HDM}, one finds that the interaction potential for the 
two Higgs doublets is\footnote{To provide some insight also valid for bottom-up models, we retain 
here $\lambda_3\neq\lambda_4$ as in such constructions, even though they are in principle 
equal in string-derived models. This should not alter the final results.}
\begin{equation}
\begin{split}
\label{eq:VH}
V_\phi(\phi_1,\phi_2) & = \mu^2_{11} |\phi_1|^2 + \mu^2_{22} |\phi_2|^2  
+ \lambda_1 |\phi_1|^4 + \lambda_2  |\phi_2|^4 + \lambda_3 |\phi_1|^2 |\phi_2|^2 
+ \lambda_4 (\phi^\dagger_1 \phi_2) (\phi^\dagger_2 \phi_1)\\
& +  \left[\,\mu^2_{12} \phi^\dagger_1 \phi_2 + \,\lambda_5 (\phi^\dagger_1 \phi_2)(\phi^\dagger_1 \phi_2) +  \,\lambda_6 |\phi_1|^2(\phi^\dagger_1 \phi_2) +\,\lambda_7 |\phi_2|^2(\phi^\dagger_1 \phi_2) + \text{h.c.}\right], 
\end{split}
\end{equation}
where we can assume that \CP is conserved, allowing us to take 
$\lambda_5,\lambda_6,\lambda_7$ and $\mu_{12}$ as real parameters. For the 
scalar DM candidate we choose the singlet denoted by $S$ from the scalar 
spectrum shown in table~\ref{tab:fulls} of appendix~\ref{appsect:spect2HDM}. 
This scalar is a singlet under $\mathcal{G}_\text{SM}\x\mathcal{G}'$. 
We assume that $S$ does not acquire a VEV so that it does not mix with the 
detected Higgs field(s). The potential for the scalar DM candidate $S$ is
again
\begin{equation}
\label{eq:Vs}
V_S(S) ~=~ \mu^2_S |S|^2 + \lambda_S |S|^4\,,
\end{equation}
and the interaction potential of the scalar DM with the two Higgs doublets 
takes the form (see
\begin{equation}
\label{eq:Vhdm}
V_{\phi S}(\phi,S) ~=~ \lambda_{1S} |\phi_1|^2|S|^2 + \lambda_{2S} |\phi_2|^2|S|^2 
                    + \lambda_{12S}\left[(\phi^\dagger_2 \phi_1)|S|^2 + \text{h.c.}\right]\,,
\end{equation}
where we take $\lambda_{12S}$ to be real since we assume a \CP-conserving 
scalar sector. In the Higgs-DM scalar sector we have then 15 real parameters, 
ten from eq.~\eqref{eq:VH}, two from eq.~\eqref{eq:Vs} and three from eq.~\eqref{eq:Vhdm}.
Note that the counting of real parameters exceeds by one the number of 
free parameters in the model with six Higgses. This arises from considering 
$\lambda_3$ and $\lambda_4$ independent here, which we do for generality.

Recalling that the cancellation of tadpoles implies conditions to minimize
the potential at tree level, we study the expansion of $V_\phi$ in eq.~\eqref{eq:VH}
close to the vacuum defined by the VEVs of the Higgses. In the vacuum the Higgs 
doublets are parametrized as
\begin{equation}
\label{eq:Higgsesn}
\phi_1 \longrightarrow 
\begin{pmatrix} \phi_1^+ \\ \frac{1}{\sqrt{2}}(v_1 + \sigma_1) \end{pmatrix}
\qquad\text{and}\qquad\,
\phi_2 \longrightarrow 
\begin{pmatrix} \phi^+_2 \\ \frac{1}{\sqrt{2}}(v_2 + \sigma_2) \end{pmatrix},
\end{equation}
where $v_{1,2}$ are taken to be real, and $\sigma_{1,2} = \rho_{1,2} + \I\,\eta_{1,2}$. 
We substitute eq.~\eqref{eq:Higgsesn} in eq.~\eqref{eq:VH} and split it as
$V_\phi=V_{\phi,1} + V_{\phi,2}$, where $V_{\phi,1}$ includes only the linear terms
in $\sigma_{1,2}$, and $V_{\phi,2}$ denotes the quadratic terms for the charged fields 
$\phi^\pm_{1,2}$, the neutral scalars $\rho_{1,2}$ (\CP-even) and the neutral pseudoscalars 
$\eta_{1,2}$ (\CP-odd). Then, the tadpole-cancellation conditions are obtained by 
requiring the absence of linear terms 
in $\sigma_{1,2}$, and the physical masses for the Higgses are obtained by diagonalizing 
the mass matrices that enter in the quadratic terms for the fields $\phi^\pm_{1,2}$, 
$\rho_{1,2}$, and $\eta_{1,2}$. 

The linear terms in $\sigma_{1,2}$ of the potential contain the tadpole contributions
\begin{equation}
V_{\phi,1} ~=~ (v_1 \zeta + v_2 \gamma) \rho_1 + (v_1 \gamma + v_2 \kappa) \rho_2\,,
\end{equation}
which must be cancelled by imposing the conditions (analogous to eq.~\eqref{eq:tadpoleCancel})
\begin{equation}
\label{eq:constraintsn}
v_1 \zeta  + v_2 \gamma ~\stackrel{!}{=}~ 0
\qquad\text{and}\qquad\,
v_1 \gamma + v_2 \kappa ~\stackrel{!}{=}~ 0\,,
\end{equation}
where, equivalently to our findings in the six-Higgs case, see eq.~\eqref{eq:tadpoleTerms},
\begin{subequations}
\label{eq:expressionsn}
\begin{eqnarray} 
\zeta  & = & \mu^2_{11} + \lambda_1 v^2_1 + \tfrac{1}{2}\lambda_3 v^2_2 + \lambda_6 v_1 v_2\,,\\
\gamma & = & \mu^2_{12} + \tfrac{1}{2}(\lambda_{4} + 2\lambda_5) v_1 v_2 + \tfrac{1}{2}\lambda_6 v^2_1 + \tfrac{1}{2}\lambda_7 v^2_2\,,\\
\kappa & = & \mu^2_{22} + \lambda_2 v^2_2 + \tfrac{1}{2}\lambda_3 v^2_1 + \lambda_7 v_1 v_2\,.
\end{eqnarray}
\end{subequations}
From eq.~\eqref{eq:constraintsn} we conclude that $\zeta\kappa=\gamma^2$ and 
$\zeta/\kappa=v_2^2/v_1^2$. 

The mass matrix for the charged fields is obtained from the quadratic terms in $\phi^{\pm}_{1,2}$,
\begin{equation}
\label{eq:Vquadcharged}
V_{\phi,2}(\phi^\pm_1,\phi^\pm_2) = \left(-\mu^2_{12} - \tfrac12(\lambda_4 + 2\lambda_5) v_1 v_2 - \tfrac12\lambda_6 v_1^2 - \tfrac12\lambda_7 v^2_2\right) (\phi^-_1,\phi^-_2)
\begin{pmatrix} \frac{v_2}{v_1} & -1 \\ -1 & \frac{v_1}{v_2} \end{pmatrix}
\begin{pmatrix} \phi^+_1 \\ \phi^+_2 \end{pmatrix}.
\end{equation}
The mass or physical eigenstates correspond to a massless charged Goldstone boson $G^+$, 
which gets eaten by the $W^+$ boson, and a massive charged Higgs $H^+$ with squared mass 
\begin{equation}
\label{eq:mHp}
m^2_{H^+} ~=~ \left(\frac{-\mu^2_{12}}{v_1 v_2}-\tfrac{1}{2}(\lambda_4 + 2\lambda_5) 
           - \tfrac{1}{2}\lambda_6 \frac{v_1}{v_2}-\tfrac{1}{2}\lambda_7 \frac{v_2}{v_1}\right)(v^2_1 + v^2_2)\,.
\end{equation}
These physical states are given by the orthogonal combinations
\begin{equation}
G^+ ~=~ \frac{1}{\sqrt{v^2_1 + v^2_2}}(v_1 \phi^\pm_1 + v_2 \phi^\pm_2)\qquad\text{and}\qquad
H^+ ~=~ \frac{1}{\sqrt{v^2_1+v^2_2}}(-v_2 \phi^\pm_1 + v_1 \phi^\pm_2)\,,
\end{equation}
where $\tan\beta:=v_2/v_1$ defines the Higgs mixing angle $\beta$. 

For the neutral pseudoscalar fields we obtain
\begin{equation}
V_{\phi,2}(\eta_1,\eta_2) ~=~ \frac12\left(-\mu^2_{12} - 2\lambda_5 v_1 v_2 - \tfrac{1}{2}\lambda_6 v^2_1 - \tfrac{1}{2}\lambda_7 v^2_2\right) (\eta_1,\eta_2)
\begin{pmatrix} \frac{v_2}{v_1} & -1 \\ -1 & \frac{v_1}{v_2} \end{pmatrix}
\begin{pmatrix} \eta_1 \\ \eta_2 \end{pmatrix}.
\label{eq:V2neutral2n}
\end{equation}
We get one massless Goldstone boson $G$, which gives mass to the $Z$ boson, 
and one massive neutral pseudoscalar $A$ whose squared mass is
\begin{equation}
\label{eq:mA}
m^2_{A} ~=~ \left(\frac{-\mu^2_{12}}{v_1 v_2}- 2\lambda_5 - \tfrac{1}{2}\lambda_6 \frac{v_1}{v_2}-\tfrac{1}{2}\lambda_7 \frac{v_2}{v_1}\right)
            (v^2_1 + v^2_2)\,.
\end{equation}
The physical states are given by
\begin{equation}
G ~=~ (\cos\theta) \eta_1 + (\sin\theta) \eta_2
\qquad\text{and}\qquad
A ~=~ -(\sin\theta) \eta_1 + (\cos\theta) \eta_2\,.
\end{equation}
Further, for the neutral scalars we have
\begin{equation}
V_{\phi,2}(\rho_1,\rho_2) ~=~ \frac{1}{2}(\rho_1,\rho_2) 
\begin{pmatrix} B & C \\ C & D \end{pmatrix}
\begin{pmatrix} \rho_1 \\ \rho_2 \end{pmatrix},
\label{eq:V2neutral1n}
\end{equation}
where
\begin{subequations}
\label{eq:BDC}
\begin{eqnarray}
 B & = & -\mu^2_{12}\frac{v_2}{v_1} + 2\lambda_1 v_1^2 + \tfrac{3}{2}\lambda_{6}v_1 v_2 - \tfrac{1}{2}\lambda_{7}\frac{v^3_2}{v_1}\,,\\
 D & = & -\mu^2_{12}\frac{v_1}{v_2} + 2\lambda_2 v_2^2 + \tfrac{3}{2}\lambda_{7}v_1 v_2 - \tfrac{1}{2}\lambda_{6}\frac{v^3_1}{v_2}\,,\\ 
 C & = & \mu^2_{12} + \lambda_3 v_1 v_2 + \lambda_4 v_1 v_2 + 2\lambda_5 v_1 v_2 + \tfrac{3}{2}\lambda_{6}v^2_1 + \tfrac{3}{2}\lambda_{7}v^2_2\,.
\end{eqnarray}
\end{subequations}
In this case, we obtain two physical massive neutral scalars 
\begin{equation}
h ~=~ (\cos\alpha) \rho_1 - (\sin\alpha) \rho_2
\qquad\text{and}\qquad
H ~=~ (\sin\alpha) \rho_1 + (\cos\alpha) \rho_2\,.
\end{equation}
The angle $\alpha$ is defined by $\tan 2\alpha = 2C/(D-B)$. The squared masses are
\begin{equation}
\label{eq:mhH}
m^2_{h(H)} ~=~ \frac{1}{2}\left((B+D)\mp\sqrt{(B-D)^2 + 4C^2}\right)\,,
\end{equation} 
where $\sqrt{(B-D)^2 + 4C^2}=(D-B)/\cos 2\alpha = 2C/\sin 2\alpha$. The 
expressions~\eqref{eq:mHp},~\eqref{eq:mA} and~\eqref{eq:mhH} give the tree-level 
masses for the charged scalars $H^\pm$, the pseudoscalar $A$, and the light $h$ 
and heavy $H$ scalars, respectively. We identify the light scalar 
$h$ as the SM Higgs whose mass must be constrained to be 
$m_h \approx 125$\,GeV, as observed at the LHC.  

Using eq.~\eqref{eq:Vs} and eq.~\eqref{eq:Higgsesn} in the DM-Higgs potencial~\eqref{eq:Vhdm},
we obtain the squared mass for the scalar DM candidate $S$,
\begin{equation}
m^2_S ~=~ \mu^2_S + \tfrac{1}{2}\lambda_{1S} v_1^2 + \tfrac{1}{2}\lambda_{2S} v_2^2 + \lambda_{12S} v_1 v_2\,. 
\end{equation}

This analytical treatment can bring us just this far. In order to 
arrive at phenomenological predictions, involving both the Higgs and DM sectors,
we must proceed with a numerical study, which is straightforward for this
model.

\section{Fitting DM and Higgs data in string-derived Higgs portals}
\label{sec:results}
Given the properties of the model described in the previous section, we are
now ready to analyze its phenomenological viability. The study presented in this 
section can be considered to be valid also for other bottom-up inspired models 
with analogous properties. We perform an increasingly 
detailed scan of the parameter space of the model in order to arrive at a best 
fit to some observable data. In particular, we aim at the best parameters that 
yield
\begin{enumerate}
\item Higgs-vacuum stability at 1-loop, avoiding metastable (and unstable) vacua;
\item Admissibly heavier states in the Higgs sector; and
\item Compatibility with the Higgs and DM observables given by~\cite{ParticleDataGroup:2022pth,Planck:2018vyg} 
\begin{equation}
\label{eq:observables}
m_h ~=~ 125.25(17)\,\text{GeV}, \quad 
v ~=~ 246.219640(63)\,\text{GeV} \quad \text{and} \quad 
\Omega_\mathrm{DM} h^2 ~=~ 0.120(1)\,, 
\end{equation}
where the electroweak VEV arises from the Fermi constant via 
$v=(\sqrt{2}G_F)^{-1/2}$ and in our model is given by $v^2=v_1^2+v_2^2$.
\end{enumerate}
For the stability of the Higgs potential, we must verify that with
the chosen parameters the potential exhibits a global minimum, so that
quantum transitions do not affect the vacuum. 
Note that this demands us to identify the parameter region free of tachyons. 
On the other hand, for our goal number 2, we impose for the scalar masses 
$m_H,m_{H^\pm}$ and for the pseudoscalar mass $m_A$
a conservative lower limit of 50 GeV above the mass $m_h$ of the lightest 
(observed) Higgs field $h$, as admissible in most scenarios contrasted
with data~\cite{ParticleDataGroup:2022pth}.

Let us now make a couple of useful observations about the Yukawa sector. 
Spontaneous electroweak symmetry breakdown, parametrized by eq.~\eqref{eq:Higgsesn}, 
implies that the leading-order interactions of eq.~\eqref{eq:VYuk2} yield large 
masses for SM fermions of the third generation. Incidentally, at this level the 
two heavier up-type quarks are degenerate, which can only be prevented by 
carefully choosing the $\mathcal Y_{u,IJ}$ couplings, the Higgs VEVs and
subleading couplings arising \`a la Froggatt-Nielsen~\cite{Froggatt:1978nt}
to due some scalar VEVs.
The details of flavor phenomenology are beyond the scope of this work and
will be assumed in the following to be suitable. For our discussion on
Higgs portals, it will be important to regard only the interactions
between the two Higgs doublets and the heaviest fermions governed by their
associated $\mathcal Y$ couplings, in turn fixed by the observed fermion 
masses. Since third-generation couplings are the largest in the SM, it is safe
to consider in the annihilation process of the DM candidate $S$ to SM fermions
that these couplings dominate the contributions to the DM annihilation cross section.

Let us examine the parameters to scan. As in the 
models with six Higgs doublets, we consider the parameters of the Higgs potential
eq.~\eqref{eq:VH} and the Higgs portal eq.~\eqref{eq:Vhdm}:
\begin{subequations}
\label{eq:params}
\begin{eqnarray}
\label{eq:paramsH}
&&\mu_{12}^2, \, \lambda_i \hspace{2cm} \text{with} \quad i=1,\dots,7\,, \\
\label{eq:paramsS}
&&\mu_S^2, \, \lambda_S, \, \lambda_{1S}, \, \lambda_{2S}, \, \lambda_{12S}\,. 
\end{eqnarray}
\end{subequations}
Notice that $\mu_{11}^2$ and $\mu_{22}^2$ have been omitted here. Instead, we include 
$\tan\beta=v_2/v_1$ and the electroweak VEV $v$ as parameters. An advantage of this 
choice is that $v$ is directly fixed (is not free) by observations and, as we shall see, 
is numerically accessible too. Further, the tadpole-cancellation 
conditions~\eqref{eq:constraintsn} constrain the original parameters at tree level.
Since in addition we demand \CP invariance, we end up with nine real free parameters 
arising from the Higgs potential, and five more from the Higgs portal of the model.

Prior to our scan, we can set a couple of constraints over some of the parameters.
We shall explore the parameter space with $\mu_{12}^2<0$ and $\lambda_4<0$. The former 
condition is required to arrive at non-trivial VEVs for the Higgs doublets while the 
latter is found to be relevant to prevent tachyonic masses in the Higgs sector.
In addition, the constraints $\mu_S^2,\,\lambda_S>0$ have to be imposed so that 
the scalar DM candidate has a trivial VEV and its potential is bounded from below. 

In order to numerically analyze the parameter space of our stringy 2HDM, we 
define its symmetries, matter content, parameters and Lagrangian in the 
\texttt{SARAH 4.14.5}~\cite{Staub:2008uz,Staub:2013tta,Goodsell:2018tti}
package for \texttt{Mathematica}. This package produces input files for\footnote{The 
relevant files used to define the model in \texttt{SARAH}, the input files and 
routines generated for \texttt{SPheno}, \texttt{micrOMEGAs} and \texttt{VevaciousPlusPlus} 
are available upon request. Please, send your inquiries preferably to \texttt{omar\_perfig@ciencias.unam.mx}.}
\begin{itemize}
\item \texttt{VevaciousPlusPlus}~\cite{JoseEliel:Vpp2014} (the \texttt{C++} version of 
\texttt{Vevacious}~\cite{Camargo-Molina:2013qva}), which we use to determine
the Higgs VEVs $(v_1,v_2)$ at several identified vacua (minima of the Higgs potential)
and their stability status;
\item \texttt{SPheno 4.0.5}~\cite{Porod:2003um,Porod:2011nf}, used here to 
compute the mass spectrum, in particular the masses of the Higgs sector, 
including 1-loop corrections; and
\item \texttt{micrOMEGAs 5.3.35}~\cite{Belanger:2018ccd}, that helps us 
compute the freeze-out DM relic abundance, including 1-loop annihilation processes.
\end{itemize}
In order for the files produced by \texttt{SARAH} to be useful for these
other packages and for our analysis, a few relevant flags must be 
activated~\cite{Staub:2015kfa}: 
\begin{itemize}
\item RGEs are calculated to 1-loop order,
\item Calculate 1-loop corrections to masses, 
\item Assume real mixing matrices (\texttt{micrOMEGAs} does not handle complex entries),
\item Print tree-level values of parameters involved in tadpole equations 
(\texttt{VevaciousPlusPlus} makes use of those values), and
\item Check if tree-level unitarity constraints are respected. 
\end{itemize}

In our scan, the routines have to be executed as follows. First, with a dedicated
routine we assign random values for the parameters to set a point $\boldsymbol{p}$
in the parameter space; these values are given to \texttt{VevaciousPlusPlus} and 
\texttt{SPheno}. \texttt{VevaciousPlusPlus} yields the minimum characterized by 
the values $(v_1,v_2)$, the stability status of the identified vacuum, and its lifetime 
(for metastable vacua). The conditions to consider admissible an outcome are: 
i) the quotient $v_2/v_1$ must be consistent with the input random value of $\tan\beta$,
ii) the value of $v^2=v_1^2+v_2^2$ must be within a few $\sigma$ close to the 
observable value~\eqref{eq:observables}, and 
iii) the vacuum must be stable.
Taking the values of $\boldsymbol{p}$ and the VEVs provided by \texttt{VevaciousPlusPlus},
\texttt{SPheno} is run to yield the masses and mixings of the various Higgs fields, 
and the mass of the DM candidate. 
The parameters are considered still suitable if $m_h$ is within a few $\sigma$ 
from its central observable value and $m_{H,H^\pm,A}$ are at least 50 GeV above $m_h$.
Finally, if these conditions are met, the data is run by \texttt{micrOMEGAs}, 
which reads from the spectrum file generated by \texttt{SPheno} all promising
parameters, and utilizes the routines built automatically by \texttt{SARAH} to 
compute the freeze-out DM relic density. 
In the case that it is found close to the observed value of 
$\Omega_\mathrm{DM}h^2$ of eq.~\eqref{eq:observables}, $\boldsymbol{p}$ is
considered a viable point. The masses $m_{H,H^\pm,A}$ and $m_S$ produced
by \texttt{SPheno} as well as the chosen value of $\tan\beta$ are considered
predictions of the scheme.

Our scan is split in two stages. The first stage consists in a coarse scan that should
allow us in a limited amount of time to roughly identify the most promising parameter 
regions, focused on the Higgs parameters and observables. In the second fine scan, 
we perform a global search in promising areas, including now the parameters of
eq.~\eqref{eq:paramsS}, with the use of algorithms of $\chi^2$ minimization to arrive 
at models that reproduce the observable values in eq.~\eqref{eq:observables} and 
predict additional observables in the Higgs and DM sectors. In our analysis, we
define $\chi^2$ as usual in terms of its components
\begin{equation}
\label{eq:chi_i}
\chi_\mu ~=~ \frac{y_\mu^\text{meas} - y_\mu^\text{model}(\boldsymbol{p})}{\sigma_\mu}\,,
\qquad\text{such that}\qquad
\chi^2 ~:=~ \sum_{\mu=1}^3 \chi_\mu^2\,,
\end{equation}
where $\mu$ counts the three observables in eq.~\eqref{eq:observables},
$y_\mu^\text{meas}$ denote their measured values, $y_\mu^\text{model}(\boldsymbol{p})$ 
are the values predicted by the model associated with the point $\boldsymbol{p}$ in 
the parameter space, and $\sigma_\mu$ are the corresponding experimental errors.

The first step of our coarse scan is to randomly set values to the parameters 
in eq.~\eqref{eq:paramsH} and $\tan\beta$, leaving the Higgs-portal 
parameters in eq.~\eqref{eq:paramsS} at arbitrary (though reasonable) values. 
In detail, after choosing random values for all free parameters but $\tan\beta$, 
we use a bisection method to find values of $\tan\beta$ yielding stable 
vacua and admissible values of $v$ and $m_h$. At this stage we aim at the $6\sigma$ 
region ($125.25\pm1\,$GeV) of the Higgs mass.
\begin{figure}[t!]
\centering
\subfigure{
\includegraphics[width=0.48\textwidth]{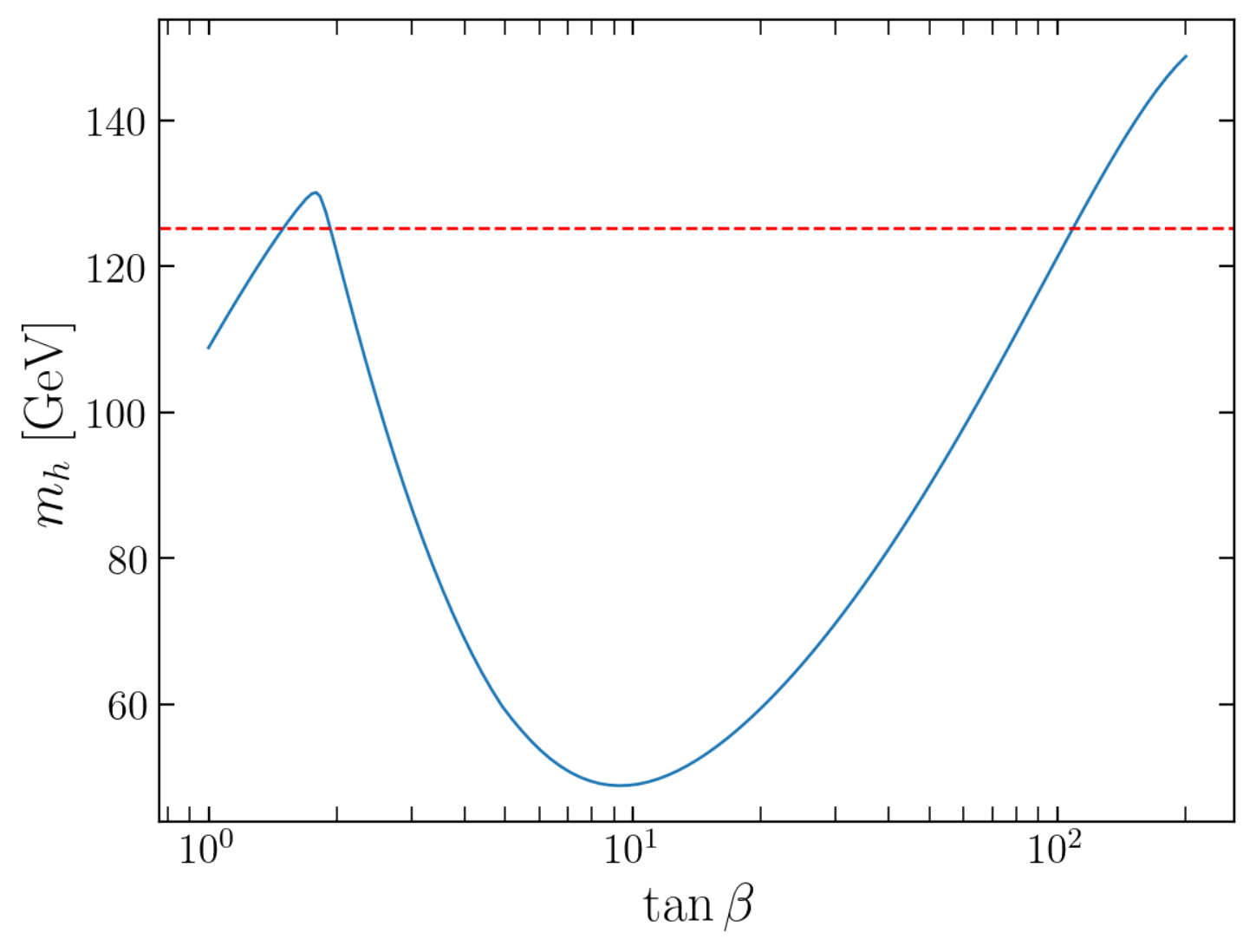}
}
\subfigure{
\includegraphics[width=0.48\textwidth]{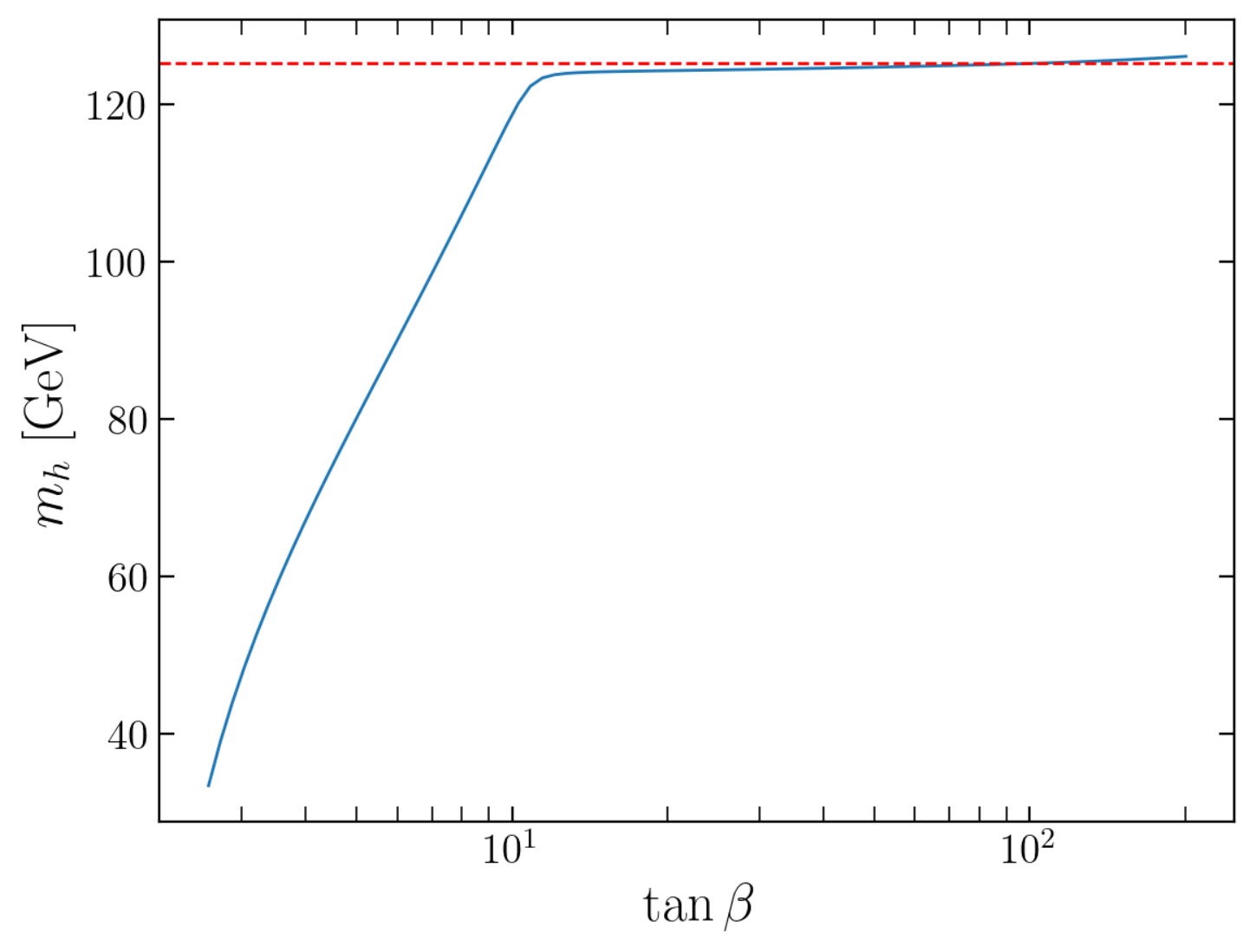}
}
\caption{Comparison of the predicted (continuous curve) and the observed (dashed line)
Higgs mass for different $\tan\beta$ values for two choices of parameters. For a generic 
point in the Higgs parameters~\eqref{eq:paramsH} (left panel), we observe a proper Higgs 
mass for $\tan\beta$ of either order one or order 100. In our best-fit scenario (right panel),
the correct Higgs mass is predicted for $\tan\beta>10$.
\label{fig:mHiggs_tanbeta}}
\end{figure}
Two relevant observations arise at this stage. First, as illustrated in the left 
panel of figure~\ref{fig:mHiggs_tanbeta}, there are points $\boldsymbol{p}$ for 
which the dependence on $\tan\beta$ of the lightest neutral Higgs scalar $m_h$ 
is such that it is compatible with observations for some values around 
$\tan\beta\sim1$ and $\tan\beta\sim100$. Yet $\chi^2$ can be as large as $10^4$ 
in such cases. For $\boldsymbol{p}$ yielding $\chi^2\lesssim9$, we find that the observed 
and predicted values of $m_h$ are compatible as long as $\tan\beta\gtrsim10$, cf.\ right panel of 
figure~\ref{fig:mHiggs_tanbeta} illustrating the result of a point with $\chi^2\sim10^{-4}$.
The second observation of the coarse scan is the constraint
$0.1\leq\lambda_2\leq0.2$, so that stable vacua with correct $m_h$ are 
possible (lower bound), and all heavier Higgs fields are well above $m_h$ (upper bound).

As mentioned earlier, in the fine scan we inspect all the parameters 
of the model, eq.~\eqref{eq:params} and $\tan\beta$, with greater precision. 
We make use of the optimization methods available in the \texttt{Python} package 
\texttt{LMFIT}~\cite{matt_newville_2022_7370358}, evaluating $\chi$ on each
chosen point in the preferred region obtained from the coarse scan. Most 
methods available with \texttt{LMFIT} minimize the $\chi_\mu$ components as 
they explore the parameter space. We performed several iterations of scans 
choosing different methods since it is easy to switch between them. 
By alternating between the {\it differential evolution} and the {\it least 
squares} methods, we are able to transit from $\chi^2\sim10^4$ to $7.6$.
The former method consists on an algorithm to let ``evolve'' 
a population of candidate solutions. New elements are added to the population 
and older ones are discarded by comparing their fitness to the solution. 
The latter method uses the trust region reflective least squares algorithm. 
These two approaches proved to be complementary: with the differential evolution 
method we perform a global exploration of the manually constrained parameter 
space, while the least squares method is used to search locally for a better 
solution starting from an initial point which is previously chosen. 

To refine the search around the best-fit point so far, we chose the ``emcee'' 
method which is a Markov Chain Monte Carlo (MCMC) ensemble sampler~\cite{Foreman-Mackey:2012any}. 
Its objective is not the optimization of the parameters to minimize $\chi^2$, 
but the exploration of the neighborhood of promising points found before. 
After refining the initial point for the MCMC scans a couple of times 
and collecting the points being explored by the sample, we obtained the best fit 
with $\chi^2=1.23\x10^{-4}$ at the parameter point defined by the values given 
in table~\ref{tab:bestfit}. 
\begin{table}[t!]
\begin{center}
\resizebox{\textwidth}{!}{
\begin{tabular}{|c|c|c|c|c|c|c|c|}
\cline{1-2}\cline{4-5}\cline{7-8}
Higgs & \multirow{2}{*}{value}  && Higgs-portal  & \multirow{2}{*}{value} && \multirow{2}{*}{observable} &  \multirow{2}{*}{value}\\
parameter &                     && parameter     &                        && & \\
\cline{1-2}\cline{4-5}\cline{7-8}
\noalign{\vskip\doublerulesep
         \vskip-\arrayrulewidth}
\cline{1-2}\cline{4-5}\cline{7-8}
$\lambda_1$ & $3.63\x10^{-4}$	&& $\mu_S^2$       & $7.77\x10^{4}\,\text{GeV}^2$ && $m_h$ & $125.25$\,GeV \rule{0pt}{2.5ex}\\
$\lambda_2$ & $1.24\x10^{-1}$	&& $\lambda_S$     & $8.33\x10^{-3}$              && $m_H$ & $381.41$\,GeV\\
$\lambda_3$ & $5.03\x10^{-1}$	&& $\lambda_{1S}$  & $4.52\x10^{-1}$              && $m_{H^\pm}$ & $386.82$\,GeV\\
$\lambda_4$ & $-1.7\x10^{-1}$	&& $\lambda_{2S}$  & $1.18\x10^{-1}$              && $m_A$ & $379.82$\,GeV\\
$\lambda_5$ & $9.35\x10^{-3}$	&& $\lambda_{12S}$ & $9.95\x10^{-4}$              && $\Omega_\mathrm{DM}h^2$ & $0.12$\\
\cline{4-5}
$\lambda_6$ & $5.52\x10^{-4}$	&\multicolumn{3}{c}{}                             && $m_S$ & $286.01$\,GeV\\
\cline{7-8}
$\lambda_7$ & $3.76\x10^{-4}$	&\multicolumn{3}{c}{}             && $\chi^2$ & $1.23\x10^{-4}$\rule{0pt}{2.5ex}\\
\cline{7-8}
$\mu_{12}^2$& $-1.45\x10^{3}\,\text{GeV}^2$	&\multicolumn{6}{c}{} \\
$v_1$		& $2.394\,\text{GeV}$		&\multicolumn{6}{c}{} \\
$v_2$		& $246.21\,\text{GeV}$		&\multicolumn{6}{c}{} \\
\cline{1-2}
\end{tabular}
}
\caption{Best fit of a 2HDM arising from heterotic orbifolds. In this case, $\tan\beta=102.84$.
\label{tab:bestfit}}
\end{center}
\end{table}

We plot the ensemble of points with a $\chi^2\leq10$ in 2D slices of the parameter space.
We display first in figure~\ref{fig:emceeplots_L2tanb} the values of $\lambda_2$
and $\tan\beta$, which are the best constrained Higgs parameters. We can regard our 
observations in particular for $\tan\beta$ as predictions about the mixing of both
Higgs fields. We see that $102.75\lesssim\tan\beta\lesssim103.75$ is preferred to 
obtain phenomenological results within $\chi^2\lesssim4$.
Acceptable values for other Higgs-potential parameters are found in 
figure~\ref{fig:emceeplots1}. In figure~\ref{fig:emceeplots3} we display the values
of the best values of the Higgs-portal parameters. Colors are used to illustrate 
the $\chi^2$ precision of each associated fit. Points in green correspond to better
fits to the observables than yellow and red points. In every plot the best-fit point 
is marked with a black star.

\begin{figure}[b!]
\centering
\subfigure{\includegraphics[width=0.55\textwidth]
	{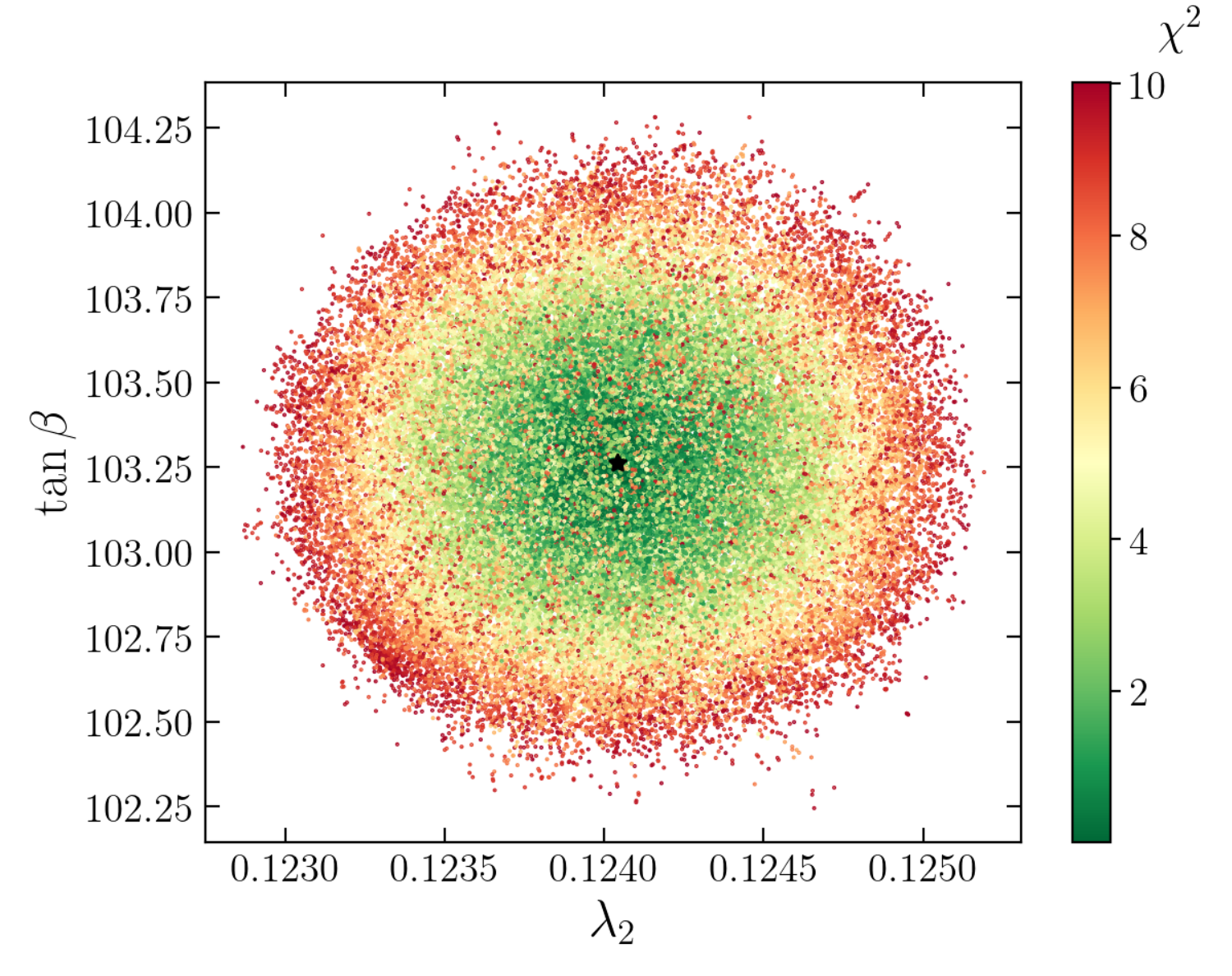}}
\caption{Best values of the Higgs parameters $\lambda_2$ and
$\tan\beta$ of a stringy 2HDM with $\chi^2\leq10$ w.r.t.\ the observables
in eq.~\eqref{eq:observables}. The black star corresponds to our best 
fit with $\chi^2=1.23\x10^{-4}$, given in table~\ref{tab:bestfit}.
\label{fig:emceeplots_L2tanb}}
\end{figure}

For most plots green and red points are neighbors because of the high 
dimensionality of the parameter space. However, close to the best-fit point
we find $\chi^2<4$ in general. Interestingly, even though there is a markedly 
preferred region for $\lambda_2$ and $\tan\beta$, our model does
not seem to set clear preferences for other parameters. 

\enlargethispage{2\baselineskip}

Once the best-fit point is identified, we find a number of predictions.
In figure~\ref{fig:emceeplots4} we plot the predicted values of the heavier
Higgs sector, the DM relic abundance and $\tan\beta$ of our stringy 2HDM.
We observe a general quasi-linear correlation, anticipated by eqs.~\eqref{eq:mHp} 
and~\eqref{eq:mA}, between the masses of all heavier Higgs fields, 
$m_{H,H^\pm,A}$, which lie between 370 and 400\,GeV. Their best-fit values, as presented
in table~\ref{tab:bestfit}, are found to be $m_H=381.41\,$GeV, $m_{H^\pm}=386.82\,$GeV 
and $m_A=379.82\,$GeV, compatible
with observational bounds for a very large value of $\tan\beta=102.84$.
Finally, figure~\ref{fig:emceeplots5} shows the preferred 
range of scalar DM mass in our scan which is roughly 270--300 GeV, with a best
fit value for 286.01 GeV. Note finally that the relic abundance coincides 
precisely with observations, lying in the interval 
$0.118\lesssim\Omega_\mathrm{DM}h^2\lesssim0.122$ for $\chi^2\leq4$.

\begin{figure}[t!]
\centering

\subfigure{\includegraphics[width=0.49\textwidth]
	{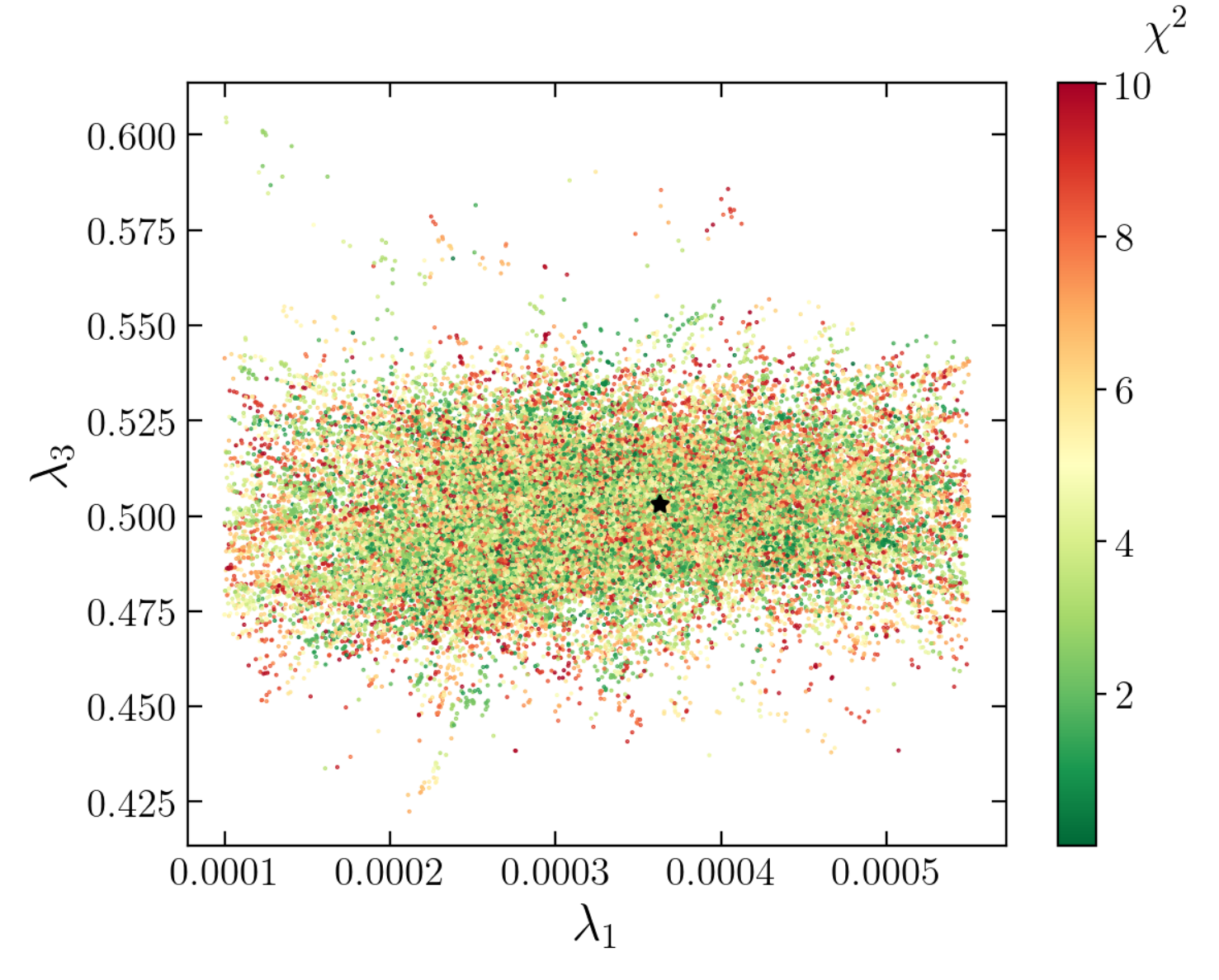}}
\subfigure{\includegraphics[width=0.49\textwidth]
	{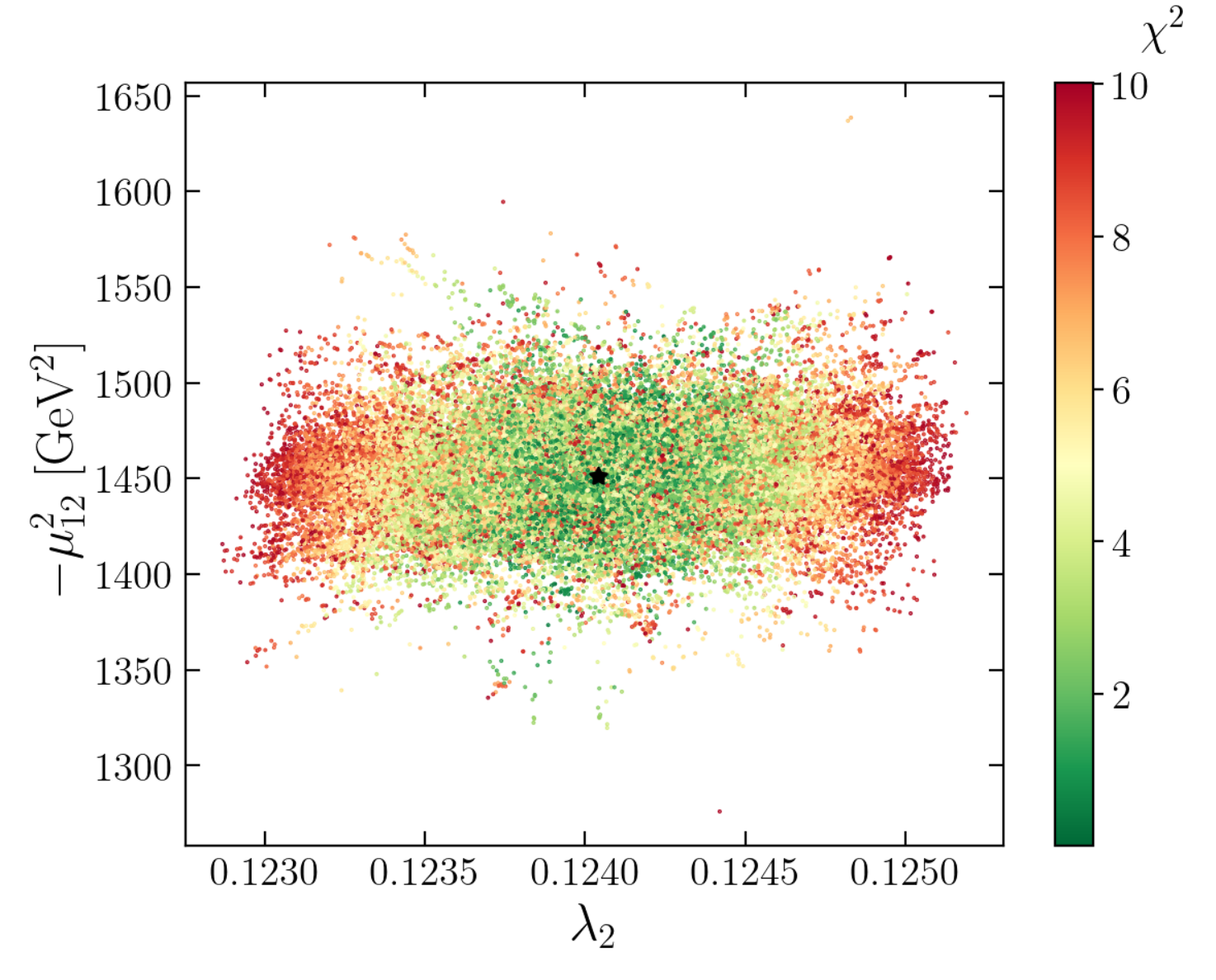}}

\subfigure{\includegraphics[width=0.49\textwidth]
	{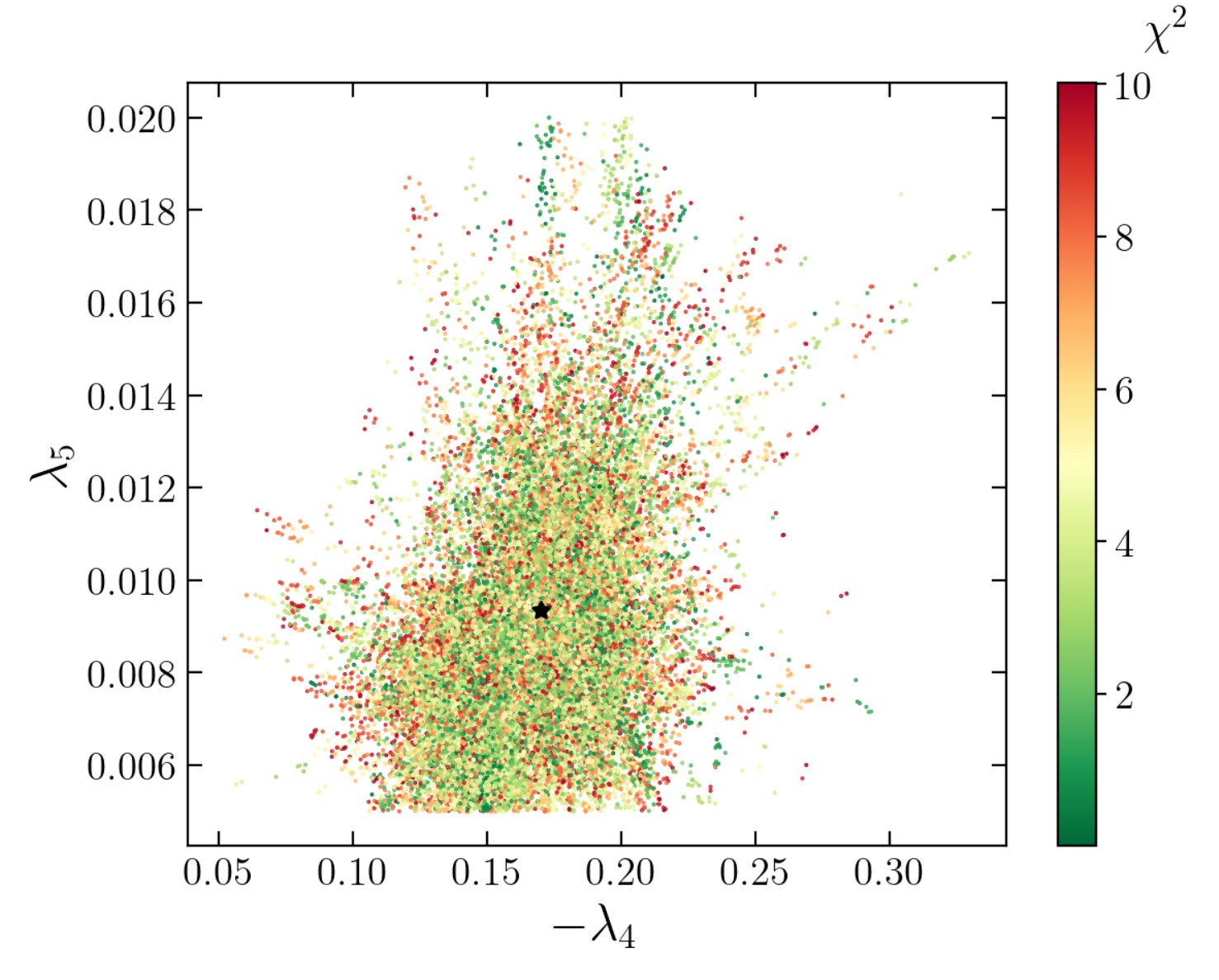}}
\subfigure{\includegraphics[width=0.49\textwidth]
{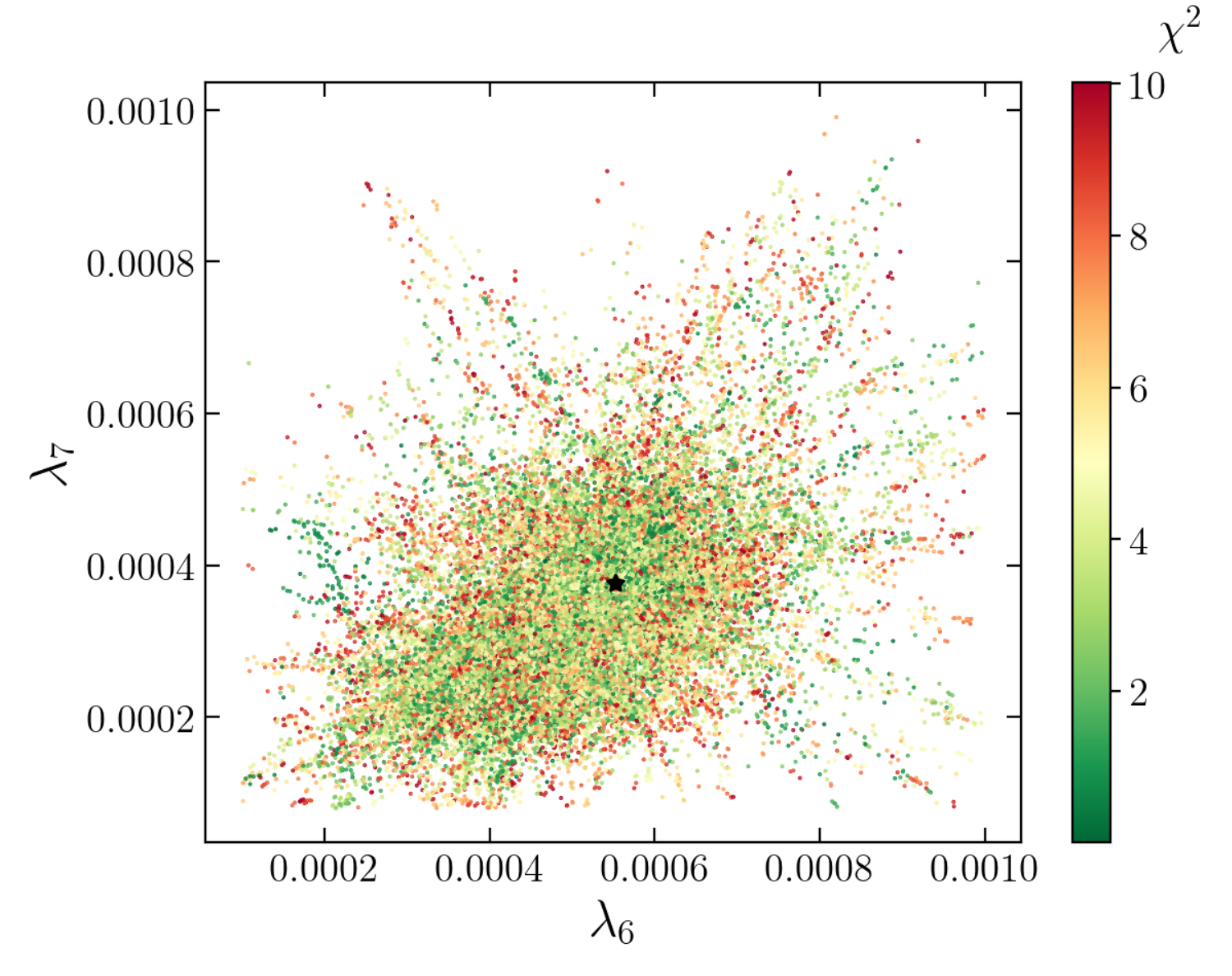}}

\caption{Some 2D projections of the most promising Higgs-potential 
parameter region of a stringy 2HDM, with $\chi^2\leq10$ w.r.t.\ the observables
in eq.~\eqref{eq:observables}. The black star corresponds to 
our best fit with $\chi^2=1.23\x10^{-4}$. 
\label{fig:emceeplots1} }
\end{figure}

\begin{figure}[t!]
\centering
\subfigure{\includegraphics[width=0.45\textwidth]
{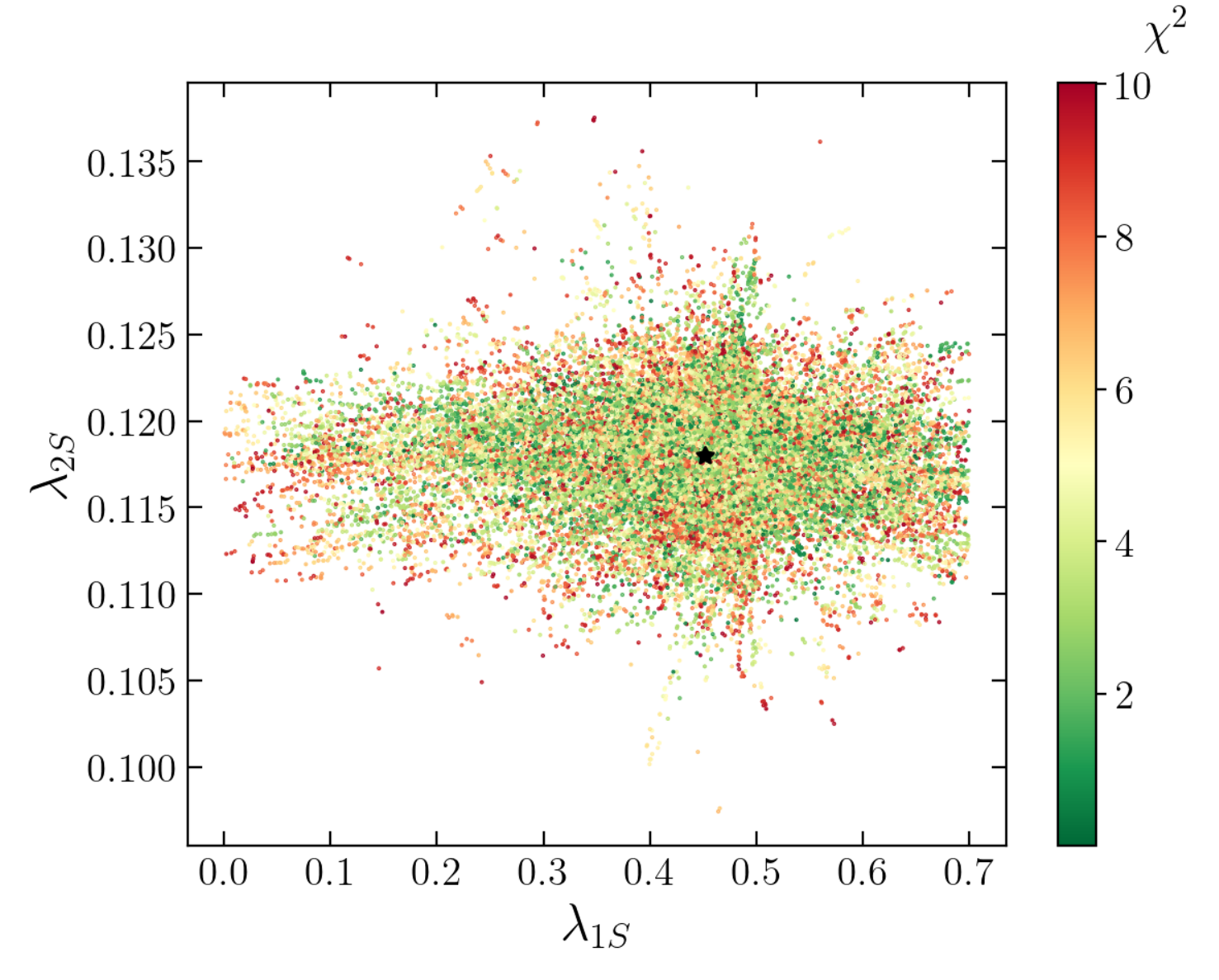}}
\subfigure{\includegraphics[width=0.45\textwidth]
{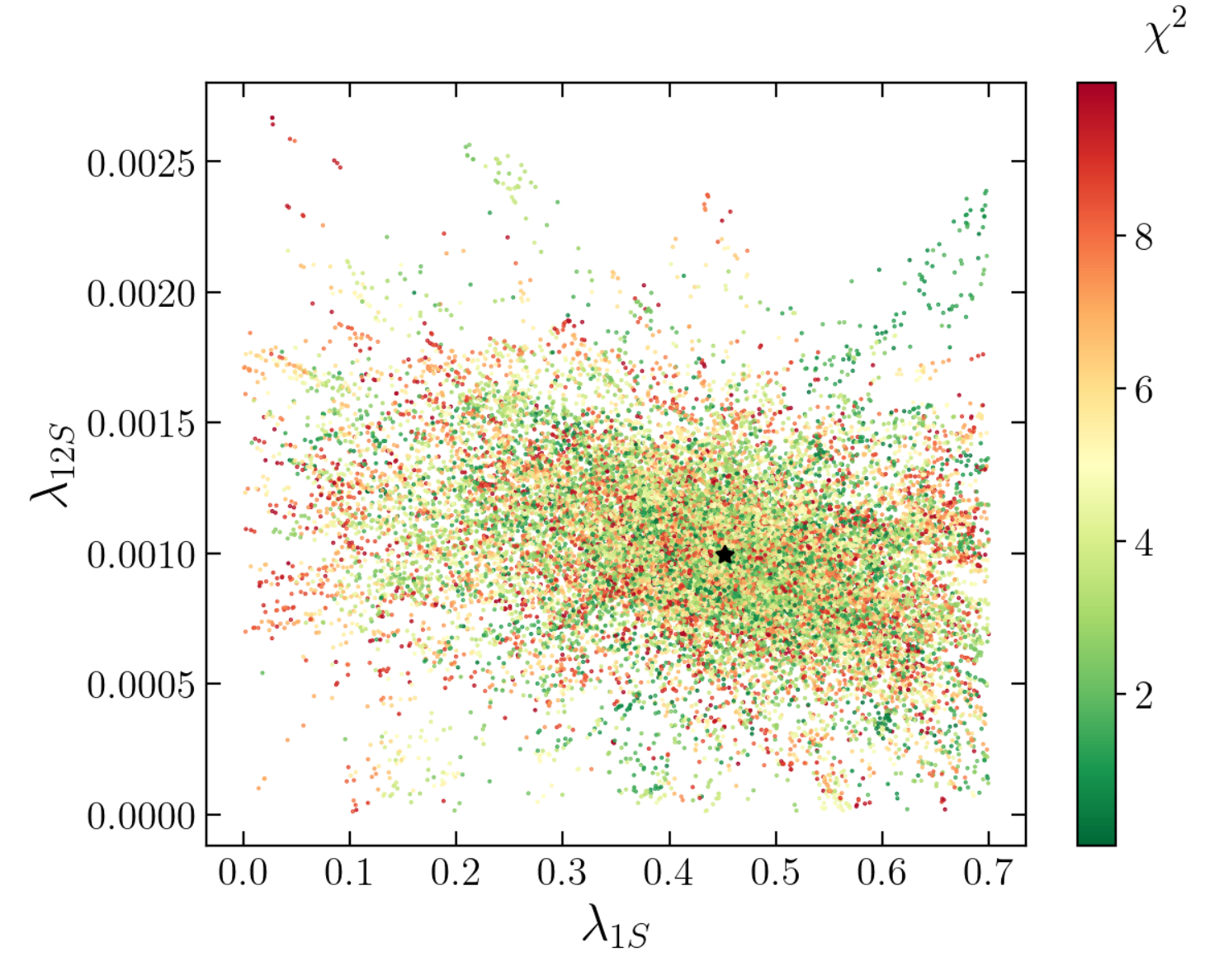}}
\subfigure{\includegraphics[width=0.45\textwidth]
{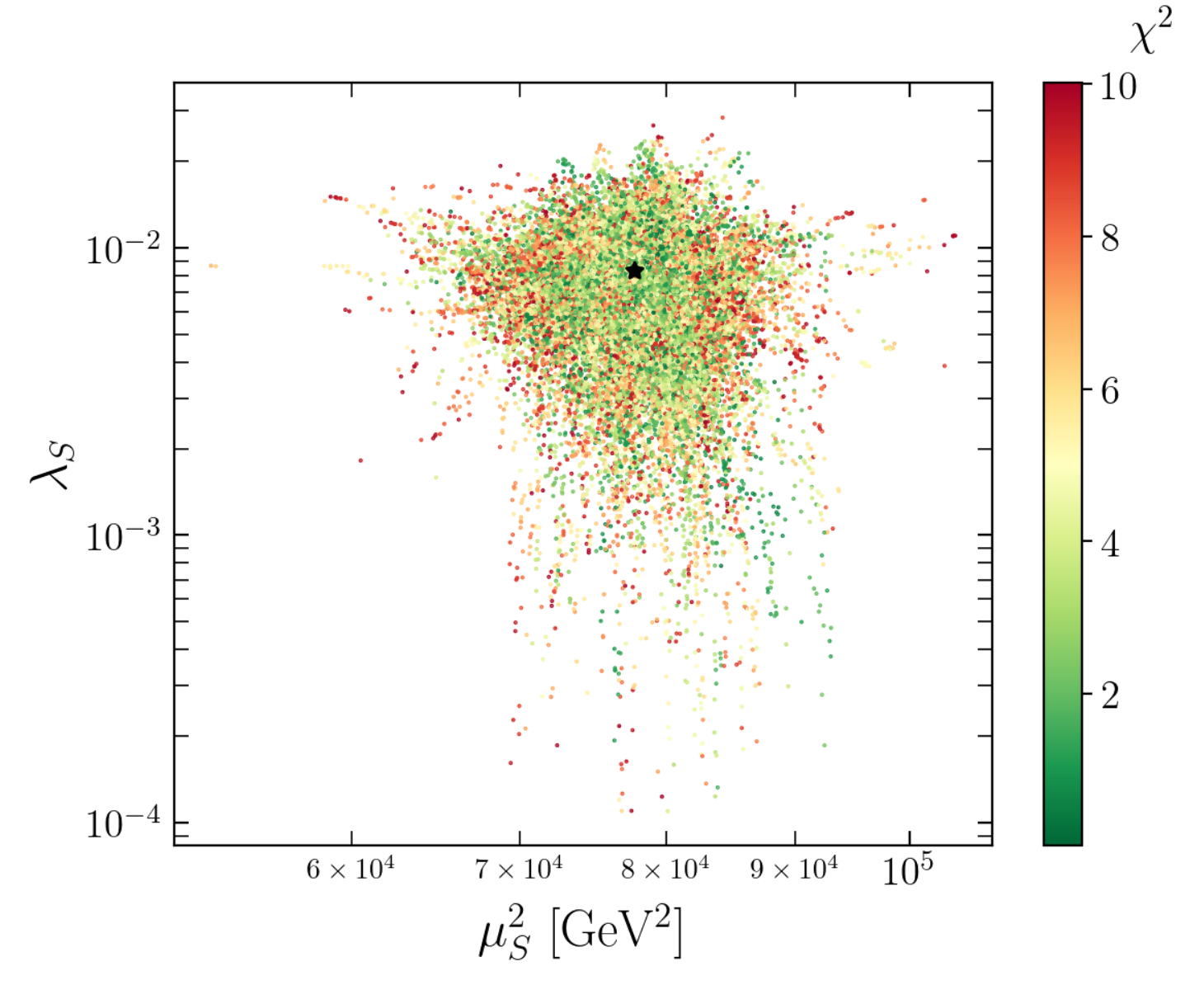}}

\caption{2D projections of the viable Higgs-portal 
parameter region of our model with $\chi^2\leq10$ compatibility
with observations. For some 
parameters the validity region is much wider than for others.
The best fit is given by the point highlighted with the black star.
\label{fig:emceeplots3}}
\end{figure}

\begin{figure}[h!]
\centering
\subfigure{\includegraphics[width=0.45\textwidth]
{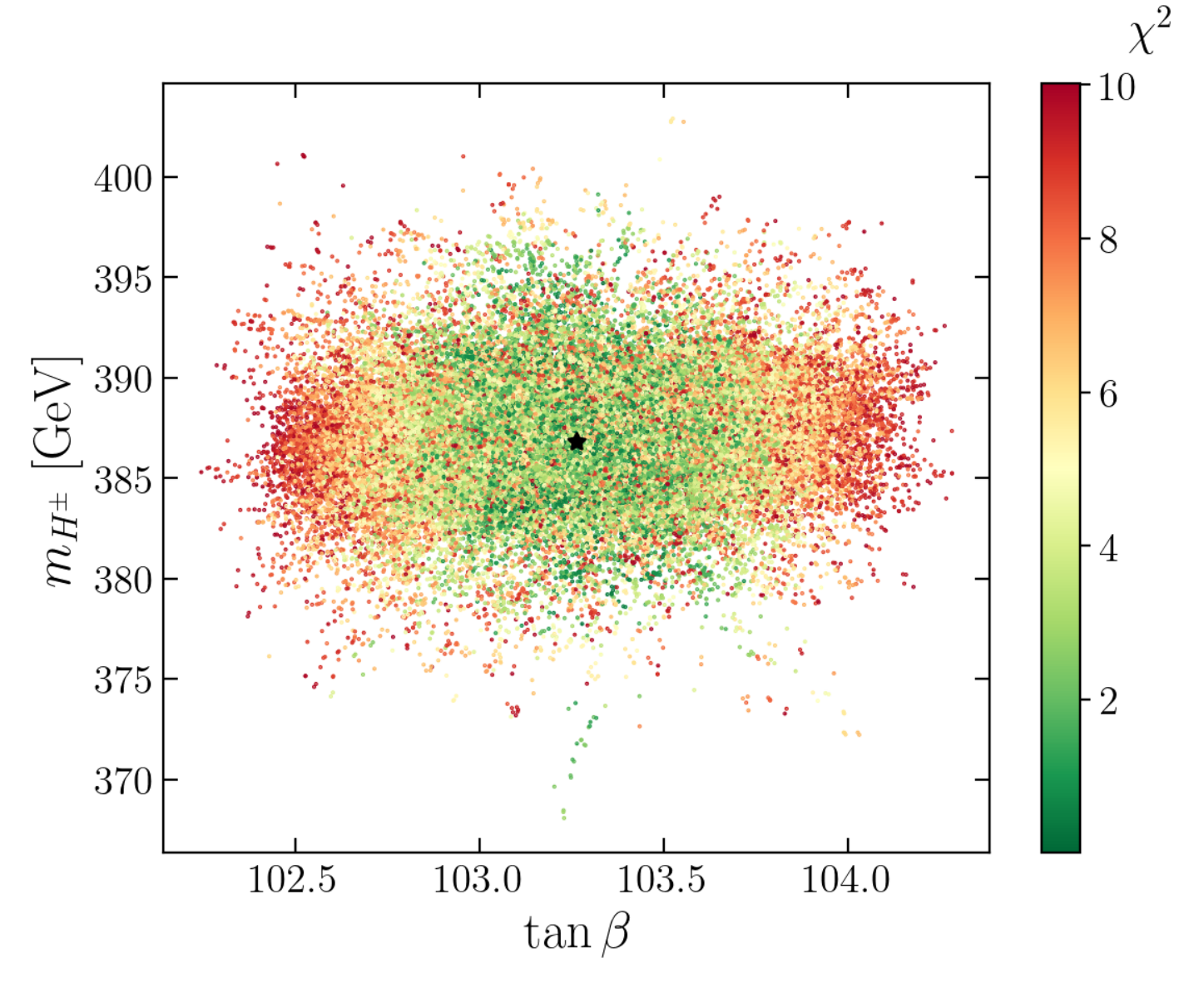}}
\subfigure{\includegraphics[width=0.45\textwidth]
{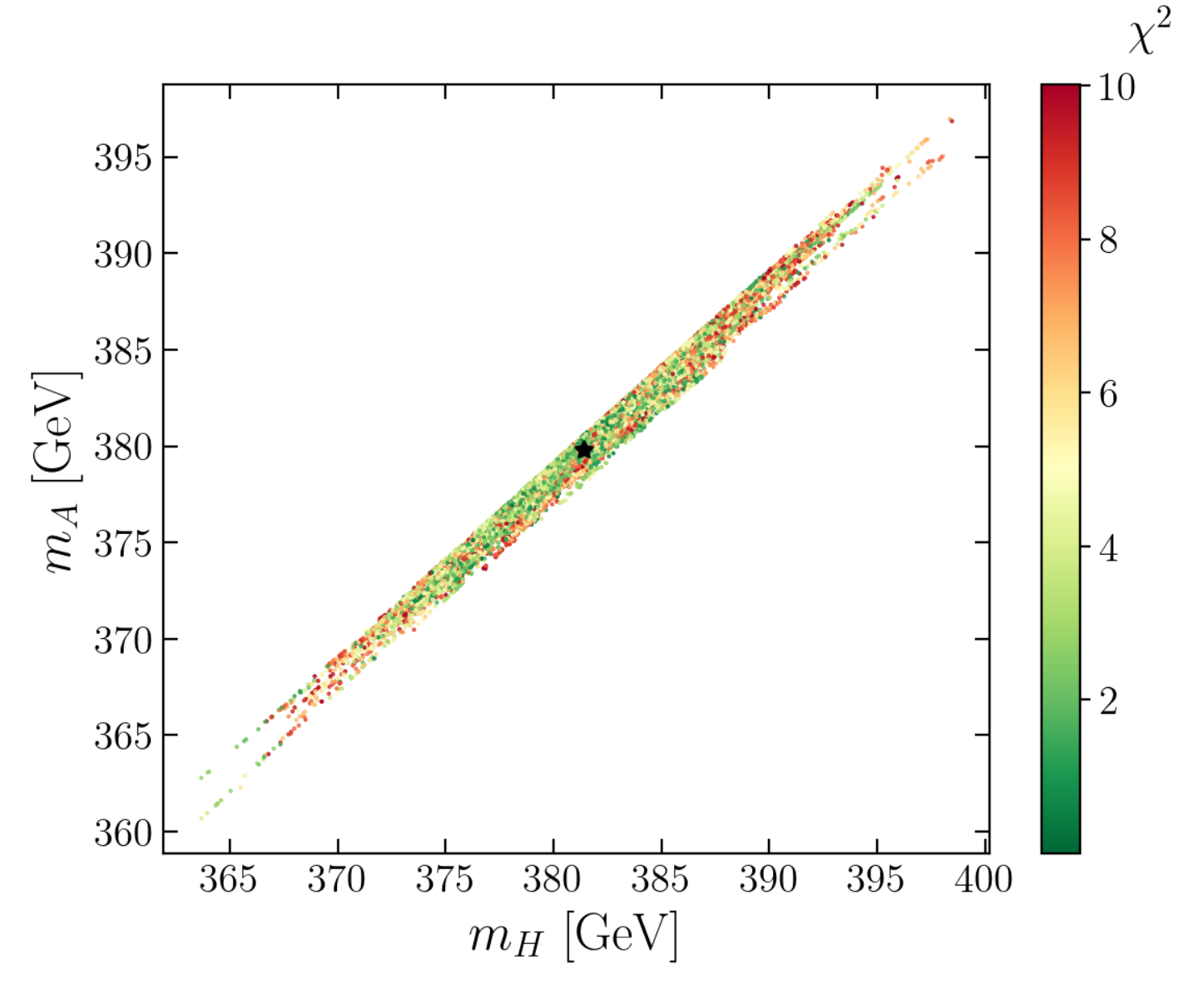}}

\caption{Predictions of a stringy 2HDM with Higgs portals on 
$\tan\beta$ and the masses of the heavier Higgs sector. We observe 
a quasi-linear dependence in the masses of the heavier (scalar and 
pseudoscalar) Higgs fields, which are constrained to lie between 370 
and 400 GeV for an accuracy of $\chi^2\leq10$. The best fit is 
highlighted by the black star.
\label{fig:emceeplots4}}
\end{figure}

\clearpage 

\begin{figure}[t!]
\centering
\subfigure{\includegraphics[width=0.49\textwidth]
{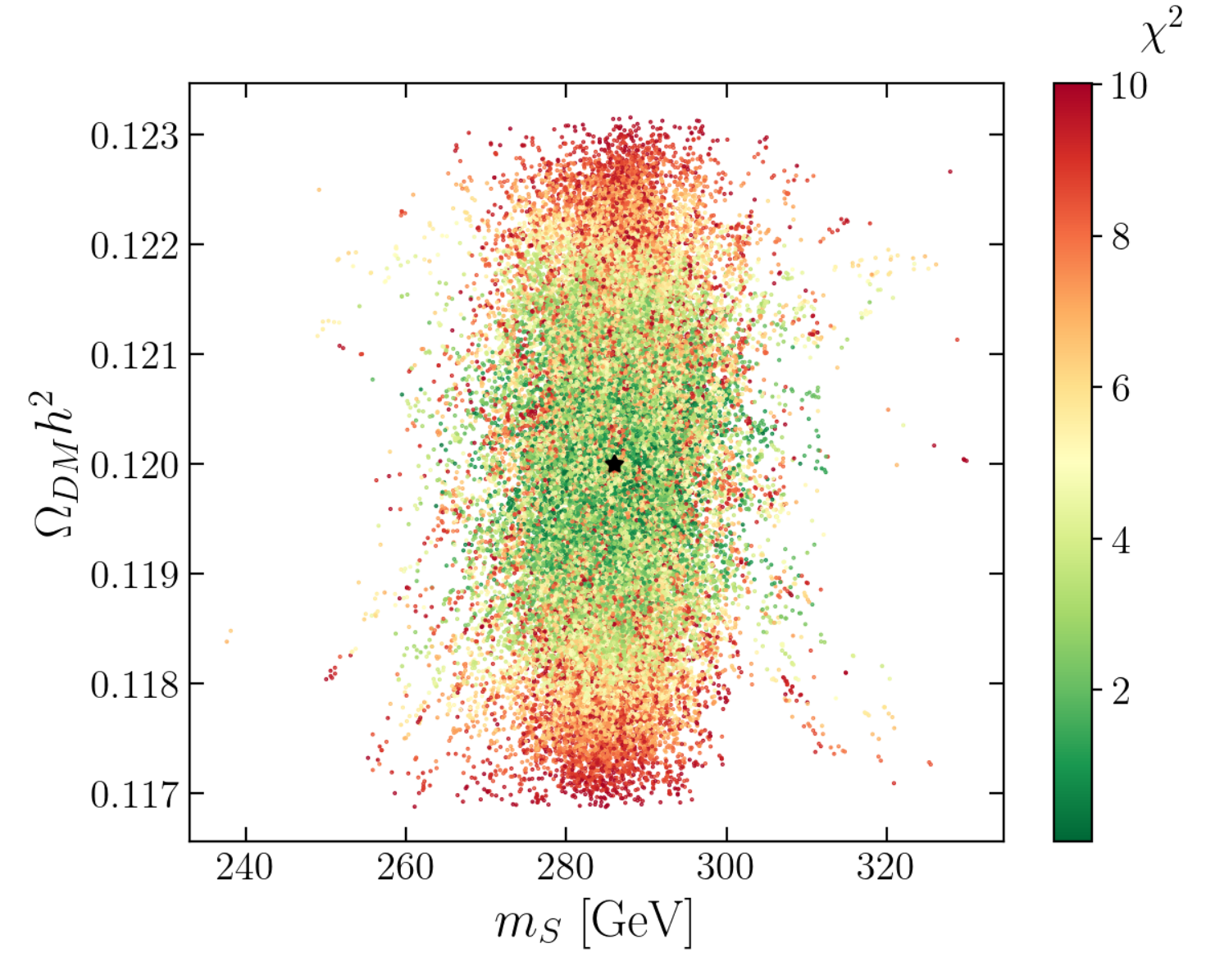}}
\caption{Predictions for the mass $m_S$ of the scalar DM candidate
and its relic abundance. The preferred mass is $286.01\,$GeV and
the best fit of $\Omega_\mathrm{DM}h^2$ coincides with the observed
value $0.12$.
\label{fig:emceeplots5}}
\end{figure}

\section{Conclusions}
\label{sec:conclusions}

In order to arrive at a proof of concept of interesting DM
phenomenology in non-SUSY orbifold compactifications of the heterotic string,
we have shown that Higgs-portal scenarios arise naturally in this context and
that they can readily accommodate observations on DM and Higgs physics. 

We recall that the most common SM-like string models without leptoquarks and 
a minimum of vectorlike exotics exhibits six Higgs fields and some continuous 
gauge flavor symmetry. Even though the number of Higgses is large, the flavor
symmetry reduces the degrees of freedom. We have inspected the structure 
of a sample model of this type. It displays promising Higgs-portal DM scenarios, 
i.e.\ scalar DM candidates coupling only to multi-Higgs sectors at leading 
order. This feature is quite general in this type of constructions. However, 
the complexity of the structure of Higgs vacua forces one to adopt very strong 
{\it ad hoc} constraints that weaken the predictivity of the model.

Hence, we have studied in more detail a simpler sample 2HDM arising from 
heterotic orbifolds. We find that the appearance of Higgs portals is generic 
in this kind of models too. The resulting effective field theory has the advantage 
that the Higgs vacuum is more tractable. Hence, we have used various numeric 
tools to scan the parameter space for phenomenologically viable regions,
including 1-loop corrections. 
We have identified vast promising neighborhoods (with $\chi^2\leq10$)
endowed with i) stable Higgs vacua, ii) lightest Higgs mass and VEV compatible 
with experiments, and iii) freeze-out DM relic abundance consistent with 
observational data. Further, we have found that in those cases, as illustrated
in figure~\ref{fig:emceeplots4}, $\tan\beta\sim100$,
the heavier (scalar and pseudoscalar) Higgs fields have masses around 
380 GeV, and the scalar DM candidate displays $m_S\sim285$ GeV. We have 
identified a best fit to data of our stringy 2HDM with $\chi^2=1.23\x10^{-4}$.
Its details are given in table~\ref{tab:bestfit}. All of these
results can be considered predictions of models like ours, independently
of whether they are motivated by top-down or bottom-up considerations.

There are various challenges to be addressed elsewhere in order to complete the 
study of this and similar models in the context of non-SUSY string compactifications.
They include the phenomenology of the leptoquarks appearing in our 2HDM and other 
similar string models; their flavor structure and phenomenology, involving all gauge, 
traditional and modular flavor symmetries of the model (see
e.g.~\cite{Baur:2019iai,Baur:2020jwc,Nilles:2020gvu,Almumin:2021fbk,Baur:2021mtl,Baur:2022hma});
the possibility of Higgs portals with multiple DM candidates, including the 
likely existence of extra pseudoscalars; and the stringy computation 
of the couplings and their RGE running from the compactification scale.
In particular, the appearance of a few leptoquarks and their (possibly negligible) 
couplings to the SM must be further studied as they may be relevant for extant 
questions in the SM~\cite{Keung:2021rps} and even for richer DM scenarios~\cite{Choi:2018stw}.

\section*{Acknowledgments}
It is a pleasure to thank J.~Armando Arroyo for insightful discussions at
early stages of this project. This work was supported by UNAM-PAPIIT IN113223,
CONACYT grant CB-2017-2018/A1-S-13051, and Marcos Moshinsky Foundation.
E.C. is supported by the National Science Centre (Poland) under the research 
Grant No. 2021/42/E/ST2/00009.


\begin{appendix}
\newpage
\section{Spectrum of the stringy model with six Higgses}
\label{ap:6Hspectrum}

\vspace{-2mm}
{\small
\begin{longtable}{|c|l|c|c|c|c|c|c|c|c|c|c|c|c|}
\caption{Spectrum of massless fermions for the non-SUSY stringy model with six Higgs
doublets introduced in section~\ref{sec:sixHiggses}. The first column corresponds to
the multiplicity of the fields. We show the quantum numbers with respect to the 4D gauge 
group, $\mathcal{G}_{\text{4D}} = \mathcal{G}_{\text{SM}} \x \mathcal{G}' \x \U1'^8$, 
where $\mathcal{G}_{\text{SM}} = \SU3_c\x \SU2_L\x \U1_Y$ and $\mathcal{G}'=\SU3_\mathrm{flavor}\x\SU2\x\SU2$. 
The first $\U1'$, associated with the charges $q_1$, is (pseudo-)anomalous. In the column 
labeled as $\SU3$ we list the $\SU3_\mathrm{flavor}$ representations.
\label{tab:complete4902f}}\\
\hline
\# & $\mathcal{G}_{\text{SM}}$ & \SU3 & \SU2 & \SU2 & $q_1$ & $q_2$ & $q_3$ & $q_4$ & $q_5$ & $q_6$ & $q_7$ & $q_8$ & \text{label} \\ \hline
\endfirsthead
\hline
\# & $\mathcal{G}_{\text{SM}}$ & \SU3 & \SU2 & \SU2 & $q_1$ & $q_2$ & $q_3$ & $q_4$ & $q_5$ & $q_6$ & $q_7$ & $q_8$ & \text{label} \\ \hline
\endhead
\hline \multicolumn{14}{|r|}{{continued...}} \\ 
\hline
\endfoot
\hline
\endlastfoot

1 & $(\rep{1},\rep{2})_{\nicefrac{-1}{2}}$  & $\rep{1}$  & $\rep{1}$ & $\rep{1}$ & -3   & 3   & -27 & -9  & -9  & 9   & -9  & 42  & $\ell_{\mathrm{L},1}$    \\
2 & $(\rep{1},\rep{2})_{\nicefrac{-1}{2}}$  & $\rep{1}$  & $\rep{1}$ & $\rep{1}$ & 2    & 3   & 1   & 25  & -13 & 13  & -13 & 32  & $\ell_{\mathrm{L} ,2i}$  \\
1 & $(\rep{1},\rep{2})_{\nicefrac{-1}{2}}$  & $\rep{1}$  & $\rep{1}$ & $\rep{1}$ & 0    & -18 & -6  & -2  & -2  & 2   & -2  & -48 & 
$\ell_{\mathrm{L} ,3}$       \\
1 & $(\rep{1},\rep{2})_{\nicefrac{-1}{2}}$  & $\rep{1}$  & $\rep{1}$ & $\rep{1}$ & 2    & 0   & -56 & 6   & 6   & -6  & 6   & -28 & $\ell_{\mathrm{L} ,4}$    \\
1 & $(\rep{1},\rep{2})_{\nicefrac{1}{2}}$   & $\rep{1}$  & $\rep{1}$ & $\rep{1}$ & 0    & 18  & 6   & 2   & 2   & -2  & 2   & 48  & 
$\overline{\ell}'_{\mathrm{L},1}$ \\
1 & $(\rep{1},\rep{2})_{\nicefrac{1}{2}}$   & $\rep{1}$  & $\rep{1}$ & $\rep{1}$ & 0    & 0   & 0   & -74 & 2   & -2  & 2   & 48  & 
$\overline{\ell}'_{\mathrm{L},2}$  \\
1 & $(\rep{1},\rep{1})_1$     & $\rep{3}$ & $\rep{1}$ & $\rep{1}$ & 1    & -1  & 9   & 3   & 3   & -3  & 3   & -14 & $\overline{e}_{\mathrm{L}}$      \\
1 & $(\rep{3},\rep{2})_{\nicefrac{1}{6}}$   & $\rep{3}$ & $\rep{1}$ & $\rep{1}$ & 1    & -1  & 9   & 3   & 3   & -3  & 3   & -14 & $q_{\mathrm{L}}$        \\
1 & $(\crep{3},\rep{1})_{\nicefrac{-2}{3}}$ & $\rep{3}$ & $\rep{1}$ & $\rep{1}$ & 1    & -1  & 9   & 3   & 3   & -3  & 3   & -14 & 
$\overline{u}_{\mathrm{L}}$  \\
1 & $(\crep{3},\rep{1})_{\nicefrac{1}{3}}$  & $\rep{1}$  & $\rep{1}$ & $\rep{1}$ & -3   & 3   & -27 & -9  & -9  & 9   & -9  & 42  & 
$\overline{d}_{\mathrm{L},1}$ \\
4 & $(\crep{3},\rep{1})_{\nicefrac{1}{3}}$  & $\rep{1}$  & $\rep{1}$ & $\rep{1}$ & 2    & 3   & 1   & 25  & -13 & 13  & -13 & 32  & $\overline{d}_{\mathrm{L},2i}$ \\
1 & $(\crep{3},\rep{1})_{\nicefrac{1}{3}}$  & $\rep{1}$  & $\rep{1}$ & $\rep{1}$ & 2    & 0   & -56 & 6   & 6   & -6  & 6   & -28 & 
$\overline{d}_{\mathrm{L},3}$  \\
1 & $(\crep{3},\rep{1})_{\nicefrac{1}{3}}$  & $\rep{1}$  & $\rep{1}$ & $\rep{1}$ & 0    & 0   & 0   & 0   & 76  & 2   & -2  & -48 & $\overline{d}_{\mathrm{L},4}$ \\
2 & $(\rep{3},\rep{1})_{\nicefrac{-1}{3}}$  & $\rep{1}$  & $\rep{1}$ & $\rep{1}$ & -2   & -3  & -1  & -25 & 13  & -13 & 13  & -32 & 
$d'_{\mathrm{L},1i}$     \\
1 & $(\rep{3},\rep{1})_{\nicefrac{-1}{3}}$  & $\rep{1}$  & $\rep{1}$ & $\rep{1}$ & 0    & 0   & 0   & -74 & 2   & -2  & 2   & 48  & 
$d'_{\mathrm{L},2}$       \\
1 & $(\rep{3},\rep{1})_{\nicefrac{-1}{3}}$  & $\rep{1}$  & $\rep{1}$ & $\rep{1}$ & 0    & 0   & 0   & 0   & -76 & -2  & 2   & 48  & 
$d'_{\mathrm{L},3}$       \\
1 & $(\rep{1},\rep{1})_0$     & $\rep{1}$  & $\rep{1}$ & $\rep{1}$ & -1   & -9  & 25  & -41 & 35  & -35 & -47 & -10 & $\nu_{\mathrm{R},1}$         \\
1 & $(\rep{1},\rep{1})_0$     & $\rep{1}$  & $\rep{2}$ & $\rep{2}$ & 1    & 9   & -25 & 41  & -35 & -4  & 4   & 10  & $\nu_{\mathrm{R},2}
$         \\
1 & $(\rep{1},\rep{1})_0$     & $\rep{1}$  & $\rep{1}$ & $\rep{1}$ & -1   & -9  & 25  & -41 & 35  & 43  & 39  & -10 & $\nu_{\mathrm{R},3}$         \\
1 & $(\rep{1},\rep{1})_0$     & $\rep{1}$  & $\rep{1}$ & $\rep{1}$ & 1    & -3  & 27  & 9   & 9   & -9  & 9   & 130 & $\nu_{\mathrm{R},4}$         \\
1 & $(\rep{1},\rep{1})_0$     & $\rep{1}$  & $\rep{1}$ & $\rep{1}$ & 1    & -9  & -31 & 39  & 39  & 39  & -39 & 10  & $\nu_{\mathrm{R},5}$         \\
1 & $(\rep{1},\rep{1})_0$     & $\rep{1}$  & $\rep{1}$ & $\rep{1}$ & 1    & 9   & 31  & -39 & -39 & 39  & -39 & 10  & $\nu_{\mathrm{R},6}$        \\
1 & $(\rep{1},\rep{1})_0$     & $\rep{1}$  & $\rep{2}$ & $\rep{2}$ & 1    & -9  & -31 & -35 & -35 & -4  & 4   & 10  & $\nu_{\mathrm{R},7}$         \\
1 & $(\rep{1},\rep{1})_0$     & $\rep{1}$  & $\rep{1}$ & $\rep{1}$ & -1   & -9  & 25  & 33  & -43 & -35 & -47 & -10 & $\nu_{\mathrm{R},8}$         \\
1 & $(\rep{1},\rep{1})_0$     & $\rep{1}$  & $\rep{1}$ & $\rep{1}$ & 1    & 9   & -25 & -33 & 43  & -43 & -39 & 10  & $\nu_{\mathrm{R},9}$         \\
1 & $(\rep{1},\rep{1})_0$     & $\rep{1}$  & $\rep{2}$ & $\rep{2}$ & -1   & 9   & 31  & 35  & 35  & 4   & -4  & -10 & $\nu_{\mathrm{R},10}$       \\
1 & $(\rep{1},\rep{1})_0$     & $\rep{1}$  & $\rep{1}$ & $\rep{1}$ & -1   & -9  & 25  & 33  & -43 & 43  & 39  & -10 & $\nu_{\mathrm{R},11}$      \\
1 & $(\rep{1},\rep{1})_0$     & $\rep{1}$  & $\rep{1}$ & $\rep{1}$ & 1    & 9   & -25 & -33 & 43  & 35  & 47  & 10  & $\nu_{\mathrm{R},12}$        \\
2 & $(\rep{1},\rep{1})_0$     & $\crep{3}$  & $\rep{1}$ & $\rep{1}$ & 0    & -5  & 17  & -19 & 19  & -19 & 19  & -60 & $\nu'_{\mathrm{R},13,i}$        \\
2 & $(\rep{1},\rep{1})_0$     & $\rep{1}$  & $\rep{1}$ & $\rep{1}$ & 5/2  & 0   & 14  & -20 & -39 & -39 & -2  & -5  & $\nu_{\mathrm{R},14,i}$        \\
2 & $(\rep{1},\rep{1})_0$     & $\rep{1}$  & $\rep{1}$ & $\rep{1}$ & 3/2  & 0   & 14  & 54  & 35  & -35 & -6  & -15 & $\nu_{\mathrm{R},15,i}$        \\
2 & $(\rep{1},\rep{1})_0$     & $\rep{1}$  & $\rep{2}$ & $\rep{1}$ & 5/2  & 0   & 14  & -20 & 37  & 2   & -2  & -5  & $\nu_{\mathrm{R},16,i}$        \\
2 & $(\rep{1},\rep{1})_0$     & $\rep{1}$  & $\rep{1}$ & $\rep{2}$ & 5/2  & 0   & 14  & -20 & 37  & 2   & -2  & -5  & $\nu_{\mathrm{R},17,i}$        \\
2 & $(\rep{1},\rep{1})_0$     & $\rep{1}$  & $\rep{1}$ & $\rep{1}$ & 5/2  & 0   & 14  & -20 & -39 & 0   & 41  & -5  & $\nu_{\mathrm{R},18,i}$       \\
2 & $(\rep{1},\rep{1})_0$     & $\rep{1}$  & $\rep{1}$ & $\rep{1}$ & 3/2  & 0   & 14  & 54  & 35  & 4   & 37  & -15 & $\nu_{\mathrm{R},19,i}$       \\
2 & $(\rep{1},\rep{1})_0$     & $\rep{1}$  & $\rep{1}$ & $\rep{1}$ & 3/2  & 6   & 16  & 30  & 11  & -11 & -30 & -75 & $\nu_{\mathrm{R},20,i}$       \\
2 & $(\rep{1},\rep{1})_0$     & $\crep{3}$  & $\rep{1}$ & $\rep{1}$ & 1/2  & -2  & -24 & -8  & -27 & 27  & 14  & 35  & $\nu'_{\mathrm{R},21,i}$       \\
2 & $(\rep{1},\rep{1})_0$     & $\rep{1}$  & $\rep{1}$ & $\rep{1}$ & 3/2  & 6   & 16  & 30  & 11  & 28  & 13  & -75 & $\nu_{\mathrm{R},22,i}$       \\
2 & $(\rep{1},\rep{1})_0$     & $\crep{3}$  & $\rep{1}$ & $\rep{1}$ & 1/2  & -2  & -24 & -8  & -27 & -12 & -29 & 35  & $\nu'_{\mathrm{R},23,i}$       \\
1 & $(\rep{1},\rep{1})_0$     & $\crep{3}$  & $\rep{1}$ & $\rep{1}$ & 0    & -2  & 74  & 0   & 0   & 0   & 0   & 0   & $\nu'_{\mathrm{R},24}$       \\
1 & $(\rep{1},\rep{1})_0$     & $\rep{3}$ & $\rep{1}$ & $\rep{1}$ & -2   & 2   & -18 & 68  & -8  & 8   & -8  & -20 & $\nu''_{\mathrm{R},25}$        \\
1 & $(\rep{1},\rep{1})_0$     & $\crep{3}$  & $\rep{1}$ & $\rep{1}$ & 2    & 16  & 24  & 8   & 8   & -8  & 8   & 20  & $\nu'_{\mathrm{R},26}$        \\
1 & $(\rep{1},\rep{1})_0$     & $\rep{3}$ & $\rep{1}$ & $\rep{1}$ & -2   & -16 & -24 & -8  & -8  & 8   & -8  & -20 & $\nu''_{\mathrm{R},27}$        \\
4 & $(\rep{1},\rep{1})_0$     & $\rep{1}$  & $\rep{1}$ & $\rep{1}$ & 1    & 6   & -26 & -58 & -20 & 20  & -20 & -50 & $\nu_{\mathrm{R},28,i}$        \\
2 & $(\rep{1},\rep{1})_0$     & $\rep{1}$  & $\rep{1}$ & $\rep{1}$ & -1   & -6  & 26  & 58  & 20  & -20 & 20  & 50  & $\nu_{\mathrm{R},29,i}$        \\
2 & $(\rep{1},\rep{1})_0$     & $\rep{1}$  & $\rep{1}$ & $\rep{1}$ & 1    & -6  & -30 & -10 & 28  & -28 & 28  & 70  & $\nu_{\mathrm{R},30,i}$       \\
2 & $(\rep{1},\rep{1})_0$     & $\rep{1}$  & $\rep{1}$ & $\rep{1}$ & 3/2  & -3  & 13  & 29  & -28 & -11 & -30 & -75 & $\nu_{\mathrm{R},31,i}$        \\
2 & $(\rep{1},\rep{1})_0$     & $\crep{3}$  & $\rep{1}$ & $\rep{1}$ & 1/2  & 7   & -21 & -7  & 12  & 27  & 14  & 35  & $\nu'_{\mathrm{R},32,i}$       \\
2 & $(\rep{1},\rep{1})_0$     & $\rep{1}$  & $\rep{1}$ & $\rep{1}$ & 3/2  & -3  & 13  & 29  & -28 & 28  & 13  & -75 & $\nu_{\mathrm{R},33,i}$       \\
2 & $(\rep{1},\rep{1})_0$     & $\crep{3}$  & $\rep{1}$ & $\rep{1}$ & 1/2  & 7   & -21 & -7  & 12  & -12 & -29 & 35  & $\nu'_{\mathrm{R},34,i}$        \\
2 & $(\rep{1},\rep{1})_0$     & $\rep{1}$  & $\rep{1}$ & $\rep{1}$ & 3/2  & -9  & 11  & 53  & -4  & -35 & -6  & -15 & $\nu_{\mathrm{R},35,i}$        \\
2 & $(\rep{1},\rep{1})_0$     & $\rep{1}$  & $\rep{2}$ & $\rep{1}$ & 5/2  & -9  & 11  & -21 & -2  & 2   & -2  & -5  & $\nu_{\mathrm{R},36,i}$     \\                                                                                                           
2 & $(\rep{1},\rep{1})_0$     & $\rep{1}$  & $\rep{1}$ & $\rep{1}$ & 5/2  & 9   & 17  & -19 & 0   & -39 & -2  & -5  & $\nu_{\mathrm{R},37,i}$        \\
2 & $(\rep{1},\rep{1})_0$     & $\rep{1}$  & $\rep{1}$ & $\rep{2}$ & 5/2  & -9  & 11  & -21 & -2  & 2   & -2  & -5  & $\nu_{\mathrm{R},38,i}$        \\
2 & $(\rep{1},\rep{1})_0$     & $\rep{1}$  & $\rep{1}$ & $\rep{1}$ & 3/2  & -9  & 11  & 53  & -4  & 4   & 37  & -15 & $\nu_{\mathrm{R},39,i}$       \\
2 & $(\rep{1},\rep{1})_0$     & $\rep{1}$  & $\rep{1}$ & $\rep{1}$ & 5/2  & 9   & 17  & -19 & 0   & 0   & 41  & -5  & $\nu_{\mathrm{R},40,i}$       \\ 
\end{longtable}

\begin{longtable}{|c|c|c|c|c|c|c|c|c|c|c|c|c|c|}
\caption{Stringy spectrum of massless complex scalars for a model with six Higgses. 
Note that our DM candidate $S$ has the same charges as $s_{10}$. 
We follow the same notation as in table~\ref{tab:complete4902f}.
\label{tab:complete4902s}}\\
\hline
\# & $\mathcal{G}_{\text{SM}}$ & \SU3 & \SU2 & \SU2 & $q_1$ & $q_2$ & $q_3$ & $q_4$ & $q_5$ & $q_6$ & $q_7$ & $q_8$ & \text{label} \\ \hline
\endhead
2 & $(\rep{1},\rep{2})_{\nicefrac{1}{2}}$   & $\rep{3}$ & $\rep{1}$ & $\rep{1}$ & -2   & 2   & -18 & -6  & -6  & 6   & -6  & 28  & $\phi_i$         \\
1 & $(\rep{1},\rep{1})_0$     & $\rep{1}$  & $\rep{1}$ & $\rep{1}$ & 3/2  & 9   & 45  & 15  & -4  & -35 & -6  & -15 & $S$         \\
1 & $(\rep{1},\rep{1})_0$     & $\rep{1}$  & $\rep{1}$ & $\rep{1}$ & 0    & -18 & -6  & -76 & 0   & 0   & 0   & 0   & $s_1$          \\
1 & $(\rep{1},\rep{1})_0$     & $\rep{1}$  & $\rep{1}$ & $\rep{1}$ & 0    & 0   & 0   & 74  & -78 & 0   & 0   & 0   & $s_2$          \\
1 & $(\rep{1},\rep{1})_0$     & $\rep{1}$  & $\rep{1}$ & $\rep{1}$ & -2   & 0   & 56  & -6  & 70  & 8   & -8  & -20 & $s_3$          \\
1 & $(\rep{1},\rep{1})_0$     & $\rep{1}$  & $\rep{1}$ & $\rep{1}$ & 2    & 18  & -50 & 8   & 8   & -8  & 8   & 20  & $s_4$          \\
2 & $(\rep{1},\rep{1})_0$     & $\rep{1}$  & $\rep{1}$ & $\rep{2}$ & 3/2  & -9  & 39  & 13  & -6  & 6   & -6  & -15 & $s_5$          \\
2 & $(\rep{1},\rep{1})_0$     & $\rep{1}$  & $\rep{1}$ & $\rep{1}$ & 5/2  & -9  & -17 & 19  & 0   & 0   & 41  & -5  & $s_6$          \\
2 & $(\rep{1},\rep{1})_0$     & $\rep{1}$  & $\rep{1}$ & $\rep{1}$ & 3/2  & 9   & 45  & 15  & -4  & 4   & 37  & -15 & $s_7$          \\
2 & $(\rep{1},\rep{1})_0$     & $\rep{1}$  & $\rep{1}$ & $\rep{1}$ & 5/2  & -9  & -17 & 19  & 0   & -39 & -2  & -5  & $s_8$          \\
2 & $(\rep{1},\rep{1})_0$     & $\rep{1}$  & $\rep{2}$ & $\rep{1}$ & 3/2  & -9  & 39  & 13  & -6  & 6   & -6  & -15 & $s_9$          \\
1 & $(\rep{1},\rep{1})_0$     & $\rep{1}$  & $\rep{1}$ & $\rep{1}$ & 3/2  & 9   & 45  & 15  & -4  & -35 & -6  & -15 & $s_{10}$         \\
2 & $(\rep{1},\rep{1})_0$     & $\rep{1}$  & $\rep{1}$ & $\rep{1}$ & 5/2  & -3  & -15 & -5  & -24 & 24  & 17  & -65 & $s_{11}$         \\
2 & $(\rep{1},\rep{1})_0$     & $\rep{3}$  & $\rep{1}$ & $\rep{1}$ & 1/2 & -7   & -7   & -27  & -8   & 8  & 33 & -25  & $s'_{12}$         \\
2 & $(\rep{1},\rep{1})_0$     & $\rep{1}$  & $\rep{1}$ & $\rep{1}$ & 5/2  & -3  & -15 & -5  & -24 & -15 & -26 & -65 & $s_{13}$         \\
2 & $(\rep{1},\rep{1})_0$     & $\rep{3}$  & $\rep{1}$ & $\rep{1}$ & 1/2 & -7   & -7   & -27  & -8   & -31  & -10  & -25  & $s'_{14}$        \\
6 & $(\rep{1},\rep{1})_0$     & $\rep{1}$  & $\rep{1}$ & $\rep{1}$ & -2   & -3  & -1  & 49  & 11  & -11 & 11  & -80 & $s_{15}$         \\
2 & $(\rep{1},\rep{1})_0$     & $\rep{1}$  & $\rep{1}$ & $\rep{1}$ & 0    & 3   & 57  & 19  & -19 & 19  & -19 & 60  & $s_{16}$         \\
2 & $(\rep{1},\rep{1})_0$     & $\rep{3}$ & $\rep{1}$ & $\rep{1}$ & 1/2  & 2   & -4  & -26 & 31  & -31 & -10 & -25 & $s'_{17}$         \\
2 & $(\rep{1},\rep{1})_0$     & $\rep{1}$  & $\rep{1}$ & $\rep{1}$ & -5/2 & -6  & 12  & 4   & -15 & 15  & 26  & 65  & $s_{18}$         \\
2 & $(\rep{1},\rep{1})_0$     & $\rep{3}$ & $\rep{1}$ & $\rep{1}$ & 1/2  & 2   & -4  & -26 & 31  & 8   & 33  & -25 & $s'_{19}$         \\
2 & $(\rep{1},\rep{1})_0$     & $\rep{1}$  & $\rep{1}$ & $\rep{1}$ & -5/2 & -6  & 12  & 4   & -15 & -24 & -17 & 65  & $s_{20}$         \\
2 & $(\rep{1},\rep{1})_0$     & $\rep{1}$  & $\rep{2}$ & $\rep{1}$ & -3/2 & 0   & -42 & -14 & -33 & -6  & 6   & 15  & $s_{21}$         \\
2 & $(\rep{1},\rep{1})_0$     & $\rep{1}$  & $\rep{1}$ & $\rep{1}$ & -3/2 & 0   & -42 & -14 & 43  & 35  & 6   & 15  & $s_{22}$         \\
2 & $(\rep{1},\rep{1})_0$     & $\rep{1}$  & $\rep{1}$ & $\rep{1}$ & -5/2 & 0   & 14  & -20 & -39 & 39  & 2   & 5   & $s_{23}$         \\
2 & $(\rep{1},\rep{1})_0$     & $\rep{1}$  & $\rep{1}$ & $\rep{2}$ & -3/2 & 0   & -42 & -14 & -33 & -6  & 6   & 15  & $s_{24}$         \\
2 & $(\rep{1},\rep{1})_0$     & $\rep{1}$  & $\rep{1}$ & $\rep{1}$ & -3/2 & 0   & -42 & -14 & 43  & -4  & -37 & 15  & $s_{25}$         \\
2 & $(\rep{1},\rep{1})_0$     & $\rep{1}$  & $\rep{1}$ & $\rep{1}$ & -5/2 & 0   & 14  & -20 & -39 & 0   & -41 & 5   & $s_{26}$        \\ \hline
\end{longtable}
}

\subsection{Hypercharge generator}
\label{sec:Ygenerator1}

The generator $Q_Y$ of $\U1_Y$ in $\mathcal{G}_{\text{SM}}\subset \mathcal{G}_{\text{4D}}$ 
is given by (cf.\ e.g.~\cite{Ramos-Sanchez:2008nwx,Vaudrevange:2008sm,Goodsell:2011wn})
\begin{equation}
Q_Y ~=~ \sum_{I=1}^{16} t^I_Y H_I\,,
\end{equation} 
in terms of the \SO{16}\x\SO{16} Cartan generators $H^I$. Frequently, the 16D vector $t_Y$ is also called 
$\U1_Y$ generator. For our six-Higgs model, we identify
\begin{equation}
t_Y=(\tfrac{1}{3}, 0, 0, 0, -\tfrac{1}{3}, -\tfrac{1}{2}, -\tfrac{1}{2}, \tfrac{1}{3}, 0, 0, 0, 0, 0, 0, 0, 0)
\end{equation}
with normalization $t_Y\cdot t_Y= \nicefrac{5}{6}$. Note that this normalization makes the hypercharge
compatible with \SU5 unification, as~\cite[eq.~(3)]{Ibanez:1993bd} $\sin^2\theta_W=1/(1+k_1)=\nicefrac38$ 
at the string scale, since the hypercharge level is given by~\cite[below eq.~(4)]{Ibanez:1993bd} $k_1=2\,t_Y\cdot t_Y= \nicefrac53$.
The hypercharge $q_Y$ of effective fields with gauge momenta $p\,\in\,\SO{16}\x\SO{16}$ is given by 
$q_Y = t_Y \cdot p$.

\section{Fermionic mass matrices of the six-Higgs-doublet model}
\label{ap:ferm_mass}

Let us provide some details on the computation of the fermionic mass eigenstates for quarks
and leptons prior to electroweak symmetry breakdown, due to the existence of exotics in the model.
To simplify the notation, we use $\braket{s_i}\to s_i$ for the singlet VEVs.
Given the mass matrix for the leptons in eq.~\eqref{mass_L} and the chosen VEV configuration
in eq.~\eqref{eq:VEVconfig}. The squared matrix is given by
\begin{align}
M_\ell^\intercal\,M_\ell = \begin{pmatrix} 0 & 0 & 0 & 0 & 0
           \\ 0 & 0 & 0 & 0 & 0
           \\ 0 & 0 & 0 & 0 & 0
           \\ 0 & 0 & 0 & c_1^2 + c_5^2 s_1^2 & c_1 c_2 s_4
           \\ 0 & 0 & 0 & c_1 c_2 s_4 & c_2^2 s_4^2 \end{pmatrix}\,.
\end{align} 
The corresponding unnormalized physical lepton states are given in terms
of the original stringy states by
\begin{equation}
\label{eq:lstates}
\begin{pmatrix}
\ell_{\mathrm{L},1}\\
\ell_{\mathrm{L},2}\\
\ell_{\mathrm{L},3}\\
\ell_{\mathrm{L},4}\\
\ell_{\mathrm{L},5}\\
\end{pmatrix}_\mathrm{phys} :=~
\begin{pmatrix}
 1 & 0 & 0 & 0 & 0 \\
 0 & 1 & 0 & 0 & 0 \\
 0 & 0 & 1 & 0 & 0 \\
 0 & 0 & 0 & w_- & 1 \\
 0 & 0 & 0 & w_+ & 1
\end{pmatrix}
\begin{pmatrix}
\ell_{\mathrm{L},1}\\
\ell_{\mathrm{L},21}\\
\ell_{\mathrm{L},22}\\
\ell_{\mathrm{L},3}\\
\ell_{\mathrm{L},4}\\
\end{pmatrix}\,,
\end{equation}
where
\begin{align}
w_\pm := \frac{c_1^2 + c_5^2 s_1^2 - c_2^2 s_4^2 \pm 
\sqrt{(c_1^2 + c_5^2 s_1^2)^2 + 2c_2^2 (c_1 - c_5 s_1)(c_1 + c_5 s_1)s_4^2 + c_2^4 s_4^4} }{2 c_1 c_2 s_4}\,.
\end{align}

We must perform an analogous procedure for down-quark singlets.
According to table~\ref{tab:tabmodel4902}, in our model there are seven 
left-handed down-quark singlets,
$\overline d := \left(\overline d_{\mathrm{L},1},\,\overline d_{\mathrm{L},21},\,\overline d_{\mathrm{L},22}
,\,\overline d_{\mathrm{L},23},\,\overline d_{\mathrm{L},24},\,\overline d_{\mathrm{L},3},\,\overline d_{\mathrm{L},4}\right)$,
and four left-handed conjugate states,
$d' := (d'_{\mathrm{L},11},\,d'_{\mathrm{L},12},\,d'_{\mathrm{L},2},\,d'_{\mathrm{L},3})^\intercal$.
Thus, we find the low-energy physical states by building the mass eigenstates.
Using the charges of these fields in table~\ref{tab:complete4902f}, 
we can find the resulting admissible couplings that lead to
\begin{equation}
\overline d\,M_d\,d'~\subset~\mathcal{L}
\,.
\end{equation}
At leading order, the down-quark mass matrix is given by
\begin{align}
M_d ~=
\begin{pmatrix}
 0 & 0 & 0 & 0 \\
 c^d_1 & c^d_2 & s_{15} & 0 \\
 c^d_3 & c^d_4 & s_{15} & 0 \\
 c^d_5 & c^d_6 & s_{15} & 0 \\
 c^d_7 & c^d_8 & s_{15} & 0 \\
 c^d_9 s_{16} & c^d_{10} s_{16} & 0 & c^d_{11} s_3\\
 0 & 0 & c^d_{12} s_2 & c^d_{13} 
\end{pmatrix}\,.
\label{Md}
\end{align}
As in the leptonic case, the coefficients $c^d_i$ are some coupling constants 
that are assumed to be real and order unity. In addition, to be consistent with 
the leptonic case, we take $s_{15}=0$ and this is why we omit the coupling 
constants of $s_{15} \overline{d}_{\mathrm{L},2i}d_{\mathrm{L},1j}$ in eq.~\eqref{Md}. 
It is easy to verify that the down-quark mass matrix has full rank 4. Hence, 
in the notation of table~\ref{tab:complete4902f}, the physical (massless) down-quark 
singlets are $\overline{d}_{\mathrm{L},1}$ and the combinations
\begin{equation}
\overline{d}_{\mathrm{L},24} - \frac{c^d_2 c^d_7 - c^d_1 c^d_8}{c^d_2 c^d_3 - c^d_1 c^d_4}
\overline{d}_{\mathrm{L},22} - \frac{c^d_4 c^d_7 - c^d_3 c^d_8}{c^d_1 c^d_4 - c^d_2 c^d_3}
\overline{d}_{\mathrm{L},21} \,,\quad 
\overline{d}_{\mathrm{L},23} - \frac{c^d_2 c^d_5 - c^d_1 c^d_6}{c^d_2 c^d_3 - c^d_1 c^d_4}
\overline{d}_{\mathrm{L},22} - \frac{c^d_4 c^d_5 - c^d_3 c^d_6}{c^d_1 c^d_4 - c^d_2 c^d_3}
\overline{d}_{\mathrm{L},21}.
\end{equation}

Now, we relabel the massless eigenstates as
\begin{equation}
\begin{split}
&\overline{d}_{\mathrm{L},1}\to \overline{d}_{\mathrm{L},1}\\
&\overline{d}_{\mathrm{L},24} - \frac{c^d_2 c^d_7 - c^d_1 c^d_8}{c^d_2 c^d_3 - c^d_1 c^d_4}
\overline{d}_{\mathrm{L},22} - \frac{c^d_4 c^d_7 - c^d_3 c^d_8}{c^d_1 c^d_4 - c^d_2 c^d_3}
\overline{d}_{\mathrm{L},21} \to \overline{d}_{\mathrm{L},2}\\
&\overline{d}_{\mathrm{L},23} - \frac{c^d_2 c^d_5 - c^d_1 c^d_6}{c^d_2 c^d_3 - c^d_1 c^d_4}
\overline{d}_{\mathrm{L},22} - \frac{c^d_4 c^d_5 - c^d_3 c^d_6}{c^d_1 c^d_4 - c^d_2 c^d_3}
\overline{d}_{\mathrm{L},21} \to \overline{d}_{\mathrm{L},3}\,.
\end{split}
\end{equation}

\section{Coefficients of the six-Higgs-doublet sector}
\label{ap:6H}
%

The quadratic terms of the neutral components in the Higgs-sector potential~\eqref{eq:VHcomponents} are
\begin{equation}
\label{eq:quadraticcharge}
\begin{split}
V_\phi(\sigma_{1I}, \sigma_{2I})~\supset~ 
& \theta_{IJ}\sigma_{1I}\sigma_{2J}^{*} + \chi_{IJ}\sigma_{1I}\sigma_{2J} + \kappa_{1,IJ}\sigma_{1I}\sigma_{1J} + \kappa_{2,IJ} \sigma_{2I}\sigma_{2J} \\
& + \omega_{1,IJ} \sigma_{1I} \sigma_{1J}^{*} + \left(\xi\delta_{IJ} + \omega_{2,IJ}\right) \sigma_{2I} \sigma_{2J}^{*} + \text{c.c.}\,,
\end{split}
\end{equation}
where $\delta_{IJ}$ is the usual Kronecker delta and
\begin{align}\label{eq:ctes}
\begin{split}
\theta_{IJ} &:=  \lambda_6 v_{1I} v_{1J} \,, \qquad \ \ 
\chi_{IJ} := \lambda_6 v_{1I} v_{1J} \,, \qquad \ \
\kappa_{1,IJ} := \lambda_1 v_{1I} v_{1J} \,, \\ 
\kappa_{2,IJ} &:= \lambda_5 v_{1I} v_{1J} \,, \qquad  
\omega_{1,IJ} := 2\lambda_1 v_{1I} v_{1J} \,, \qquad
\omega_{2,IJ} := \frac12\lambda_{4} v_{1I}v_{1J}  \,.
\end{split}
\end{align}
The entries of the mass matrix~\eqref{eq:Mphi}, which follow from 
the potential~\eqref{eq:quadraticcharge} in the limit of $\epsilon^2\sim0$, 
are given by
\begin{equation}\label{entries_Mphi}
\begin{split}
\Omega_1 &:= \frac{1}{2}\omega_{1,11} + \kappa_{1,11}\,,\qquad 
\Omega_2 := \frac{1}{2}\omega_{2,11} +\kappa_{2,11}\,,\\ 
\xi &:= \mu_{22}^2 + \frac12\lambda_3 v_{11}^2 \qquad\text{and} \qquad 
P := \theta_{11} + \chi_{11}\,.
\end{split}
\end{equation}
Here we assume the hierarchy $v_{11}\gg v_{12} \gg v_{13}$ via the parameter 
$\epsilon$, such that $\epsilon = v_{12}/v_{11} = v_{13}/v_{12}$, with 
$\epsilon^2\sim 0$.

\newpage
\section{Spectrum for the stringy 2HDM}
\label{appsect:spect2HDM}

\begin{longtable}{|c|l|c|c|c|c|c|c|c|c|c|c|c|c|} 
\caption{Spectrum of massless fermions in a stringy 2HDM. The first column corresponds 
to the multiplicity of the fields. The 4D gauge group is 
$\mathcal{G}_{\text{4D}} = \mathcal{G}_{\text{SM}} \x \mathcal{G}' \x \U1'^8$, 
where $\mathcal{G}_{\text{SM}} = \SU3_c\x \SU2_L\x \U1_Y$ and 
$\mathcal{G}'=\SU2_\mathrm{flavor}\x\SU3\x\SU2$. The first $\U1'$, associated with the 
charges $q_1$ is (pseudo-)anomalous. The first column with the label \SU2 displays the 
representations under $\SU2_\mathrm{flavor}$.\label{tab:fullf}} \\
\hline
\# & $\mathcal{G}_{\text{SM}}$ & \SU2 & \SU3 & \SU2 & $q_1$ & $q_2$ & $q_3$ & $q_4$ & $q_5$ & $q_6$ & $q_7$ & $q_8$ & \text{label} \\ \hline
\endfirsthead
\hline 
\# & $\mathcal{G}_{\text{SM}}$ & \SU2 & \SU3 & \SU2 & $q_1$ & $q_2$ & $q_3$ & $q_4$ & $q_5$ & $q_6$ & $q_7$ & $q_8$ & \text{label} \\ \hline
\endhead 
\hline \multicolumn{14}{|r|}{{continued...}} \\ 
\hline
\endfoot
\hline
\endlastfoot

1 & $(\rep{1},\rep{2})_{\nicefrac{-1}{2}}$  & $\rep{1}$ & $\rep{1}$  & $\rep{1}$ & -3 & 0  & -3 & -9  & 9   & -9  & -27  & 6 & $\ell_{\mathrm{L},3}$  \\
2 & $(\rep{1},\rep{2})_{\nicefrac{-1}{2}}$  & $\rep{1}$ & $\rep{1}$  & $\rep{1}$ & 2  & 0  & -3 & -9  & 9   & -9  & -27 & 6 & $\ell_{\mathrm{L},i}$   \\
1 & $(\rep{1},\rep{1})_1$       & $\rep{2}$ & $\rep{1}$  & $\rep{1}$ & 1  & 1  & 1  & 3   & -3  & 3   & 9    & -2  & 
$\overline{e}_{\mathrm{L}}$\\
1 & $(\rep{1},\rep{1})_1$       & $\rep{1}$ & $\rep{1}$  & $\rep{1}$ & 1  & -2 & 1  & 3   & -3  & 3   & 9    & -2  & $\overline{e}_{\mathrm{L},3}$     \\
1 & $(\rep{3},\rep{2})_{\nicefrac{1}{6}}$   & $\rep{2}$ & $\rep{1}$  & $\rep{1}$ & 1  & 1  & 1  & 3  & -3  & 3   & 9  & -2  & $q_{\mathrm{L}}$   \\
1 & $(\rep{3},\rep{2})_{\nicefrac{1}{6}}$   & $\rep{1}$ & $\rep{1}$  & $\rep{1}$ & 1  & -2 & 1  & 3   & -3  & 3   & 9    & -2  & $q_{\mathrm{L},3}$      \\
1 & $(\crep{3},\rep{1})_{\nicefrac{-2}{3}}$ & $\rep{2}$ & $\rep{1}$  & $\rep{1}$ & 1  & 1  & 1  & 3   & -3  & 3   & 9    & -2  & 
$\overline{u}_{\mathrm{L}}$\\
1 & $(\crep{3},\rep{1})_{\nicefrac{-2}{3}}$ & $\rep{1}$ & $\rep{1}$  & $\rep{1}$ & 1  & -2 & 1  & 3   & -3  & 3   & 9    & -2  & 
$\overline{u}_{\mathrm{L},3}$ \\
1 & $(\crep{3},\rep{1})_{\nicefrac{1}{3}}$  & $\rep{1}$ & $\rep{1}$  & $\rep{1}$ & -3 & 0  & -3 & -9  & 9   & -9  & -27  & 6   & 
$\overline{d}_{\mathrm{L},3}$     \\
4 & $(\crep{3},\rep{1})_{\nicefrac{1}{3}}$  & $\rep{1}$ & $\rep{1}$  & $\rep{1}$ & 2  & 0  & -3 & -9  & 9   & -9  & -27  & 6   & 
$\overline{d}_{\mathrm{L},1i}$    \\
2 & $(\rep{3},\rep{1})_{\nicefrac{-1}{3}}$  & $\rep{1}$ & $\rep{1}$  & $\rep{1}$ & -2 & 0  & 3  & 9   & -9  & 9   & 27   & -6  & $d'_{\mathrm{L},1i}$   \\
1 & $(\rep{1},\rep{1})_0$        & $\rep{1}$ & $\crep{3}$ & $\rep{2}$ & 1  & 0  & -9 & -27 & 27  & 3   & -23  & 2   & $\nu_{\mathrm{R},1}$     \\
1 & $(\rep{1},\rep{1})_0$        & $\rep{1}$ & $\rep{1}$  & $\rep{1}$ & 1  & 0  & -9 & 29  & -29 & -31 & -93  & 2   & $\nu_{\mathrm{R},2}$     \\
1 & $(\rep{1},\rep{1})_0$        & $\rep{1}$ & $\crep{3}$ & $\rep{1}$ & -1 & 0  & 9  & 27  & -27 & 27  & -47  & -2  & $\nu_{\mathrm{R},3}$      \\
1 & $(\rep{1},\rep{1})_0$        & $\rep{1}$ & $\rep{1}$  & $\rep{1}$ & -1 & 0  & 9  & 27  & -27 & -33 & 93   & -2  & $\nu_{\mathrm{R},4}$      \\
1 & $(\rep{1},\rep{1})_0$        & $\rep{1}$ & $\rep{1}$  & $\rep{1}$ & 1  & 0  & 3  & 9   & -9  & 9   & 27   & 22  & $\nu_{\mathrm{R},5}$      \\
1 & $(\rep{1},\rep{1})_0$        & $\rep{1}$ & $\crep{3}$ & $\rep{1}$ & 1  & 0  & 9  & -29 & 29  & -29 & -23  & 2   & $\nu_{\mathrm{R},6}$      \\
1 & $(\rep{1},\rep{1})_0$        & $\rep{1}$ & $\rep{1}$  & $\rep{2}$ & -1 & 0  & -9 & 29  & -29 & -1  & 93   & -2  & $\nu_{\mathrm{R},7}$      \\
1 & $(\rep{1},\rep{1})_0$        & $\rep{1}$ & $\rep{3}$  & $\rep{2}$ & 0  & 0  & -9 & -27 & -31 & 1   & 35   & 0   & $\nu_{\mathrm{R},8}$      \\
1 & $(\rep{1},\rep{1})_0$        & $\rep{1}$ & $\crep{3}$ & $\rep{2}$ & 0  & 0  & 9  & 27  & 31  & -1  & -35  & 0   & $\nu_{\mathrm{R},9}$      \\
1 & $(\rep{1},\rep{1})_0$        & $\rep{1}$ & $\rep{3}$  & $\rep{2}$ & 0  & 0  & -9 & 29  & 29  & 1   & 35   & 0   & $\nu_{\mathrm{R},10}$    \\
1 & $(\rep{1},\rep{1})_0$        & $\rep{1}$ & $\crep{3}$ & $\rep{2}$ & 0  & 0  & 9  & -29 & -29 & -1  & -35  & 0   & $\nu_{\mathrm{R},11}$     \\
4 & $(\rep{1},\rep{1})_0$        & $\rep{2}$ & $\rep{1}$  & $\rep{1}$ & 0  & -1 & 5  & 15  & -15 & 15  & 45   & -10 & $\nu'_{\mathrm{R},12,i}$     \\
2 & $(\rep{1},\rep{1})_0$        & $\rep{2}$ & $\rep{1}$  & $\rep{1}$ & 0  & 1  & -5 & -15 & 15  & -15 & -45  & 10  & $\nu'_{\mathrm{R},13,i}$     \\
2 & $(\rep{1},\rep{1})_0$        & $\rep{1}$ & $\rep{1}$  & $\rep{1}$ & 0  & 2  & 5  & 15  & -15 & 15  & 45   & -10 & $\nu_{\mathrm{R},14,i}$     \\
2 & $(\rep{1},\rep{1})_0$        & $\rep{1}$ & $\rep{1}$  & $\rep{1}$ & 2  & 0  & -6 & -18 & -11 & -19 & 39   & -11 & $\nu_{\mathrm{R},15,i}$     \\
2 & $(\rep{1},\rep{1})_0$        & $\rep{2}$ & $\rep{1}$  & $\rep{1}$ & 0  & -1 & 2  & 6   & 23  & 7   & -75  & 5   & $\nu'_{\mathrm{R},16,i}$     \\
2 & $(\rep{1},\rep{1})_0$        & $\rep{1}$ & $\rep{1}$  & $\rep{1}$ & 0  & 2  & 2  & 6   & 23  & 7   & -75  & 5   & $\nu_{\mathrm{R},17,i}$     \\
2 & $(\rep{1},\rep{1})_0$        & $\rep{1}$ & $\rep{1}$  & $\rep{2}$ & 2  & 0  & 9  & -1  & -28 & -2  & -6   & -1  & $\nu_{\mathrm{R},18,i}$     \\
2 & $(\rep{1},\rep{1})_0$        & $\rep{1}$ & $\rep{1}$  & $\rep{1}$ & 2  & 0  & -9 & 1   & -30 & 30  & -6   & -1  & $\nu_{\mathrm{R},19,i}$     \\
2 & $(\rep{1},\rep{1})_0$        & $\rep{1}$ & $\rep{1}$  & $\rep{1}$ & 2  & 0  & 9  & -1  & 30  & 30  & -6   & -1  & $\nu_{\mathrm{R},20,i}$     \\
2 & $(\rep{1},\rep{1})_0$        & $\rep{1}$ & $\rep{1}$  & $\rep{1}$ & 2  & 0  & 3  & 9   & 20  & 10  & -66  & -11 & $\nu_{\mathrm{R},21,i}$     \\
2 & $(\rep{1},\rep{1})_0$        & $\rep{2}$ & $\rep{1}$  & $\rep{1}$ & 0  & -1 & -7 & -21 & -8  & -22 & 30   & 5   & $\nu'_{\mathrm{R},22,i}$     \\
2 & $(\rep{1},\rep{1})_0$        & $\rep{1}$ & $\rep{1}$  & $\rep{1}$ & 0  & 2  & -7 & -21 & -8  & -22 & 30   & 5   & $\nu_{\mathrm{R},23,i}$     \\
2 & $(\rep{1},\rep{1})_0$        & $\rep{1}$ & $\rep{3}$  & $\rep{1}$ & 2  & 0  & 0  & 28  & 1   & -1  & 29   & -1  & $\nu_{\mathrm{R},24,i}$     \\
2 & $(\rep{1},\rep{1})_0$        & $\rep{1}$ & $\rep{1}$  & $\rep{1}$ & 2  & 0  & 0  & -28 & -1  & 1   & 99   & -1  & $\nu_{\mathrm{R},25,i}$    \\
2 & $(\rep{1},\rep{1})_0$        & $\rep{1}$ & $\rep{3}$  & $\rep{2}$ & 0  & 0  & -9 & 1   & -1  & 1   & 35   & 0   & $\nu_{\mathrm{R},26,i}$     \\
2 & $(\rep{1},\rep{1})_0$        & $\rep{1}$ & $\crep{3}$ & $\rep{2}$ & 0  & 0  & 9  & -1  & 1   & -1  & -35  & 0   & $\nu_{\mathrm{R},27,i}$     \\
2 & $(\rep{1},\rep{1})_0$        & $\rep{1}$ & $\rep{3}$  & $\rep{1}$  & 0  & 0  & 9  & -1  & 1   & -31 & 35   & 0   & $\nu_{\mathrm{R},28,i}$     \\
2 & $(\rep{1},\rep{1})_0$        & $\rep{1}$ & $\rep{1}$  & $\rep{1}$ & 0  & 0  & 9  & -1  & 1   & 29  & -105 & 0   & $\nu_{\mathrm{R},29,i}$     \\
2 & $(\rep{1},\rep{1})_0$        & $\rep{1}$ & $\crep{3}$ & $\rep{1}$ & 0  & 0  & -9 & 1   & -1  & 31  & -35  & 0   & $\nu_{\mathrm{R},30,i}$     \\
2 & $(\rep{1},\rep{1})_0$        & $\rep{1}$ & $\rep{1}$  & $\rep{1}$ & 0  & 0  & -9 & 1   & -1  & -29 & 105  & 0   & $\nu_{\mathrm{R},31,i}$     \\
2 & $(\rep{1},\rep{1})_0$        & $\rep{1}$ & $\rep{1}$  & $\rep{1}$ & 2  & 0  & -6 & 10  & 19  & -19 & 39   & -11 & $\nu_{\mathrm{R},32,i}$     \\
2 & $(\rep{1},\rep{1})_0$        & $\rep{2}$ & $\rep{1}$  & $\rep{1}$ & 0  & -1 & 2  & -22 & -7  & 7   & -75  & 5   & $\nu'_{\mathrm{R},33,i}$     \\
2 & $(\rep{1},\rep{1})_0$        & $\rep{1}$ & $\rep{1}$  & $\rep{1}$ & 0  & 2  & 2  & -22 & -7  & 7   & -75  & 5   & $\nu_{\mathrm{R},34,i}$     \\
2 & $(\rep{1},\rep{1})_0$        & $\rep{1}$ & $\rep{1}$  & $\rep{2}$ & 2  & 0  & 9  & 27  & 2   & -2  & -6   & -1  & $\nu_{\mathrm{R},35,i}$     \\
2 & $(\rep{1},\rep{1})_0$        & $\rep{1}$ & $\rep{1}$  & $\rep{1}$ & 2  & 0  & -9 & 29  & 0   & 30  & -6   & -1  & $\nu_{\mathrm{R},36,i}$     \\
2 & $(\rep{1},\rep{1})_0$        & $\rep{1}$ & $\rep{1}$  & $\rep{1}$ & 2  & 0  & 9  & -29 & 0   & 30  & -6   & -1  & $\nu_{\mathrm{R},37,i}$     \\
2 & $(\rep{1},\rep{1})_0$        & $\rep{1}$ & $\rep{1}$  & $\rep{1}$ & 2  & 0  & 3  & -19 & -10 & 10  & -66  & -11 & $\nu_{\mathrm{R},38,i}$    \\
2 & $(\rep{1},\rep{1})_0$        & $\rep{2}$ & $\rep{1}$  & $\rep{1}$ & 0  & -1 & -7 & 7   & 22  & -22 & 30   & 5   & $\nu'_{\mathrm{R},39,i}$     \\
2 & $(\rep{1},\rep{1})_0$        & $\rep{1}$ & $\rep{1}$  & $\rep{1}$ & 0  & 2  & -7 & 7   & 22  & -22 & 30   & 5   & $\nu_{\mathrm{R},40,i}$     \\
2 & $(\rep{1},\rep{1})_0$        & $\rep{1}$ & $\rep{3}$  & $\rep{1}$ & 2  & 0  & 0  & 0   & -29 & -1  & 29   & -1  & $\nu_{\mathrm{R},41,i}$     \\
2 & $(\rep{1},\rep{1})_0$        & $\rep{1}$ & $\rep{1}$  & $\rep{1}$ & 2  & 0  & 0  & 0   & 29  & 1   & 99   & -1  & $\nu_{\mathrm{R},42,i}$    \\ 
\end{longtable}

\begin{longtable}{|c|c|c|c|c|c|c|c|c|c|c|c|c|c|}
\caption{Massless scalar spectrum for a stringy 2HDM. The scalar singlet 
labeled as $S$ is chosen as our DM candidate. We use the same conventions 
as in table~\ref{tab:fullf}. \label{tab:fulls}}\\
\hline
\# & $\mathcal{G}_{\text{SM}}$ & \SU2 & \SU3 & \SU2 & $q_1$ & $q_2$ & $q_3$ & $q_4$ & $q_5$ & $q_6$ & $q_7$ & $q_8$ & \text{label} \\ \hline
\endfirsthead 
\hline 
\# & $\mathcal{G}_{\text{SM}}$ & \SU2 & \SU3 & \SU2 & $q_1$ & $q_2$ & $q_3$ & $q_4$ & $q_5$ & $q_6$ & $q_7$ & $q_8$ & \text{label} \\ \hline
\endhead 
\hline \multicolumn{14}{|r|}{{continued...}} \\ 
\hline
\endfoot
\hline
\endlastfoot

2  & $(\rep{1},\rep{2})_{\nicefrac{1}{2}}$  & $\rep{1}$ & $\rep{1}$  & $\rep{1}$ & -2 & -2 & -2  & -6  & 6   & -6  & -18 & 4   & $\phi_i$ \\
1  & $(\rep{1},\rep{1})_0$     & $\rep{1}$ & $\rep{1}$  & $\rep{1}$ & 1  & 0  & -18 & 2   & 56  & 4   & 12  & 2   & $S$  \\
2  & $(\crep{3},\rep{1})_{\nicefrac{1}{3}}$ & $\rep{2}$ & $\rep{1}$  & $\rep{1}$ & 2  & -1 & 2   & 6   & -6  & 6   & 18  & -4  & $x_i$  \\
2  & $(\rep{1},\rep{1})_0$    & $\rep{1}$ & $\rep{3}$  & $\rep{1}$ & 0  & 0  & 18  & -2  & 2   & -2  & -70 & 0   & $s_1$  \\
2  & $(\rep{1},\rep{1})_0$     & $\rep{1}$ & $\rep{1}$  & $\rep{1}$ & 0  & 0  & 0   & -56 & -60 & 0   & 0   & 0   & $s_2$  \\
4  & $(\rep{1},\rep{1})_0$     & $\rep{1}$ & $\rep{3}$  & $\rep{1}$ & 1  & 0  & 0   & -28 & 28  & 2   & -58 & 2   & $s_3$  \\
4  & $(\rep{1},\rep{1})_0$     & $\rep{1}$ & $\rep{1}$  & $\rep{2}$ & 1  & 0  & 0   & -28 & 28  & -28 & 12  & 2   & $s_4$  \\
4  & $(\rep{1},\rep{1})_0$     & $\rep{1}$ & $\rep{1}$  & $\rep{1}$ & -1 & 0  & 18  & 26  & -26 & -4  & -12 & -2  & $s_5$  \\
16 & $(\rep{1},\rep{1})_0$     & $\rep{1}$ & $\rep{1}$  & $\rep{1}$ & 0  & 0  & 0   & -28 & -30 & 0   & 0   & 0   & $s_6$ \\
1  & $(\rep{1},\rep{1})_0$     & $\rep{1}$ & $\rep{3}$  & $\rep{1}$ & 1  & 0  & 0   & -56 & -2  & 2   & -58 & 2   & $s_7$  \\ 
1  & $(\rep{1},\rep{1})_0$     & $\rep{1}$ & $\rep{1}$  & $\rep{2}$ & -1 & 0  & 0   & 56  & 2   & 28  & -12 & -2  & $s_8$ \\
1  & $(\rep{1},\rep{1})_0$     & $\rep{1}$ & $\rep{1}$  & $\rep{2}$ & 1  & 0  & 0   & 0   & 58  & -28 & 12  & 2   & $s_{9}$ \\
1  & $(\rep{1},\rep{1})_0$     & $\rep{1}$ & $\crep{3}$ & $\rep{1}$ & -1 & 0  & 0   & 0   & -58 & -2  & 58  & -2  & $s_{10}$ \\
1  & $(\rep{1},\rep{1})_0$     & $\rep{1}$ & $\rep{1}$  & $\rep{1}$ & -1 & 0  & 18  & 54  & 4   & -4  & -12 & -2  & $s_{11}$ \\
2  & $(\rep{1},\rep{1})_0$     & $\rep{1}$ & $\rep{1}$  & $\rep{1}$ & 2  & 0  & 3   & 9   & 20  & 10  & -66 & -11 & $s_{12}$ \\
2  & $(\rep{1},\rep{1})_0$     & $\rep{2}$ & $\rep{1}$  & $\rep{1}$ & 0  & -1 & -7  & -21 & -8  & -22 & 30  & 5   & $s'_{13}$ \\
2  & $(\rep{1},\rep{1})_0$     & $\rep{1}$ & $\rep{1}$  & $\rep{1}$ & 0  & 2  & -7  & -21 & -8  & -22 & 30  & 5   & $s_{14}$ \\
2  & $(\rep{1},\rep{1})_0$     & $\rep{1}$ & $\rep{3}$  & $\rep{1}$ & 2  & 0  & 0   & 28  & 1   & -1  & 29  & -1  & $s_{15}$ \\
2  & $(\rep{1},\rep{1})_0$     & $\rep{1}$ & $\rep{1}$  & $\rep{1}$ & 2  & 0  & 0   & -28 & -1  & 1   & 99  & -1  & $s_{16}$ \\
2  & $(\rep{1},\rep{1})_0$     & $\rep{1}$ & $\rep{1}$  & $\rep{1}$ & 2  & 0  & -6  & -18 & -11 & -19 & 39  & -11 & $s_{17}$ \\
2  & $(\rep{1},\rep{1})_0$     & $\rep{2}$ & $\rep{1}$  & $\rep{1}$ & 0  & -1 & 2   & 6   & 23  & 7   & -75 & 5   & $s'_{18}$ \\
2  & $(\rep{1},\rep{1})_0$     & $\rep{1}$ & $\rep{1}$  & $\rep{1}$ & 0  & 2  & 2   & 6   & 23  & 7   & -75 & 5   & $s_{19}$ \\
2  & $(\rep{1},\rep{1})_0$     &$\rep{1}$ & $\rep{1}$  & $\rep{2}$ & 2  & 0  & 9   & -1  & -28 & -2  & -6  & -1  & $s_{20}$ \\
2  & $(\rep{1},\rep{1})_0$     & $\rep{1}$ & $\rep{1}$  & $\rep{1}$ & 2  & 0  & -9  & 1   & -30 & 30  & -6  & -1  & $s_{21}$ \\
2  & $(\rep{1},\rep{1})_0$     & $\rep{1}$ & $\rep{1}$  & $\rep{1}$ & 2  & 0  & 9   & -1  & 30  & 30  & -6  & -1  & $s_{22}$ \\
2  & $(\rep{1},\rep{1})_0$     & $\rep{1}$ & $\crep{3}$ & $\rep{1}$ & -2 & 0  & 0   & 0   & 29  & 1   & -29 & 1   & $s_{23}$ \\
2  & $(\rep{1},\rep{1})_0$     & $\rep{1}$ & $\rep{1}$  & $\rep{1}$ & -2 & 0  & 0   & 0   & -29 & -1  & -99 & 1   & $s_{24}$ \\
2  & $(\rep{1},\rep{1})_0$     & $\rep{1}$ & $\rep{1}$  & $\rep{1}$ & 0  & -2 & 7   & -7  & -22 & 22  & -30 & -5  & $s_{25}$ \\
2  & $(\rep{1},\rep{1})_0$     & $\rep{2}$ & $\rep{1}$  & $\rep{1}$ & 0  & 1  & 7   & -7  & -22 & 22  & -30 & -5  & $s'_{26}$ \\
2  & $(\rep{1},\rep{1})_0$     & $\rep{1}$ & $\rep{1}$  & $\rep{1}$ & -2 & 0  & -3  & 19  & 10  & -10 & 66  & 11  & $s_{27}$ \\
2  & $(\rep{1},\rep{1})_0$     & $\rep{1}$ & $\rep{1}$  & $\rep{2}$ & -2 & 0  & -9  & -27 & -2  & 2   & 6   & 1   & $s_{28}$ \\
2  & $(\rep{1},\rep{1})_0$     & $\rep{1}$ & $\rep{1}$  & $\rep{1}$ & -2 & 0  & -9  & 29  & 0   & -30 & 6   & 1   & $s_{29}$ \\
2  & $(\rep{1},\rep{1})_0$     & $\rep{1}$ & $\rep{1}$  & $\rep{1}$ & -2 & 0  & 9   & -29 & 0   & -30 & 6   & 1   & $s_{30}$ \\
2  & $(\rep{1},\rep{1})_0$     & $\rep{1}$ & $\rep{1}$  & $\rep{1}$ & 0  & -2 & -2  & 22  & 7   & -7  & 75  & -5  & $s_{31}$ \\
2  & $(\rep{1},\rep{1})_0$     & $\rep{2}$ & $\rep{1}$  & $\rep{1}$ & 0  & 1  & -2  & 22  & 7   & -7  & 75  & -5  & $s'_{32}$ \\
2  & $(\rep{1},\rep{1})_0$     & $\rep{1}$ & $\rep{1}$  & $\rep{1}$ & -2 & 0  & 6   & -10 & -19 & 19  & -39 & 11  & $s_{33}$ \\
\end{longtable}

\subsection{Hypercharge generator}

As in appendix~\ref{sec:Ygenerator1}, the so-called $\U1_Y$ generator $t_Y$ for the two-Higgs-doublet 
model is given by 
\begin{equation}
t_Y=(-\tfrac{1}{3}, \tfrac{1}{2}, -\tfrac{1}{3}, 0, 0, -\tfrac{1}{3}, \tfrac{1}{2}, 0, 0, 0, 0, 0, 0, 0, 0, 0)\,,
\end{equation}
with normalization $t_Y\cdot t_Y= \frac{5}{6}$, implying that $\sin^2\theta_W = \nicefrac38$~\cite{Ibanez:1993bd}
at the string scale, as discussed in appendix~\ref{sec:Ygenerator1}.
The hypercharge $q_Y$ of matter fields with gauge momenta $p\,\in\,\SO{16}\x\SO{16}$ is given by $q_Y = t_Y \cdot p$.

\section{Down-quark mass matrix of the stringy 2HDM}
\label{app:Md2HDM}

Let us provide the massless physical states for the down-quark singlets. In 
table~\ref{tab:spectrumSM-m5115} we notice that we have five left-handed 
down-quark singlets and two left-handed conjugate partners. Using the charges 
in table~\ref{tab:fullf} for these fields, we construct the down-quark mass 
matrix $M_d$ from the invariant terms in $\overline d\,M_d\,d' + \text{h.c.}\,\subset\,\mathcal L$, 
where  $\overline d := \left(\overline d_{\mathrm{L},3},\,\overline d_{\mathrm{L},11},\,\overline d_{\mathrm{L},12},\,\overline d_{\mathrm{L},13},\,\overline d_{\mathrm{L},14}\right)$ and $d' := (d'_{\mathrm{L},11},\,d'_{\mathrm{L},12})^\intercal$. 
We find that
\begin{align}
M_d^\intercal ~=
\begin{pmatrix}
 0 & c_1 & c_3 & c_5 & c_7\\
 0 & c_2 & c_4 & c_6 & c_8
\end{pmatrix}\,
\label{Md0}
\end{align}
has (full) rank two. The $c_i$ are some coupling constants of order unity that 
are assumed to be real. The squared mass matrix $M_d M_d^\intercal$ encodes 
the three massless physical down-quark states through the following linear combinations
\begin{equation}
\begin{split}
&(\overline d_{\mathrm{L},1})_{\text{phys}} ~=~ \alpha_1 \overline d_{\mathrm{L},11} + \alpha_2 \overline d_{\mathrm{L},12} + \overline d_{\mathrm{L},14}\,,\\
&(\overline d_{\mathrm{L},2})_{\text{phys}} ~=~ \alpha_3 \overline d_{\mathrm{L},11} + \alpha_4 \overline d_{\mathrm{L},12} + \overline d_{\mathrm{L},13}\,,\\
&(\overline d_{\mathrm{L},3})_{\text{phys}} ~=~ \overline d_{\mathrm{L},3}\,,
\end{split}
\end{equation}
where $\alpha_1=\frac{c_4 c_7-c_3 c_8}{c_2 c_3-c_1 c_4}$,  $\alpha_2=\frac{c_1 c_8-c_2 c_7}{c_2 c_3-c_1 c_4}$,  $\alpha_3=\frac{c_4 c_5-c_3 c_6}{c_2 c_3-c_1 c_4}$ and $\alpha_4=\frac{c_1 c_6-c_2 c_5}{c_2 c_3-c_1 c_4}$.
Note that the third down-quark singlet $\overline d_{\mathrm{L},3}$ 
is an eigenstate and can be identified with the third generation
down quark, as implied by the Yukawa couplings in eq.~\eqref{eq:VYuk2}.

\end{appendix}

\newpage
{\small

\providecommand{\bysame}{\leavevmode\hbox to3em{\hrulefill}\thinspace}
\frenchspacing
\newcommand{\origttfamily}{}
\let\origttfamily=\ttfamily
\renewcommand{\ttfamily}{\origttfamily \hyphenchar\font=`\-}

}

\end{document}